\documentclass{aa}  
\usepackage{graphicx,amssymb,amsmath}
\usepackage{natbib}
\usepackage{verbatim} 
\usepackage{rotating}
\usepackage[table]{xcolor}
\usepackage{multicol}
\usepackage{txfonts}
\usepackage{float}  
\begin{document}
   \title{Merging Galaxies in Isolated Environments II.}
   \subtitle{Evolution of Star Formation and Accretion Activity during the Merging Process}

   \author{Paula Calder\'on-Castillo\inst{1}
          \and
        Rory Smith \inst{1}
          }
   \offprints{Paula Calder\'on-Castillo}

   \institute{Departamento de Física, Universidad Técnica Federico Santa María, Avenida Vicuña Mackenna 3939, San Joaquín, Santiago de Chile \\ 
              \email{pau.astro.cc@gmail.com}
             }
   \date{Received ; Accepted }

  \abstract
   {It is now well known that certain massive galaxies undergo enormous enhancements in their star formation rate (SFR) 
   when they undergo major mergers, as high as 100 times the SFR of unperturbed galaxies of the same stellar mass. 
   Previous works found that the size of this boost in star formation (SF) is  
   related to the morphology of and the proximity to the companion. The same trend has also been observed for active galactic nuclei (AGN) 
   fraction, where galaxies which are closer together tend to have higher AGN fractions. 
   }
   {We aim to analyse the SF enhancement and AGN fraction evolution during the merger process, using a more 
   timeline-like merger sequence. Additionally, we aim to determine the relation between the SF enhancement in 
   mergers and the morphology of the galaxies involved. 
   }
   {Taking advantage of the stellar masses (M$_*$) and SFR of $\sim$600 nearby isolated mergers 
   obtained in our previous study, we calculate the distance of each of our galaxies from the star-forming 
   main sequence (MS, sSFR/sSFR$_{MS}$), which we
   refer to as the SF mode. We then analyse how the SF mode varies during the merger process, 
   as a function of morphology and M$_*$. Additionally, we analyse the AGN content of our mergers, 
   using multiple diagnostics based on emission line ratios and WISE colours. 
   }
   {We observe that, overall, merging galaxies show a SF mode that is governed by their morphology. 
   Spirals typically show high SF mode values while highly-disturbed (HD) galaxies are generally even more enhanced 
   (median values of +0.8~dex and +1.0~dex above the MS, respectively). 
   On the contrary, elliptical and lenticular galaxies show the lowest SF modes, as expected. However, 
   even they show SF enhancement compared to their unperturbed counterparts. For example, their median SF mode 
   is just within the 1-sigma scatter of the MS, and this can occur even before the galaxies have coalesced. 
   We see a trend for SF mode to gradually increase with increasing merger stage. We do not find a clear dependency of the observed AGN fraction on merger stage for the majority of our classification methods. 
} 
   { We find mergers can significantly enhance SF in galaxies of all morphologies. For early-type galaxies, this could 
   suggest that some gas was present prior to the merger which may be triggered to form stars by the tidal interaction. 
   As the SF enhancement continues throughout the merger process, this suggests that the enhancement may be a 
   long-lived event, contrary to the short starbursts seen in some models. 
 }

   \maketitle
%
\section{Introduction}

In the hierarchical scenario, galaxies evolve and grow in mass as they interact and merge with other galaxies. 
Spiral galaxies can merge to form larger ellipticals \citep{TnT72} or larger spirals \citep{SpringelnHernquist05}. 
These transformations also take place in parallel with changes in colour, 
stellar mass (M$_*$), star formation rate (SFR), dust and gas content, and also super-massive black hole 
(SMBH) activity. Galaxies experience all of these changes throughout the merging process. 

As galaxies approach each other, their gas start to be disturbed and disrupted, and may even be transferred to their 
companion, creating tidal tails and bridges. Additionally, gas is funneled towards their centers, which can trigger star formation (SF) and can also activate their SMBHs. 
Contrasting conclusions have been drawn from different studies (observations and simulations) of merging galaxies. Some show that the merging process could affect the SFR in a minimal manner or could even suppress it \citep{Pearsonetal19, Renaudetal22, Lietal23}. Others show that the SFR can be strongly enhanced in both close and separated merging galaxies \citep{LarsonnTinsley78, Ellisonetal10, Scudderetal12, Pattonetal13, Daviesetal15, Pattonetal16, Gardunoetal21, Morenoetal21}. 
These results highlight the difficulty and complexity in studying galaxies that are going through the merger process as their starting properties are likely unequal and hard to determine. 
Nonetheless, the SFR enhancement is found to depend on the stellar mass ratio 
between the merging galaxies, the bulge to total luminosity ratio, the gas and dust content, the orbital parameters of the collision, and the occurrence of active galactic nuclei (AGN) feedback
\citep{MihosnHernquist94a, MihosnHernquist94b, DiMatteoetal08, Hopkinsetal08, DiMatteoetal05, Scudderetal15, Parketal17, Morenoetal19, ByrneMamahitetal23}. 

The star formation (SF) enhancement is usually studied using the SFR-M$_*$ plane as a reference point, as multiwavelength 
observations have shown 
that in this plane most galaxies form a sequence that evolves with redshift 
\citep{Elbazetal07, Noeskeetal07, Daddietal07, Pannellaetal09, Pannellaetal15, Karimetal11, Schreiberetal15, 
Schreiberetal17, Wuytsetal11, Rodighieroetal14, Whitakeretal12, Whitakeretal14, RenzininPeng15}, \citep[hereafter \textsc{Chang+15}]{Changetal15}, 
the so-called main sequence (MS) of star formation. 
This relation is followed mostly by galaxies that form stars in a secular way. Starburst galaxies are outliers from the 
sequence with an excess SFR for a given M$_*$, and these galaxies are mainly very disturbed objects undergoing 
violent merger processes. Their SFR can rise up to 10-100 times the SFR of unperturbed galaxies with the same M$_*$
\citep{Sandersetal96, Engeletal10, Elbazetal11, Rodighieroetal11, Schreiberetal15}. We aim to identify when this SF enhancement occurs 
during the merger process by using a more chronological approach as compared to many previous observational studies. 
Such studies have already shown us that SF enhancement is seen when interacting pairs have small projected 
separations. Similarly they also show that the AGN fraction increases with smaller projected separations 
\citep{Ellisonetal08, Dargetal10, Ellisonetal11, Ellisonetal13}. 
The SF enhancement and AGN fraction have also been linked with the morphology of the interacting galaxies. 
\citet{Hwangetal10} and \citet{Hwangetal11} show that luminous infrared galaxies (LIRGs) with a late-type 
companion show values of SFR, AGN fraction, and IR luminosity that are enhanced, and further enhanced when 
the two galaxies are closer together. Meanwhile, LIRGs with an early-type companion show little dependence on separation distance for 
AGN fraction and IR luminosity, and low values for SFR. Thus, galaxies can be highly affected by their close 
companion and the effects caused on the galaxy are likely to be related to the morphology of, and distance to, 
the companion.

To study the M$_*$ and SFR of mergers we have taken advantage of a new catalog of nearby (z$<$ 0.1) isolated mergers 
presented in our previous work \citep[hereafter Paper I]{PauCC24}. 
This catalogue includes photometric data, morphological and merger stage classification, as well as physical 
properties, such as M$_*$, SFR, and dust masses of nearby isolated mergers. 
These properties have been estimated using the spectral energy distribution (SED) fitting code 
MAGPHYS \citep{daCunhaetal2008}, 
using our own photometry measured on Galaxy Evolution Explorer (GALEX), Sloan Digital Sky Surveys 
(SDSS), and Wide-field Infrared Survey Explorer (WISE) imaging. This new photometry includes 
the entire light of the merging galaxy. This is in contrast to automated photometric catalogues which are not suited 
to dealing with disturbed galaxies, and often 
exclude tidal tails and star-forming regions, as these are very difficult to capture in an automated manner. 

As we mentioned above, previous studies of mergers mainly compare the properties of the merging galaxies to 
the projected distance between them. In order to analyse the merging galaxies properties throughout the merging process, 
we study this process using a more timeline-like merger sequence. We based our merger sequence in the 
classification shown by \citet{Veilleuxetal02}, plus five additional sub-classes described in our previous work and 
summarised in this study. Additionally, we analyse the relation of the SF enhancement to the morphology of both 
merging components. Finally, we identify the AGNs in our sample and study how the AGN fraction relates to the 
merger stage. 

Throughout this paper, we refer to merging galaxies as `galaxies', and for unperturbed galaxies, we specify that 
we are referring to `unperturbed galaxies'. 
The data and the methodology used to determined the properties used in this study are summarised in 
Sec. \ref{DATA}. The analysis on the SF enhancement and AGN fraction during the merger sequence can 
be found in Sec. \ref{RESULTS}. We discuss our results and present our final conclusions in Sec. 
\ref{CONCLUSIONS}. 
Throughout this paper we adopt a standard cosmology with $\rm \Omega_m=0.3$ and 
$\rm H_0=72 ~km~s^{-1}Mpc^{-1}$.

\section{Data}  \label{DATA}

\subsection{The sample} \label{sec:Sample}

The merger sample has been assembled using five parent samples that include mainly mergers: 
the Arp's Catalog of Peculiar Galaxies\footnote{\url{http://arpgalaxy.com}} \citep[ARP Galaxies]{Arp96}; 
the VV\footnote{\url{www.sai.msu.su/sn/vv}} Catalogue of Interacting Galaxies 
\citep{VVetal01}; 
the mergers classified by Nagar (private communication); 
the mergers classified by citizen scientists in the Galaxy Zoo (GZ) 
Project\footnote{\url{http://data.galaxyzoo.org}} \citep[GZ mergers]{Holinchecketal16}; 
and the mergers selected from the Great Observatories All-sky LIRG 
Survey\footnote{\url{Http://goals.ipac.caltech.edu}} \citep[GOALS]{Sandersetal03}. 
From this set of catalogues, we have only selected mergers that are isolated, showing only one or two 
components, and that do not belong to either a group or cluster. In the presence of two components, 
we have restricted our selection to systems where their 
respective velocities are $\Delta \rm v_{rel} < 500~ km/s$ ($\Delta z$ < 0.002), 
to try to exclude fly-bys or galaxies that only appear close by projection effects. In this way, we try to select 
galaxy interactions that are more likely to eventually coalesce. 
We also restricted our sample by only including mergers with available imaging in all of the following filters: 
FUV and NUV from GALEX; u, g, r, i, and z from SDSS; and W1, W2, W3, and W4 from WISE. 
Applying the above constraints results in a final sample of 540 isolated mergers (919 merging galaxies in total). 
The redshift has a range up to z = 0.1. 
For more details please refer to Sec. 2 of Paper I.

The analysis presented here aims to understand the relation between the merger stage and the morphology of the 
components. We classified galaxies in four different types: spiral, elliptical, lenticular, and, in cases where none of 
those classes were applied, the galaxies were labeled as highly-disturbed (HD). This latter classification (HD) was 
also sub-classified either as late- or early-type. HD galaxies with gas (visible in their SDSS images as streams, 
tidal tails, and/or bridges and blue colour) are considered late-type galaxies. These dominate the HD sample by number. 
On the other hand, HD galaxies without gas (visible in the SDSS images as shells and red colour) are considered as 
early-type galaxies. 

\begin{figure*}[!h]
\begin{center} 
  \includegraphics[width=\textwidth,clip]{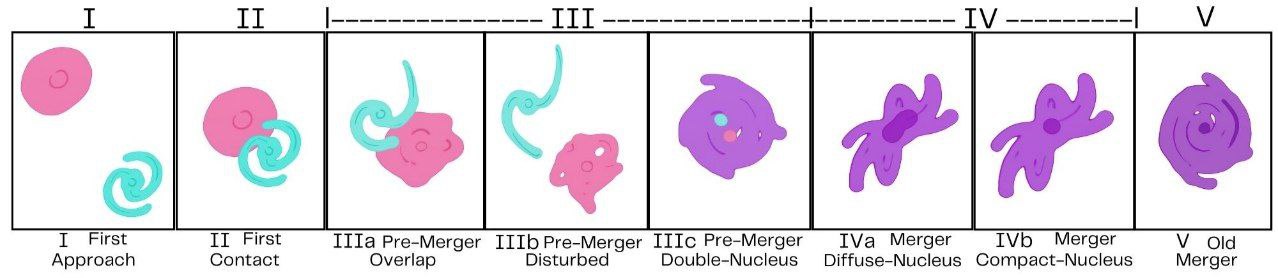}
   \caption{ Cartoon representing the Merger Stages classification used in this work. I (First Approach): the two galaxies are clearly 
separated but approaching to each other, following our $\Delta \rm v_{rel}$ condition. 
II (First Contact): the two galaxies are overlapping with 
no strong perturbations. IIIa (Pre-Merger Overlap): the two galaxies are overlapping and showing 
strong perturbations. IIIb (Pre-Merger Disturbed): the two galaxies are clearly separated but showing 
strong perturbations. IIIc (Pre-Merger Double-Nucleus): there is only one very perturbed object clearly 
showing two nuclei. IVa (Merger Diffuse-Nucleus): one very perturbed object with a very diffuse nucleus. 
IVb (Merger Compact-Nucleus): one very perturbed object with a very luminous and compact nucleus. 
V (Old Merger): one galaxy with no strong perturbations but disturbed central morphology. 
}  
\label{fig:MrgStages}
\end{center}
\end{figure*}

The merger stages are based on the merger sequence shown by \citet{Veilleuxetal02}, 
plus additional sub-classes that we introduced in Paper I. For a full description of the merger stage see the previous work. However, a summary can be found in Figure \ref{fig:MrgStages} and its caption. 
We have adopted eight different merger stages: I (First Approach): the two galaxies are clearly 
separated but approaching to each other, following our $\Delta \rm v_{rel}$ condition. 
II (First Contact): the two galaxies are overlapping with 
no strong perturbations. IIIa (Pre-Merger overlap): the two galaxies are overlapping and showing 
strong perturbations. IIIb (Pre-Merger disturbed): the two galaxies are clearly separated but showing 
strong perturbations. IIIc (Pre-Merger double-nucleus): there is only one very perturbed object clearly 
showing two nuclei. IVa (Merger diffuse-nucleus): one very perturbed object with a very diffuse nucleus. 
IVb (Merger compact-nucleus): one very perturbed object with a very luminous and compact nucleus. 
V (Old Merger): one galaxy with no strong perturbations but disturbed central morphology. 
Cartoon representation of the merger stage classification can be seen in Fig. \ref{fig:MrgStages}. For representative images of the morphology please see Figs. A1 of Paper I. 

\subsection{Merging galaxies properties}

In order to study the star-forming properties of mergers, we have estimated the M$_*$ and SFR of these 
objects using the publicly available code 
MAGPHYS\footnote{\url{http://www.iap.fr/magphys/}} \citep{daCunhaetal2008}. 
This code fits the spectral energy distribution (SED) of a galaxy based on photometric data, which can 
range from UV to submm. We have chosen the following surveys as they span a large range in wavelength, 
tracing young stars (GALEX), old stars (SDSS), and obscuration of young stars by dust (WISE). 
This allows us to determine $\rm M_*$ and SFR more accurately, taking advantage of the large, wide-field surveys 
that are publicly available. 
We have fitted SEDs to our entire sample, utilizing our own photometric measurements, 
that were measured using a semi-automated method. The semi-automated method allowed us to extract the entire 
merging galaxy's light, including faint tidal tails, and bright 
star-forming regions, which might otherwise be considered as distinct objects by a fully-automated method. 
This procedure was necessary since the automated measurements catalogued by the aforementioned 
surveys are found to be not reliable enough for merging galaxies, often showing 
systematically lower fluxes compared to our measured values (see Paper I for more details).

\section{Results} \label{RESULTS}

\subsection{Stellar masses}

\begin{figure}[!h]
\begin{center} 
  \includegraphics[width=0.5\textwidth,clip]{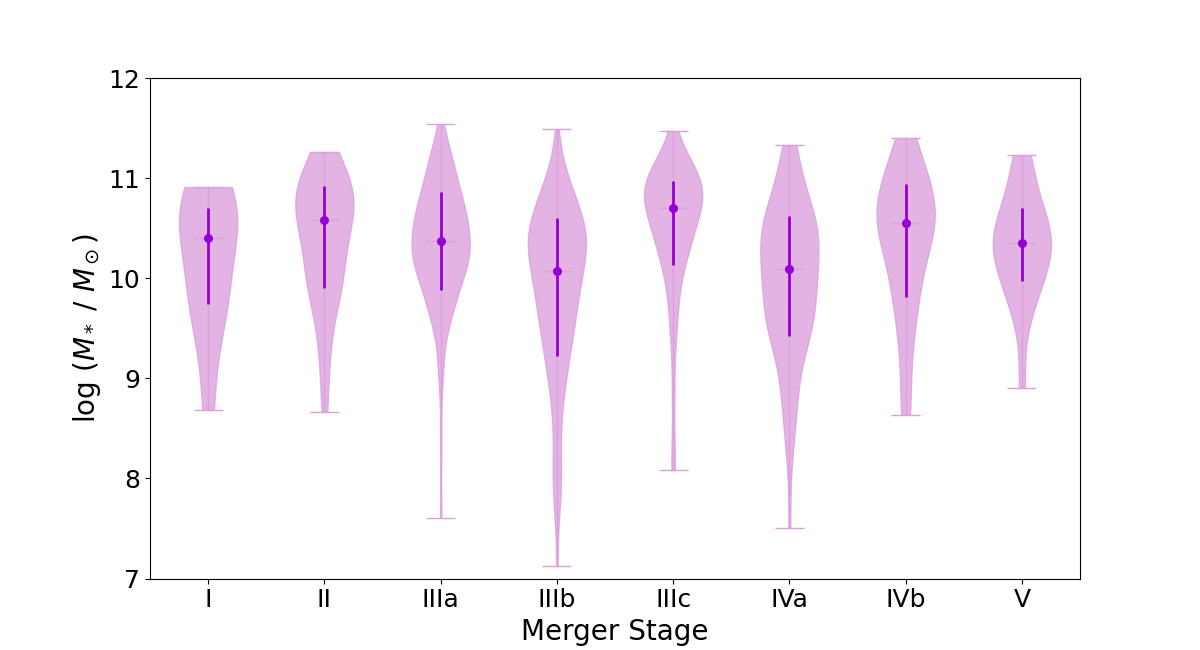}
   \caption{ Violin-plot of the stellar masses according to merger stage. \textbf{The shaded regions show the smoothed normalised number distribution.} The filled circles and bold lines indicate the median and one-sigma percentile, respectively.
}  
\label{fig:Mstellar_MrgStg_dist}
\end{center}
\end{figure} 

In order to look for biases in our study, we first considered the distribution of stellar masses 
separated by merger stages to search for any dependence. 
Fig. \ref{fig:Mstellar_MrgStg_dist} shows the M$_*$ distributions for each 
merger stage as a 'violin plot', where the filled circle indicates the median and the bold lines show the percentile including 34\% of objects from each side of the median (equivalent to one-sigma).

It seems there is a small decline in stellar mass related to the merger stage. However, the difference between 
the median M$_*$ of the merger stage showing the highest median M$_*$ (Merger Stage IIIc) and the last merger stage (Merger Stage V) is 
less than 0.5 dex, which can be considered negligible for our present study, and is smaller than the error bars on each data point. 
Thus, we can assume that there is no significant bias of stellar mass as a function of merger stage in our study. 

\begin{figure}[!h]
\begin{center} 
  \includegraphics[width=0.5\textwidth,clip]{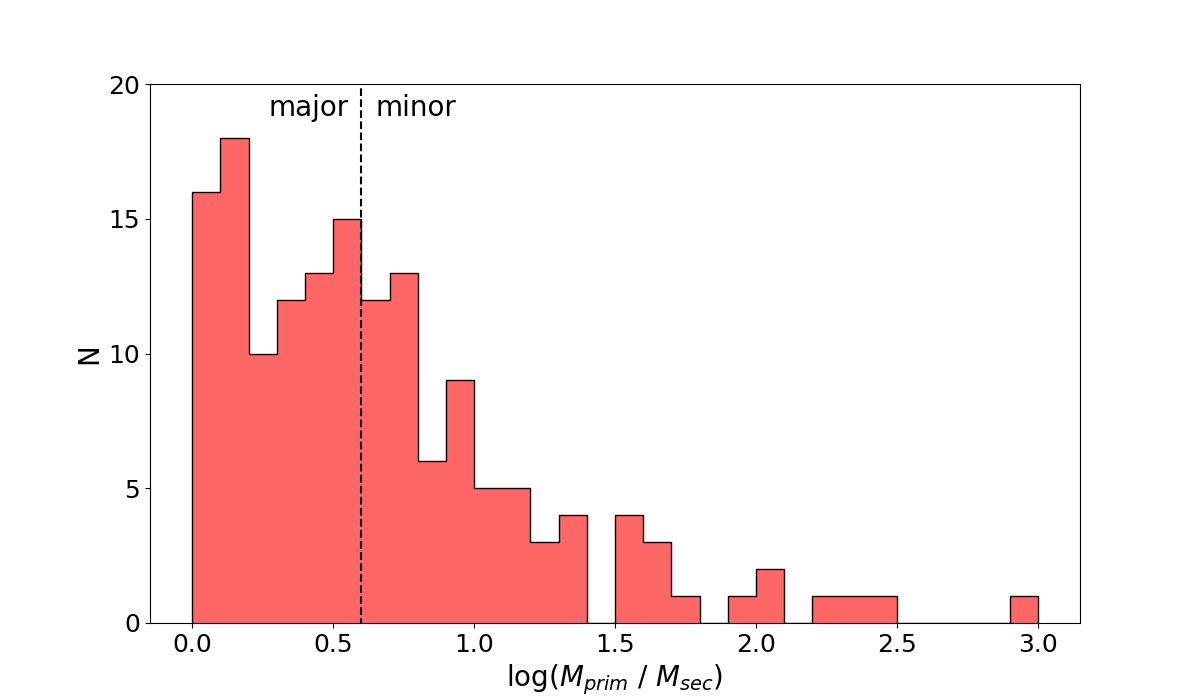}
   \caption{ Stellar mass ratio between the primary and the secondary component of the mergers 
   showing two separated components.
}  
\label{fig:Mstellar_ratio}
\end{center}
\end{figure} 

For the SF mode analysis, we have separated our sample according to the stellar mass ratio between 
the interacting pairs, since the difference in stellar mass of the components can affect the resulting 
SF enhancement \citep{DiMatteoetal08, Hopkinsetal08, Parketal17}. We have adopted the definition of 
major merger when a system shows a stellar mass ratio up to 1:4 
\citep{RodriguezGomezetal15, Weinzirl15, vandeVoortetal18}. 
Figure \ref{fig:Mstellar_ratio} shows the stellar mass ratio distribution between the 
primary (most massive) and the secondary component. The mergers shown in this figure are 
the systems where both components are separated enough to be detected as different objects by 
SExtractor (not overlapping), thus have separated photometric measurements and individual M$_*$ and SFR. 
The dashed line shows the separation between major and minor mergers.
The effect on the SF mode caused by the stellar mass ratio of the components is presented in 
Sec. \ref{sec:SFmode_Morphdep}. 

We have also separated the sample into different stellar mass bins in order to study how the stellar mass 
of a merger affects our results. For this, we have defined three stellar mass bins. 
The low-M$_*$ bin for stellar masses lower than log(M$_*$/M$_{\odot}) <$ 9.5. 
The medium-M$_*$ bin for stellar masses between  9.5 $<$ log(M$_*$/M$_{\odot}) <$ 10.5. 
The high-M$_*$ bin for stellar masses higher than log(M$_*$/M$_{\odot}) >$ 10.5. 
Since there are no clear dependencies on our results caused by the difference in stellar mass of the merger, 
we present these results in the App. \ref{App:Mstellarbins}.

\subsection{Evolution during the merger process: movement in the main sequence} \label{sec:SFmode}

\begin{figure}[!h]
\begin{center} 
  \includegraphics[width=0.5\textwidth,clip]{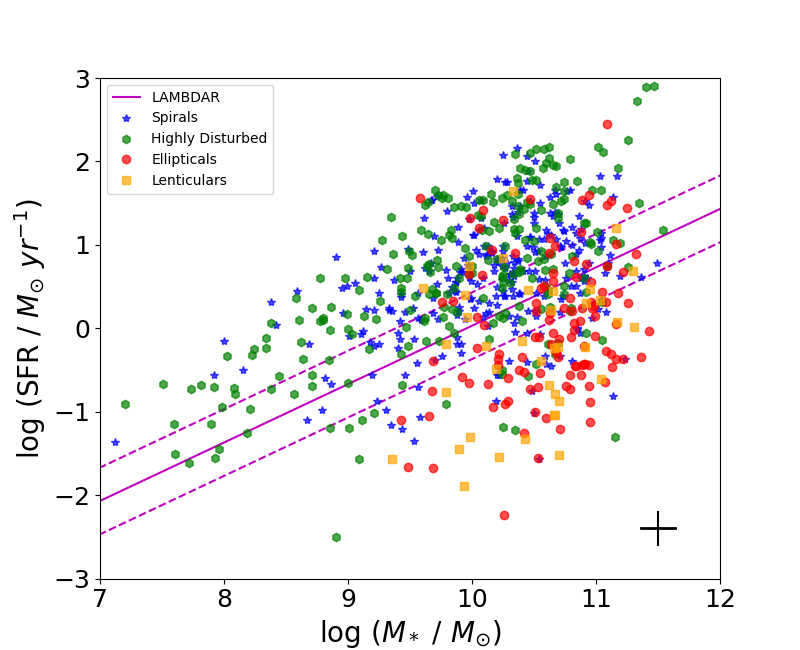}
  \includegraphics[bb=50 50 950 900, width=0.5\textwidth,clip]{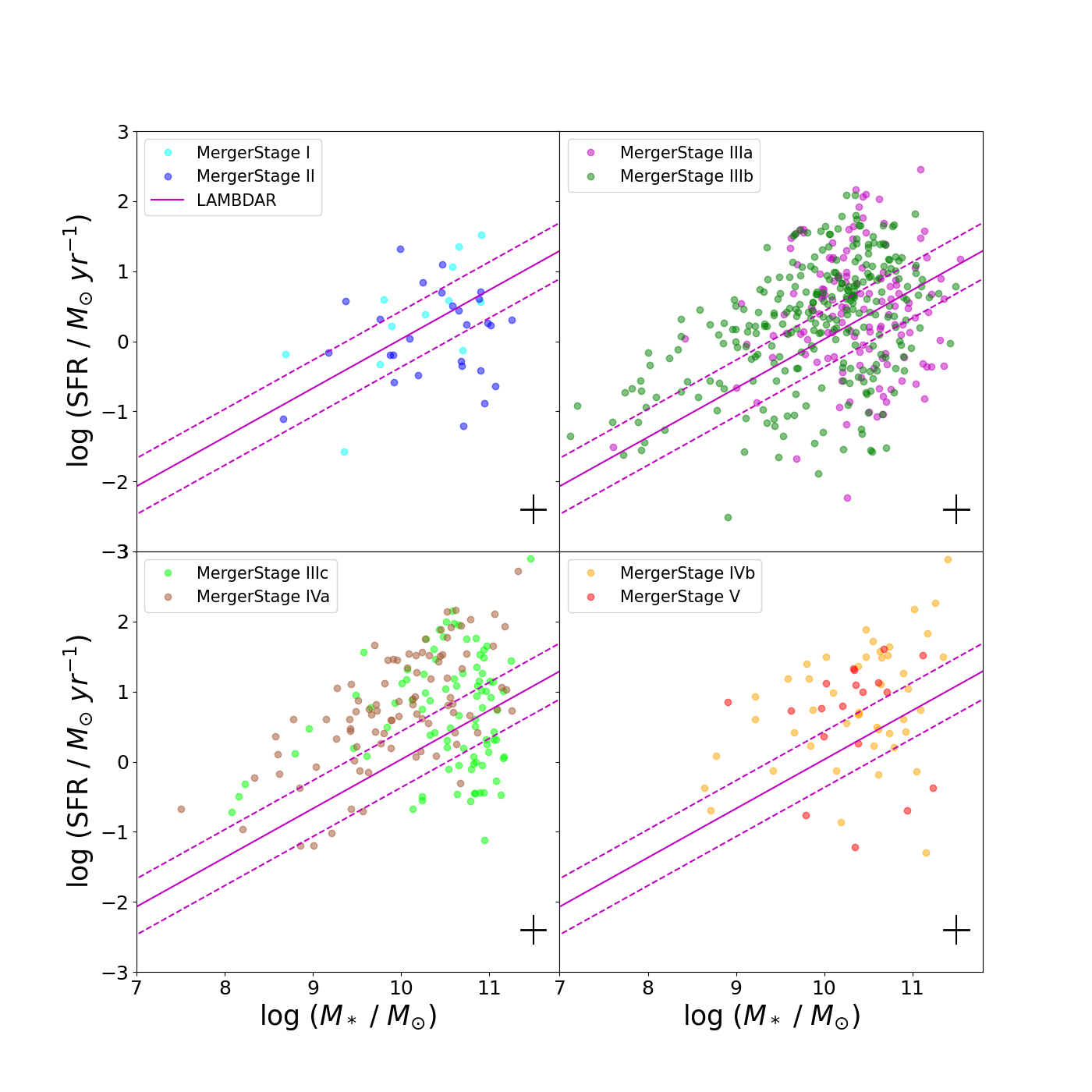}
   \caption{ SFR-M$_*$ plane separated by morphology (top panel) and merger stage (bottom panels). 
   The solid- and dashed-magenta lines represent the MS and the scatter (see text). 
   The typical error is shown on the bottom-right corner of each panel. 
}  
\label{fig:SFRMstellarPlanes}
\end{center}
\end{figure}

Figure \ref{fig:SFRMstellarPlanes} shows the SFR-M$_*$ plane of mergers. The solid-magenta line shows the 
MS determined using the pan-chromatic data from the LAMBDAR project \citep{LAMBDAR}. This project provides the photometry of more than 200,000 objects in 21 bands plus their uncertainties. We ran MAGPHYS on this data, using the same filters as in our study, in order to derive the position of the main sequence, following the process described in \textsc{Chang+15}. In this way, we can measure the location of our galaxies with respect to the main sequence. All mergers within this scatter will be referred to as mergers within the MS. 
We have adopted this choice of MS because \textsc{LAMBDAR} measures the photometry using a comparable method to our Paper I and \textsc{Chang+15} uses the same SED fitting method 
(MAGPHYS) as our study to determine it. 
Other MS are presented in the literature such as \citet{Elbazetal07, Elbazetal11}, but their 
M$_*$ and SFR values are estimated using a very different approach (based on the SDSS fiber 
and optical filters) and so are less easily comparable with our own results. We note that, all of these 
MS are determined for the nearby Universe, over a similar redshift range as our sample. Thus, we are 
not concerned about evolution of the MS in the redshift range of our sample.

The top panel shows the 
merging galaxies coloured by morphology as shown in the legend. Spiral and HD galaxies (blue stars and 
green hexagons) are mainly located within the MS or above it. On the other hand, elliptical and lenticular 
(red circles and orange squares) galaxies cover all three regions: above, within and below the MS but only at 
stellar masses higher than 10$^9$ M$_{\odot}$. 
The bottom panels show the SFR-M$_*$ plane separated by merger stage according to the legend in 
each sub-panel. Again, our MS is shown by the solid- and dashed-magenta lines. As shown 
in these sub-panels, there is no clear dependence on merger stage. Thus, to study the SF enhancement through 
the merger sequence, we calculated their SFR/M$_*$ value, and determine their distance from the MS. We refer to this distance as their 'SF mode', which can be positive for objects above the MS 
or negative for objects below the MS. We consider each galaxy's SF mode separated by their merger stage. 

\begin{figure}[!h]
\begin{center} 
  \includegraphics[width=0.5\textwidth,clip]{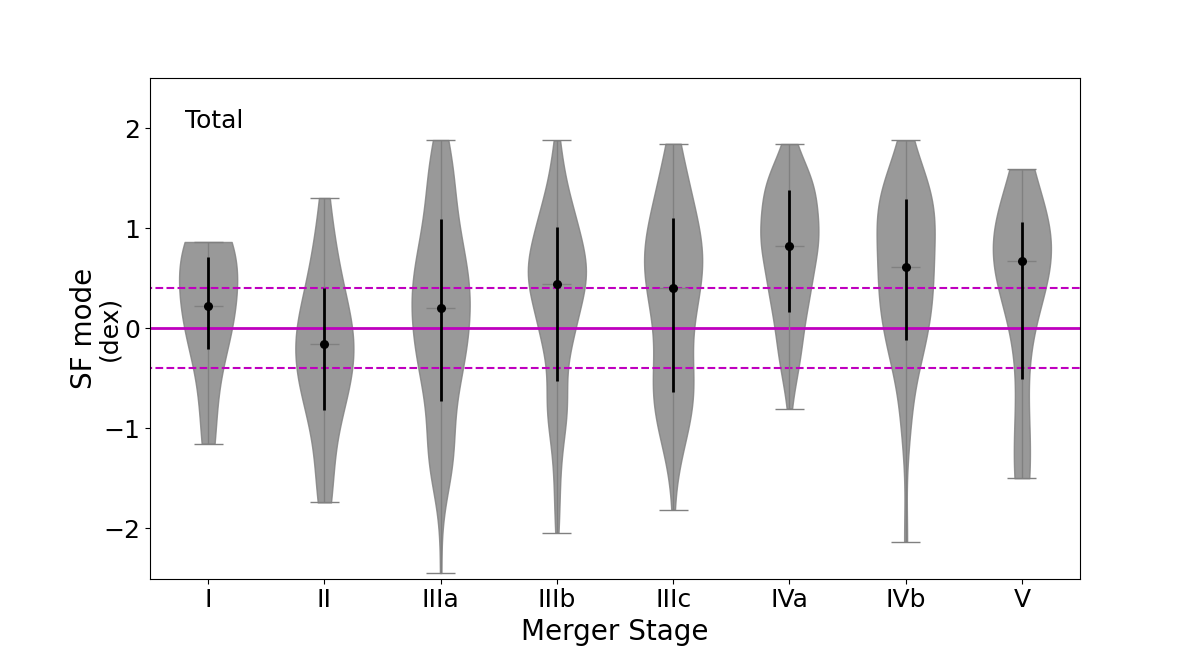}
  \includegraphics[width=0.5\textwidth,clip]{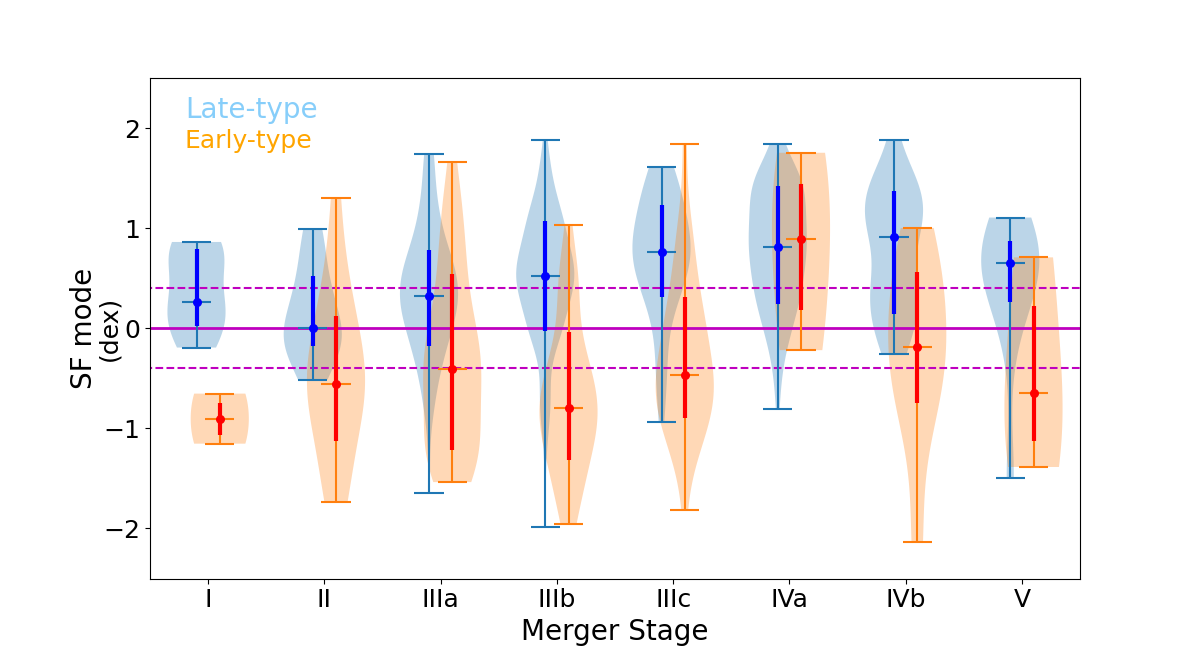}
   \caption{ Star formation mode distribution in each merger stage. The filled circle indicates the 
   median and the bold lines show the percentile including 34\% of objects from each side of the 
   median (equivalent to one-sigma). The horizontal solid- and dashed-magenta lines represent the MS. 
   The top panel shows the SF mode distribution of the entire sample. The bottom panel shows the 
   SF mode distribution of the merging galaxies separated by late-(blue) and early-(orange) type. 
   HDs are separated into the late- and early-type categories (see text for details).
}  
\label{fig:SFmode_Stg_violin_gral_lateearly}
\end{center}
\end{figure}

\begin{figure}[!h]
\begin{center} 
  \includegraphics[bb= 50 0 800 580, width=0.37\textwidth,clip]{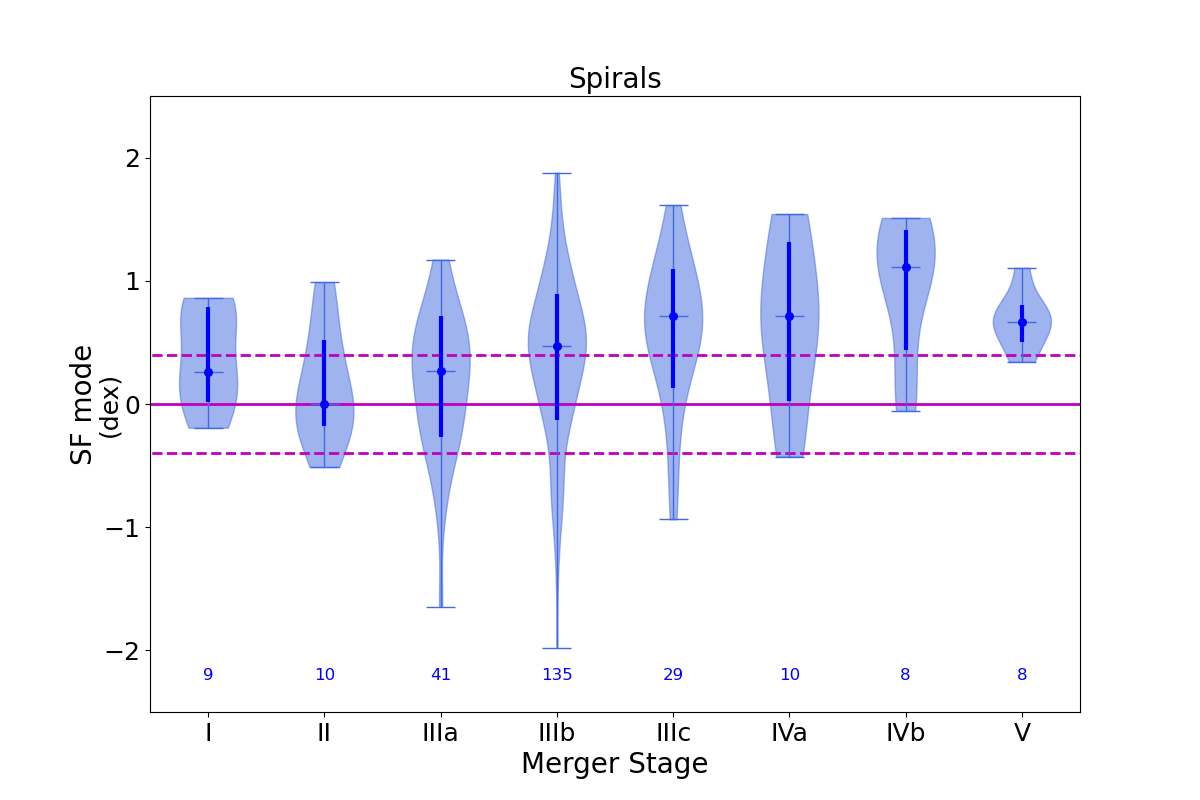}
  \includegraphics[bb= 50 0 800 580, width=0.37\textwidth,clip]{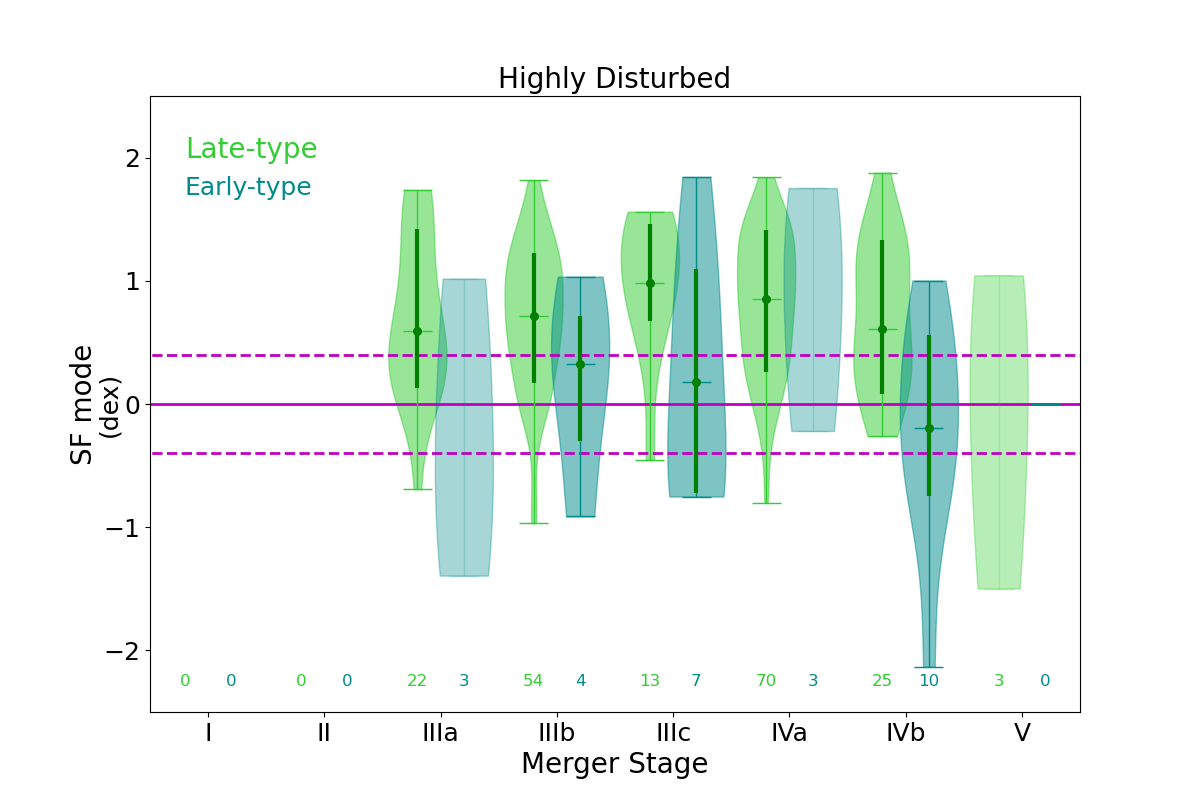}
  \includegraphics[bb= 50 0 800 580, width=0.37\textwidth,clip]{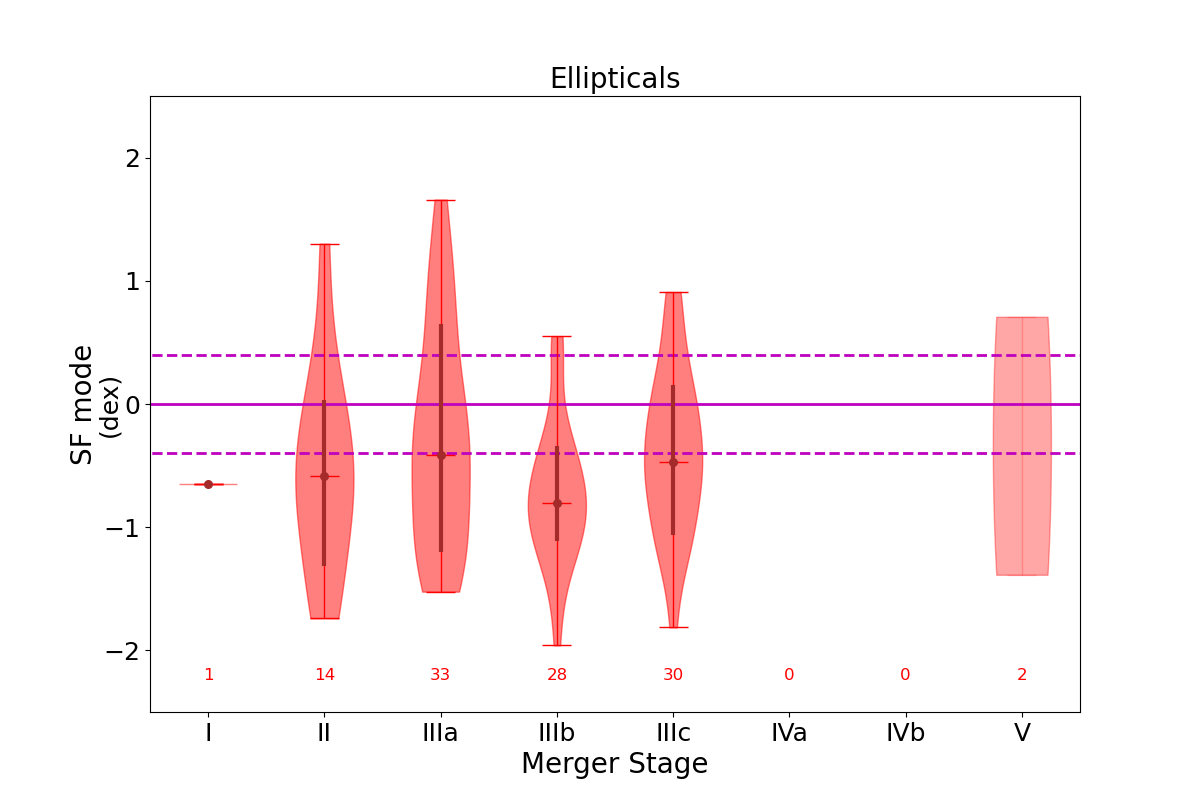}
  \includegraphics[bb= 50 0 800 580, width=0.37\textwidth,clip]{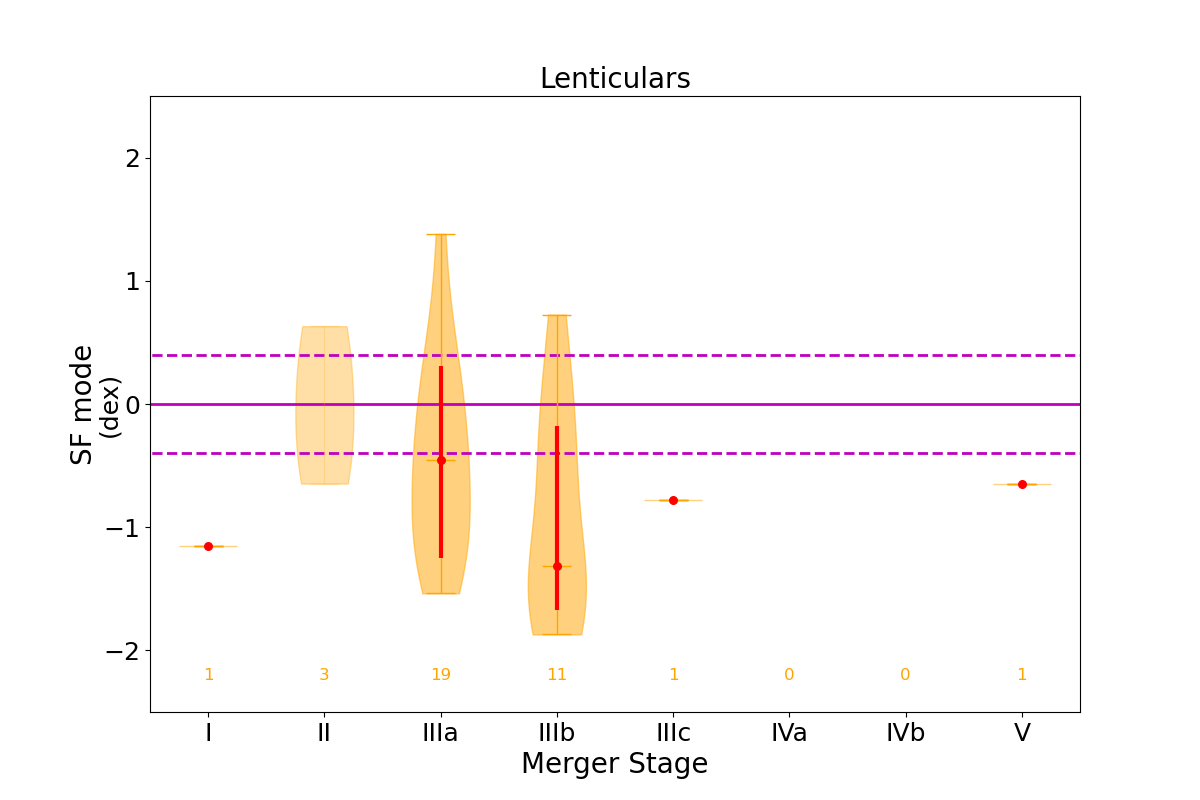}
   \caption{ Star formation mode distribution in each merger stage separated by morphology. 
   From top to bottom: Spiral, HD (late- and early-type), elliptical, and lenticular galaxies. 
   The solid- and dashed-magenta lines represent the MS. Numbers under each violin distribution shows the number of objects in each merger stage. For low N ($\leq$3), shaded regions are shown with a lighter colour and median/quartile lines removed.
}  
\label{fig:SFmode_Stg_violin_Morphsep}
\end{center}
\end{figure}      

Figure \ref{fig:SFmode_Stg_violin_gral_lateearly} shows the distribution of the SF mode in each 
merger stage, using so-called `violin-plots'. For our work, we use the $matplotlib.pyplot.violinplot$ task available in $python$ with default settings. Violin-plots are essentially histograms of the normalised 
number distribution, only the bars are smoothed out to form the shaded region, and bars extend 
symmetrically to the left and right to give the characteristic violin-like shape.
The median of the distribution is shown by the filled circle and overlayed horizontal line, found within 
each distribution. The dark-coloured lines show the percentiles including 34\% of the objects from each 
side of the median (equivalent to one-sigma of a Gaussian distribution, but additionally allowing for 
asymmetry in the distribution about the median). The MS, as described previously, is shown by 
the solid- and dashed-magenta lines. The top panel shows the SF mode distribution of 
the entire sample. 
The first thing to notice is that almost all the mergers show enhanced SF mode. Furthermore, the merging galaxies show an increase in SF as the merger evolves, peaking at about merger stage IVa. 
More generally, mergers show higher SF mode at intermediate 
and late stages (from IIIb to V) compared to early stages (I to IIIa).

Galaxies that are actually at early stages of the merging process might fall from large separation to small separation, 
and they increase their separation again after passing pericentre. Thus, our early and intermediate merger stages 
(I to IIIc) likely contain a mix of different projected distances between the merging galaxies. 
Meanwhile post-merger galaxies (e.g. as defined in \citealt{Ellisonetal13}) could be found in any of our 
late merger stages (IVa to V). \citet{Ellisonetal13} find that the SFR peaks at post-merger stages, 
and we indeed see a steady SF enhancement. 

The bottom panel shows the SF mode 
distribution for late-(blue) and early-(orange) type galaxies separately. Following our morphological classification, 
spirals are considered late-type galaxies, and lenticulars and ellipticals are considered early-types. 
As explained in Sec. \ref{sec:Sample}, HD galaxies have been sub-classified as late- and early-type galaxies. 
These galaxies have been included in the blue and orange distributions, respectively. 
It is clear that late-type galaxies show higher (+1.0 dex on average) SF mode 
compared to early-type galaxies. In fact, we find this result is independent of stellar mass 
(see Fig. \ref{fig:SFmode_dists_Mstellarbins}). 
This suggests that the SF mode principally depends on the gas content of the merging galaxy. 
   
Figure \ref{fig:SFmode_Stg_violin_Morphsep} shows the SF mode distribution separated by morphology, 
as shown in the title of each sub-panel. Spiral and HD (i.e. late-type) merging galaxies show high SF. 
The median of each merger stage is above the MS. 
In fact it is the HD galaxies which show the highest SF mode values, with median SF modes around +1 
(ten times higher SFR than the MS at that mass) and many being enhanced by as much as a factor of 50 in SFR.
As shown in the HD panel, both late-(light green) and early-(dark green) type mergers show SF enhancement, 
suggesting that mergers which have suffered such strong disturbances enhance their SF either by consuming 
their own gas reservoir or possibly using gas acquired from their companion. 
On the other hand, elliptical and lenticular galaxies show lower SF mode compared to spiral and HD galaxies, 
but still show higher values compared to unperturbed galaxies with the same morphology (which are typically found significantly more than 1 dex below the MS, e.g. \citealt{CanoDiazetal2019}). This suggests that this type of galaxy could form new stars from their own small 
($<$5\%; \citealt{Catinellaetal18}) gas reservoirs  
or possibly collect enough cold gas from their companion to form new stars. 

A key feature of Figure \ref{fig:SFmode_Stg_violin_gral_lateearly} and \ref{fig:SFmode_Stg_violin_Morphsep} 
is that enhanced SFR are seen over multiple merger stages, from stage IIIa through to stage V. This might 
be considered somewhat at odds with the evolving SFR seen in merger simulations, which often show short 
peaks of enhanced SF at critical moments such as first passage, second passage or final coalescence 
\citep{MihosnHernquist94b, DiMatteoetal05, Parketal17}. As a result of the long duration of the SF enhancement 
that we observe, it raises the possibility that the enhanced SF could contribute significantly to the overall 
stellar mass growth of the galaxies involved. To get some feeling for the significance of this effect, we employ 
a back-of-the-envelope calculation. First, we consider two typical MS galaxies; a low-mass 
(M$_*$ = 10$^9$ M$_{\odot}$) and a high-mass (M$_*$ = 10$^{11}$ M$_{\odot}$) galaxy. We then consider 
how much they would increase their own stellar mass by star formation alone (excluding stars accreted from the companion) over the time period from first passage until final coalescence 
(roughly corresponding to the merger stage IIIa through to merger stage V). We consider two cases; if they 
(a) remain on the MS and continue star-forming at the same rate, and (b) suffers a SF enhancement of 1 dex 
above the MS (starburst) over the whole time period. We then compare case (b) to case (a) and calculate by what percentage the stellar mass of case (b) is higher than case (a) at the time of final coalescence. Actually, the time period between first passage and final coalescence 
is, in part, a function of mass ratio with more major mergers tending to merge more quickly. 
We took the duration 
of the time period from the simulations of \citet{Parketal17} for a 1:6 (minor) merger and a 1:1 (major) 
merger with a parabolic orbit, and found values of 2.3 and 1.1 Gyr, respectively. 

\begin{figure*}
\centering
   \includegraphics[width=17cm]{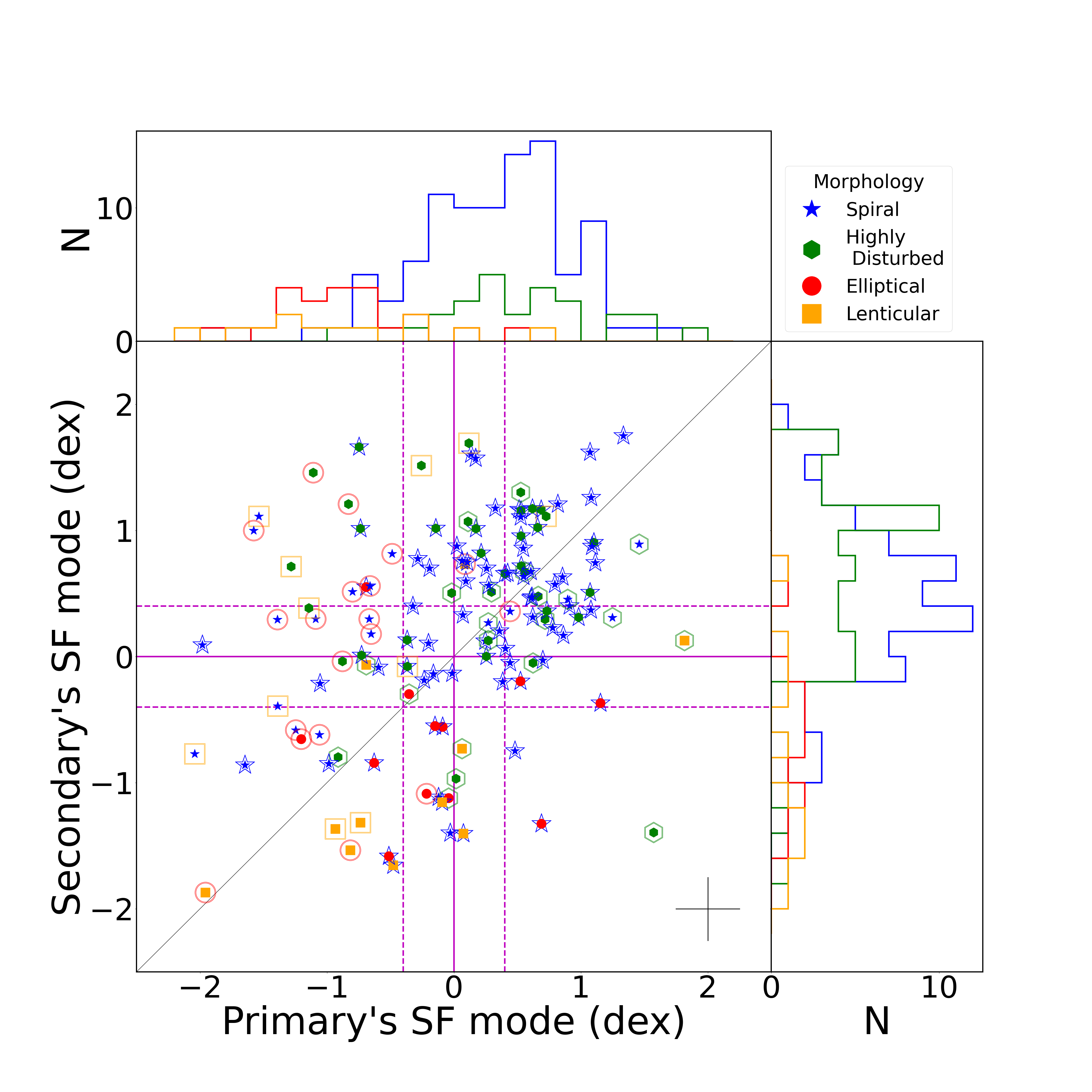}
     \caption{Comparison between the SF mode of the primary and the secondary component. 
   The coloured symbols show the morphology of the primary (unfilled symbols) and secondary (filled symbols) components as shown in the legend. The solid- and dashed-magenta lines show 
   the MS and the black line shows the one-to-one relation. 
   The SF mode distributions of the primary and secondary component are shown in the top and right panel, 
   respectively. The SF mode distributions are coloured by morphology as shown in the legend. }
     \label{fig:SFmode_PrimSec}
\end{figure*}

The results of the back-of-the-envelope calculation can be summarised as follows: 
The low-mass MS galaxy increases its stellar mass by 23\% (11\%) due to the star formation enhancement alone for the 1:6 (1:1) merger. In comparison, 
the low-mass starburst galaxy increases its stellar mass due to the star formation enhancement alone by 230\% (110\%) for the 1:6 (1:1) merger. Meanwhile, 
the high-mass MS galaxy increases its stellar mass due to the star formation enhancement alone by 1.3\% (0.6\%) for the 1:6 (1:1) merger, while 
the high-mass starburst galaxy increases its stellar mass by 13\% (6\%) for the 1:6 (1:1) merger. We note that this percentage increases do not consider the additional mass that the galaxy would have gained by accreting stars from the secondary galaxy, as we are trying to asses the significance of the starburst only.

In summary, and noting that we have neglected the limitations driven by the available gas reservoir, for low-mass galaxies, 
the starburst alone can be very significant to the overall stellar mass growth, and can 
more than double their mass, even neglecting the mass that would be accreted from the secondary galaxy. We note that our choice of a 1 dex enhancement in SFR is not even very extreme, 
with some galaxies showing as much as 1.5 dex enhancement. However, for higher mass galaxies, the enhancement 
in stellar mass is much less significant for the overall growth. This is a natural consequence of the fact that 
high-mass MS galaxies tend to have low sSFR, meaning their current SF is not as significant for their stellar mass growth 
than in low-mass MS galaxies.

\subsection{SF mode dependence on the companion's morphology} 
\label{sec:SFmode_Morphdep}

In order to study how the different morphologies affect the SF mode of the merger before coalescence, we 
first selected a subsample of systems that show both merger components separately. Thus, the following analysis is based on systems 
at merger stages I to IIIb. Because of the low number of mergers at early stages, we study these 
four merger stages combined. 
Figure \ref{fig:SFmode_PrimSec} shows the comparison between the SF mode of the 
primary ($x$-axis) and secondary ($y$-axis) component of the merger in the main panel. 
The solid- and dashed-magenta lines show the MS location for each axis and the black line shows the 
one-to-one relation, i.e. where the SF mode of the primary is equal to the SF mode of the secondary component. 
Coloured symbols show the morphology (see legend) of each component, where unfilled-large symbols show the 
morphology of the primary and filled-small symbols show the morphology of the secondary component. 
For example, a merger that has an elliptical as a primary component and a HD galaxy as a secondary component 
will be shown as a big red circle with a green hexagon inside (see some examples of this case in the top-left corner 
of the main panel). Distributions of the SF mode coloured by morphology are shown at the top for the primary 
component and at the right for the secondary component. 

As noted previously, the spirals and HD galaxies show higher SF modes compared to elliptical and lenticular galaxies. 
Thus, as a rule of thumb, the SF mode of a merging galaxy seems to be primarily dictated by the galaxy's morphology. 
For example, primary spirals (large blue stars) typically appear on the right-hand side of the diagram, 
similarly for HD galaxies. Similarly, primary elliptical and lenticular galaxies (large red and orange symbols) are mostly 
found towards the left of the panel. The same is true for secondaries, e.g. secondary spirals 
(small filled-blue stars) are found towards the top, while secondary lenticulars (small filled-orange squares) 
generally are found towards the bottom of the diagram.

\begin{figure}[ht!]
\begin{center} 
  \includegraphics[width=0.5\textwidth,clip]{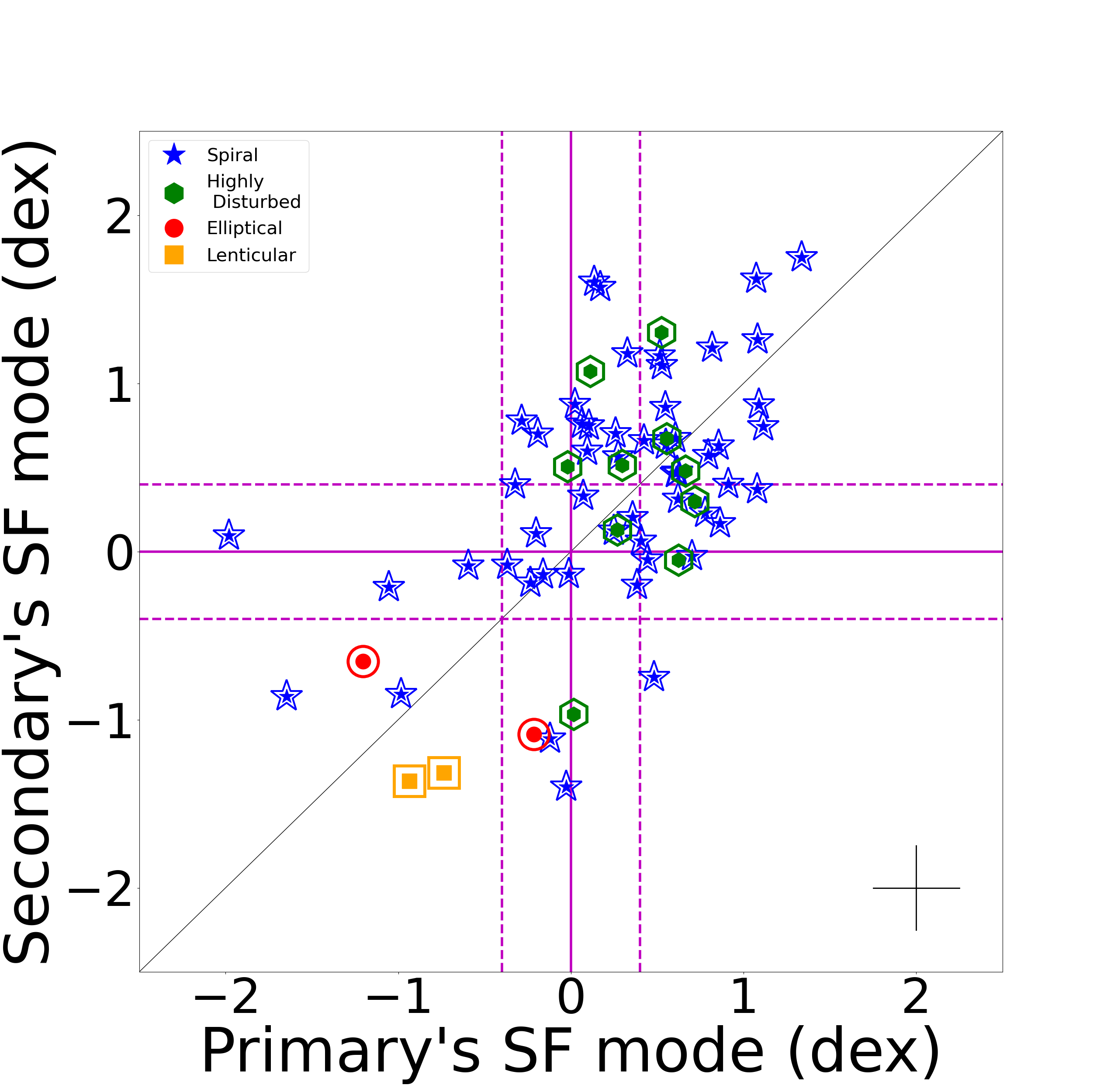}
   \caption{ Comparison between the SF mode of the primary and the secondary component, 
   for mergers with components of the same morphology. 
   The coloured symbols show the morphology of the primary (unfilled symbols) and the 
   secondary (filled symbols) components as shown in the legend. The solid- and dashed-magenta lines represent 
   the MS and the black line shows the one-to-one relation. The typical error is shown in the bottom-right corner. 
}  
\label{fig:SFmode_PrimSec_SameMorph}
\end{center}
\end{figure}

Simplifying the previous plot, we now consider only systems where both components share the same morphology, 
so both the large surrounding symbol (the primary) and the small filled symbol (the secondary) will match, 
as shown in Fig. \ref{fig:SFmode_PrimSec_SameMorph}. It can now be seen that the distribution of points is not 
symmetrical about the one-to-one line as would be expected if both primary and secondary responded equally to 
the tidal interactions. Instead, an overpopulation of points can be seen above the one-to-one line 
compared to 
the population below (57\% compared to 43\%, respectively). This shows that the SF mode of the secondary 
component tends to be more enhanced than the primary component. 
We also tested this in both major and minor mergers separately, and found the same trend in both cases 
(see figures in App. \ref{App:MajorandMinor}). 
We have so far only considered SF mode calculated with respect to the MS derived as described in Sec \ref{sec:SFmode}. 
In fact, this is a sensible choice as this MS also uses the same SED fitting tool, and so is 
a fairer comparison. However, we were curious to see how sensitive these results are to our choice of MS. 
Therefore, we also reproduced Figures 3-7 using the MS derived in \textsc{Chang+15} and \citet{Elbazetal07}. Overall, we find the 
results show the same main features, and the same general distribution. One minor difference is that the spiral 
galaxies show a hint of being slightly closer to the MS with the Elbaz MS, although they are still clearly enhanced, 
but for other galaxy morphologies the results are very similar. Comparing our results to \textsc{Chang+15}'s MS, we find that the median of all merger stages are slightly higher. Therefore, in general, we find our main conclusions 
are not strongly dependent on our choice of MS, over the \textsc{Chang+15} MS and the Elbaz MS.

Spiral-spiral and HD-HD mergers typically show enhancement of their SF mode.   
We will consider if having matched morphologies impacts the amount of SF enhancement later in this section.

\begin{figure*}
\centering
   \includegraphics[width=8.5cm]{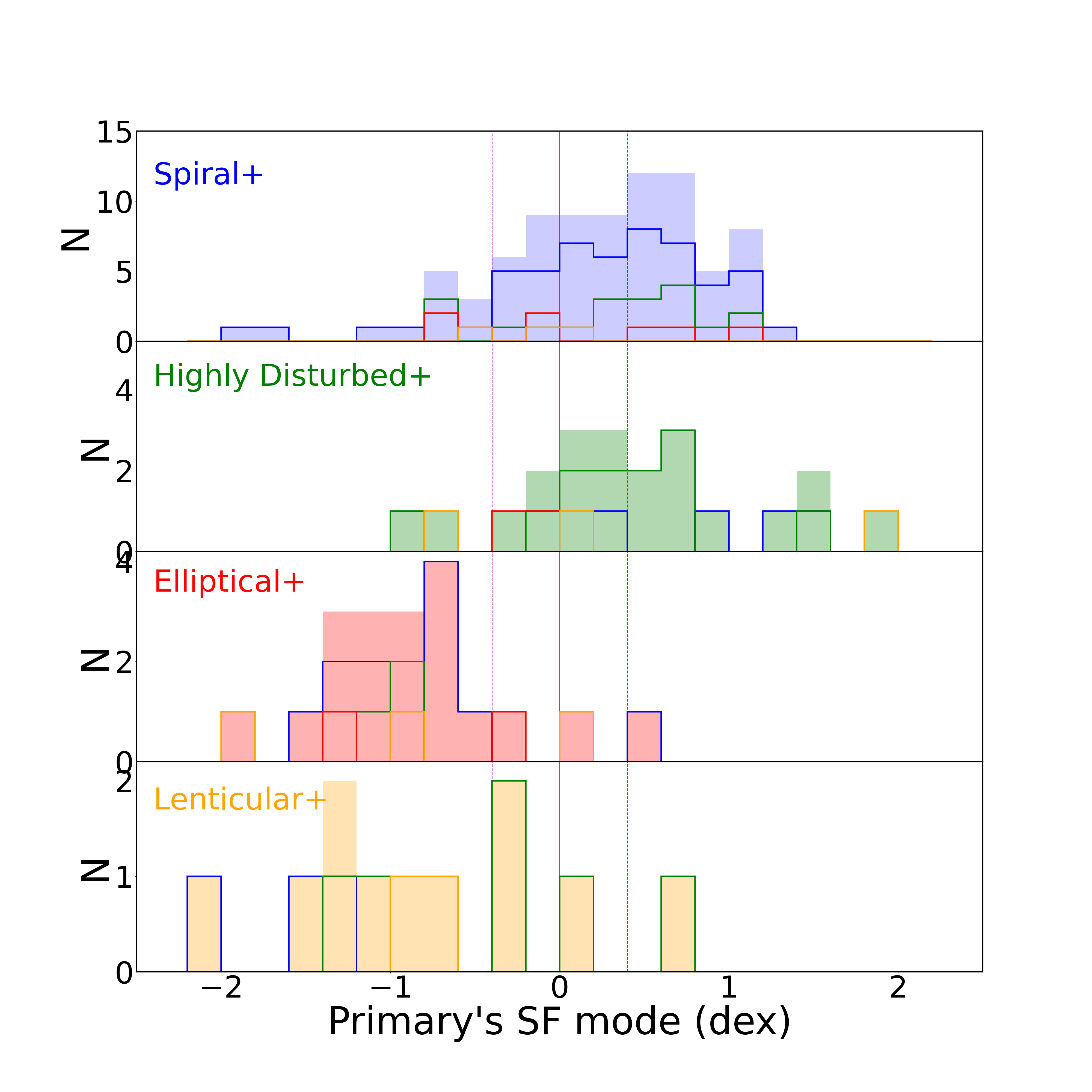}
   \includegraphics[width=8.5cm]{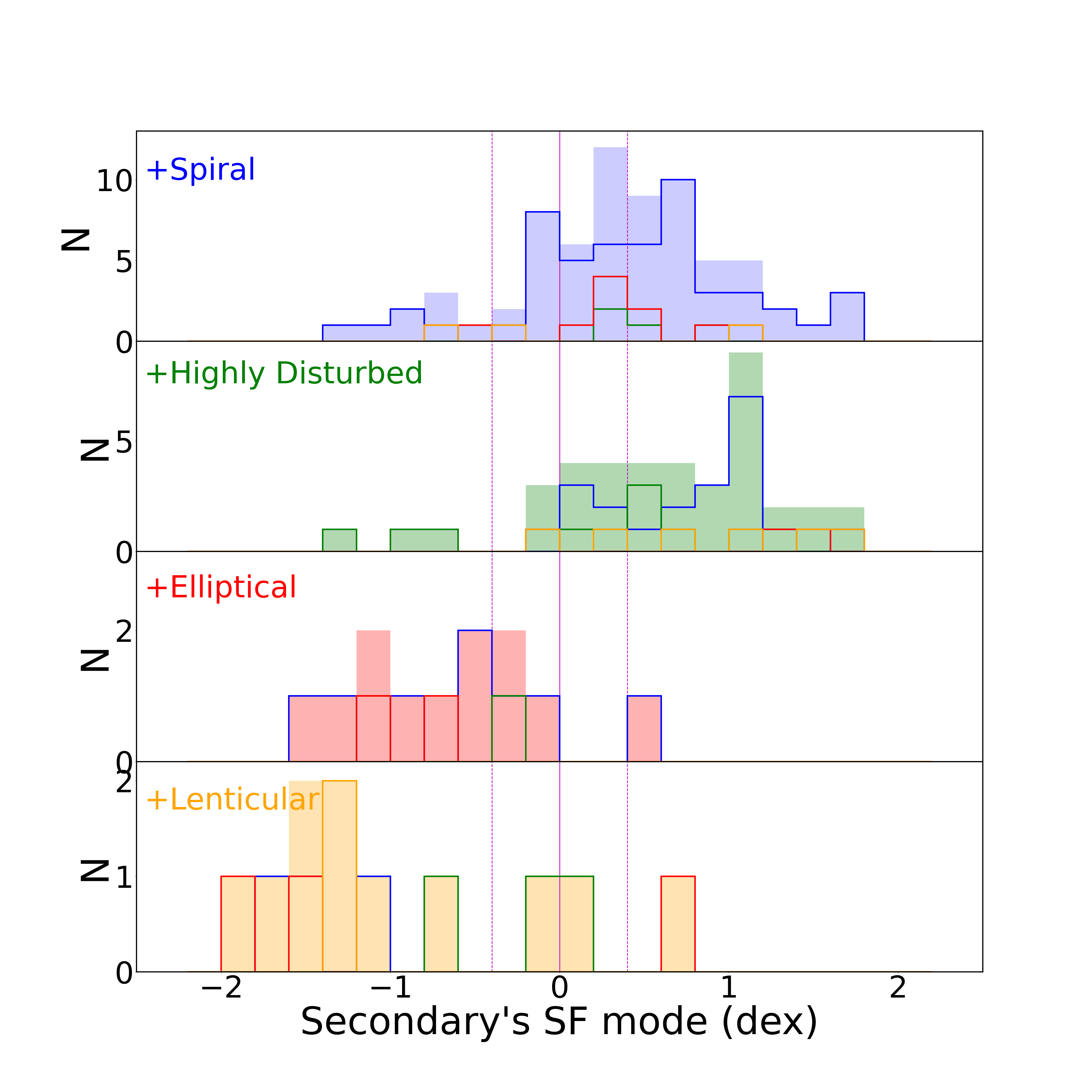}
     \caption{SF mode of the primary (left panel) and secondary (right panel) component separated by 
   their morphology. The filled distributions show the total 
   distribution of the morphology. The unfilled distributions show the 
   distribution of the companion's morphology. }
     \label{fig:SFmode_PrimSec_histos}
\end{figure*}

We now attempt to understand if the morphology of the companion has any impact on the SF properties of a merging 
galaxy. To do this we start with the histogram shown at the top of Fig. \ref{fig:SFmode_PrimSec}), but now separate 
each of the histograms (separated by primary morphology) shown into a different sub-panel, as shown in the left 
column of Figure \ref{fig:SFmode_PrimSec_histos}. The total sample in each sub-panel is given by a shaded histogram. 
We then colour the histogram lines based on the morphology of the companion. For example, to see how a primary spiral 
(top panel) responds to having a secondary elliptical (red lines), we can see where the red bars fall on the x-axis 
(the primary's SF mode). We now repeat this process in the right column of Figure \ref{fig:SFmode_PrimSec_histos} 
but adjusted so that we can instead see the effect of the primary's morphology (histogram bar colour) on the 
secondary galaxy's SF mode (x-axis).  The labeling of the sub-panels uses 
the following system: 'Primary morphology' + 'Secondary morphology'. e.g. the sub-panel labeled as 
'Spiral+' shows the mergers that have a spiral galaxy as primary, and '+Elliptical' shows galaxy's with a secondary 
elliptical. The colour of the labels in each subpanel matches the colour of the histograms. 

\begin{figure*}
\centering
    \textbf{Major Mergers} \\
   \includegraphics[width=8.5cm]{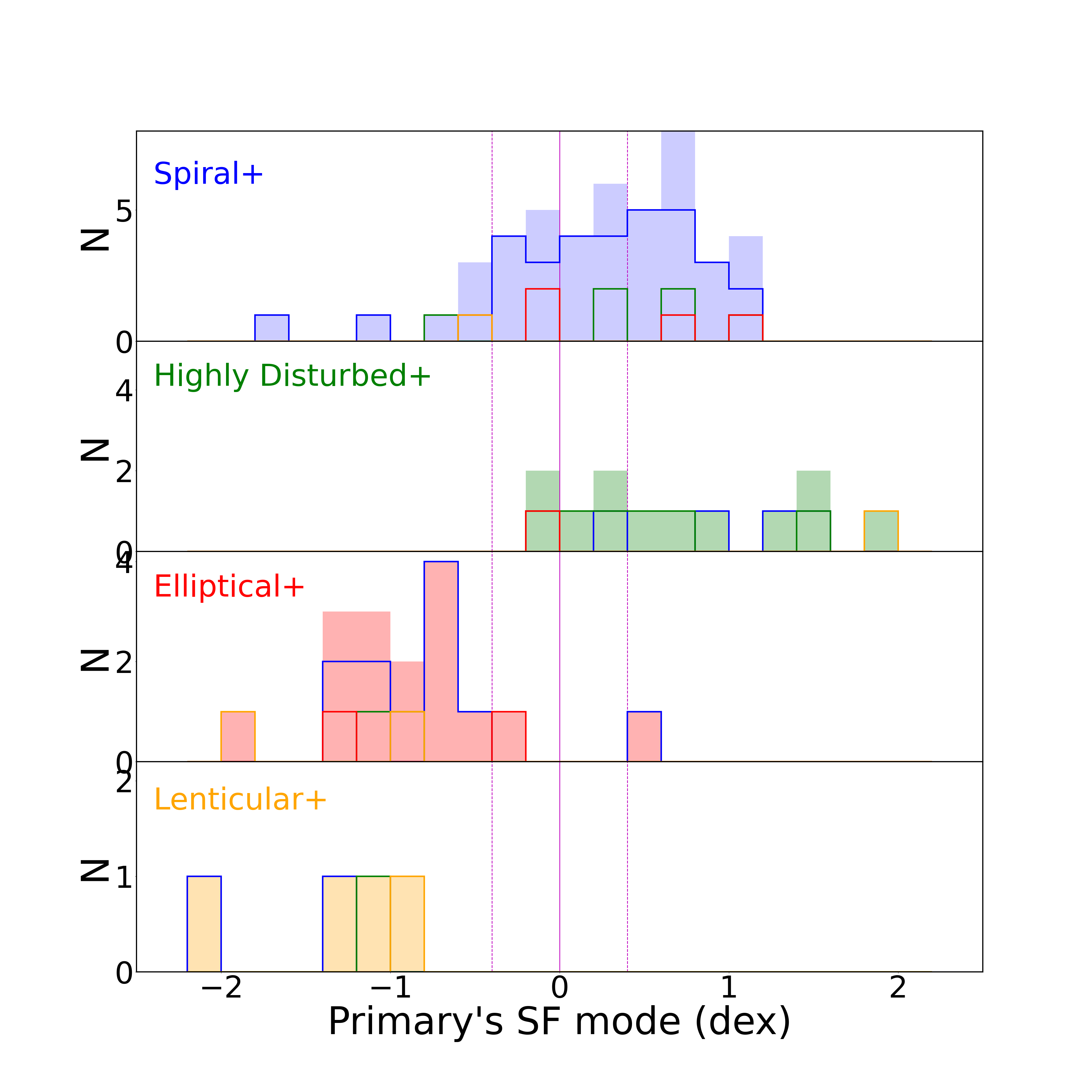}
   \includegraphics[width=8.5cm]{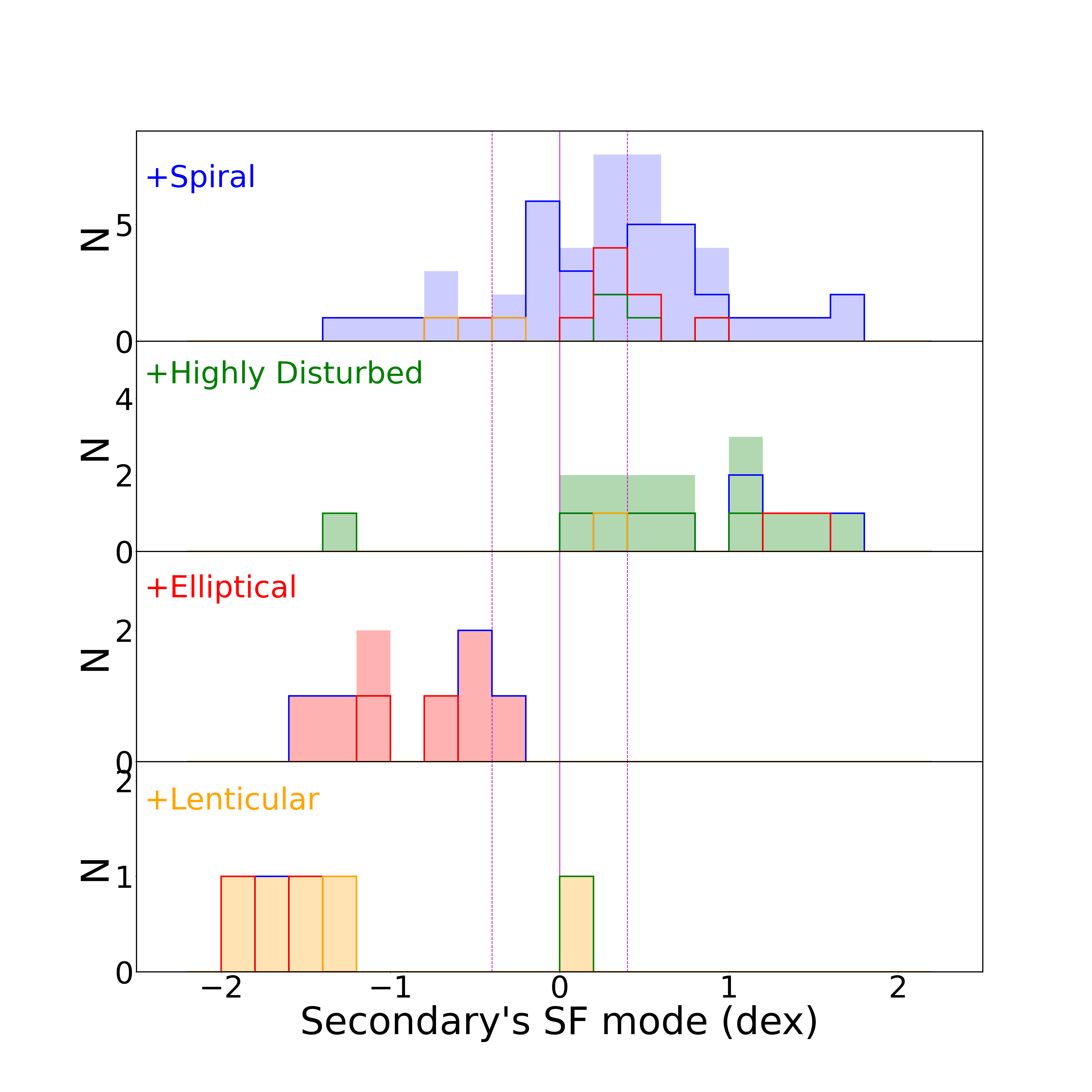} \\
    \textbf{Minor Mergers} \\
   \includegraphics[width=8.5cm]{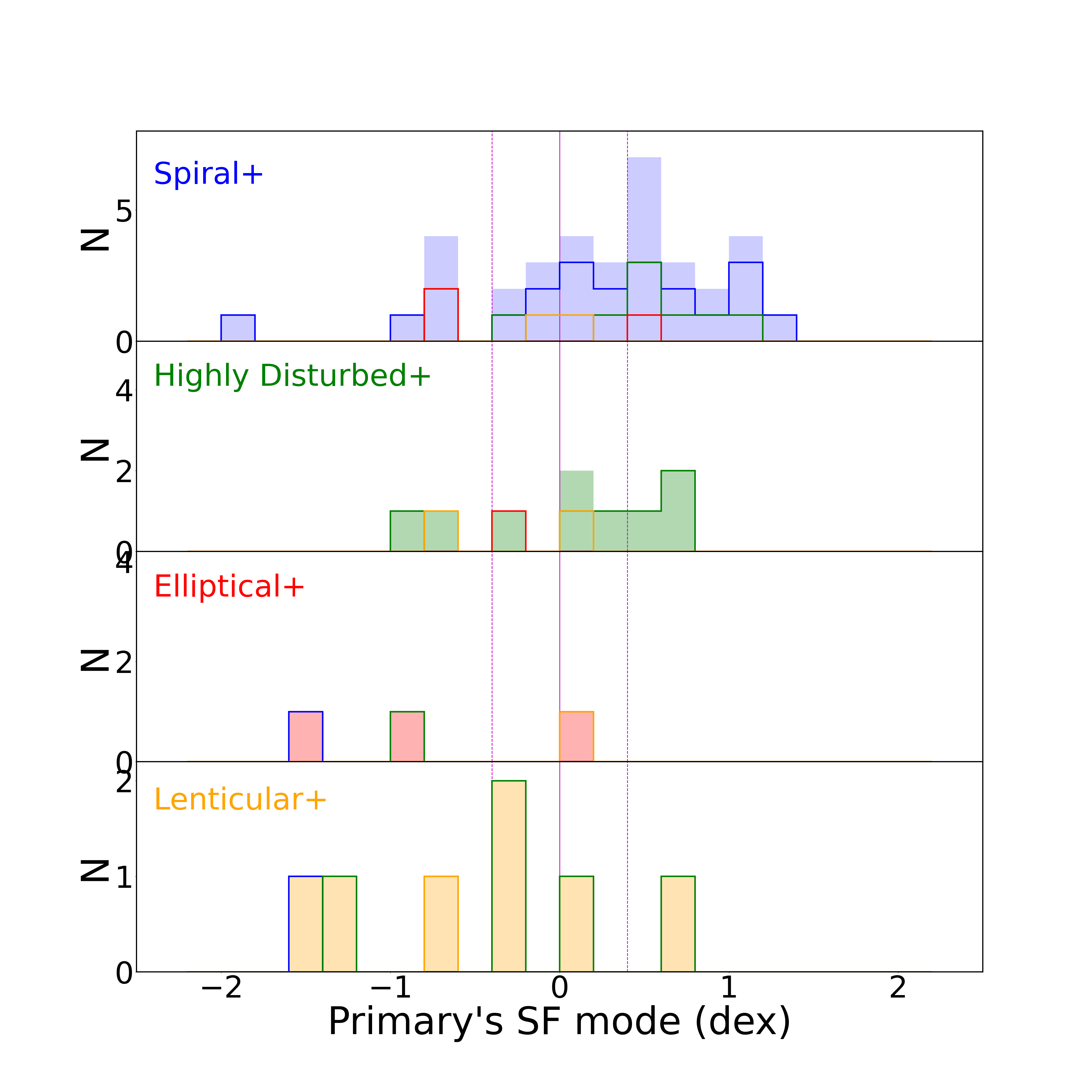}
   \includegraphics[width=8.5cm]{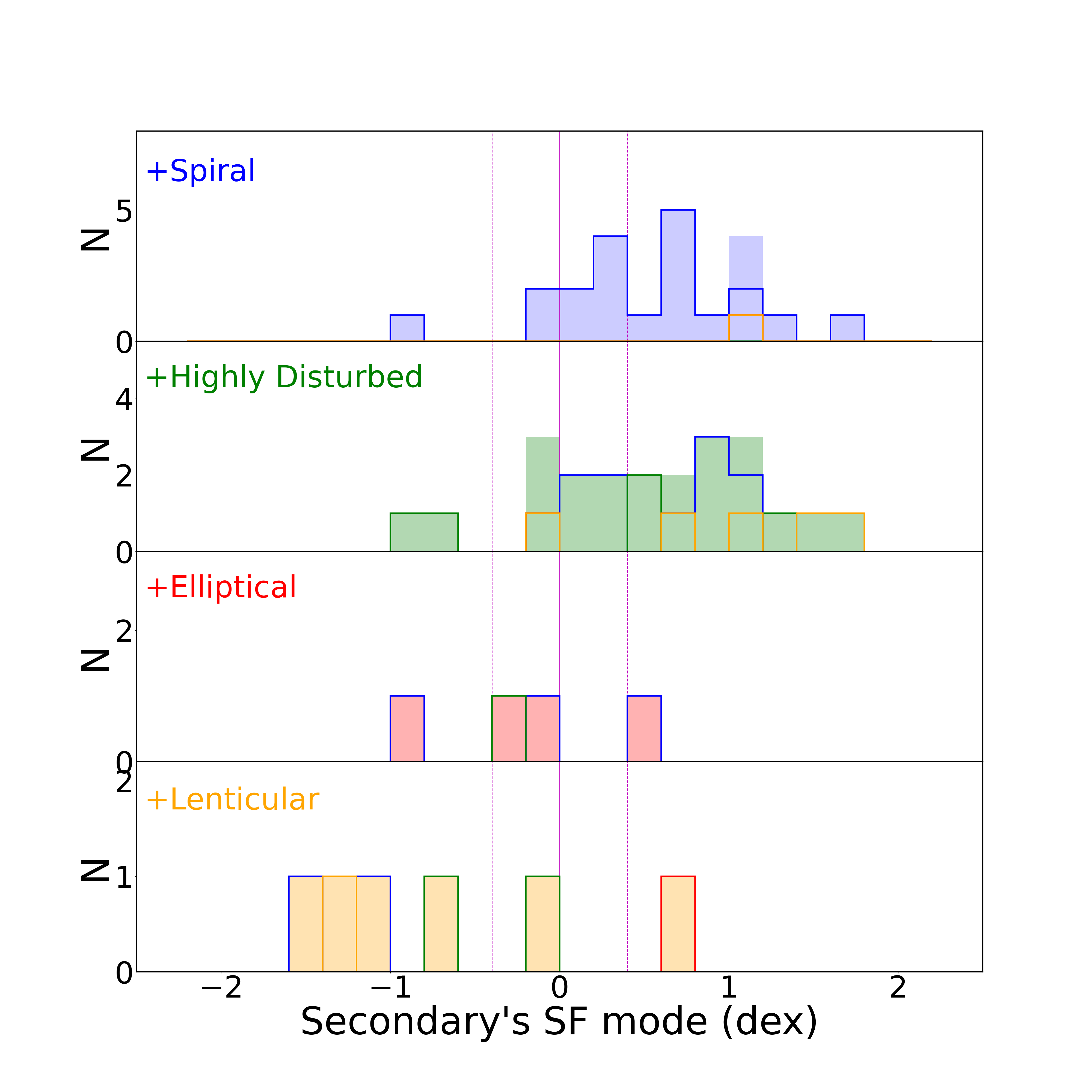}
     \caption{As in Fig. \ref{fig:SFmode_PrimSec_histos}. The left panels show the SF mode of the 
  primary and the right panels show the SF mode of the secondary component, 
  separated by major (top panels) and minor (bottom panels) mergers. 
  The filled distributions show the total 
   distribution of the morphology shown by the label in each panel. The unfilled distributions show the 
   distribution of the companion's morphology. }
     \label{fig:SFmode_PrimSec_histos_MajorMinor}
\end{figure*}

As mentioned before, spiral and HD galaxies continue to show clear SF enhancements,  with HD galaxies showing slightly higher SF modes. 
One exception is when the HD galaxy is a primary and has an elliptical as secondary. In this case, both galaxies seem 
to approach the MS from both sides, with HD galaxies showing decreasing SF mode and ellipticals showing 
increased SF mode. For example, the 'HD+' panel shows red bars on the left (i.e. HD suppression), while the same 
system can be seen in the '+Elliptical' panel as green bars on the right (i.e. elliptical enhancement), although clearly 
this result has the caveat of small number statistics.

\begin{figure*}[h!]
\begin{center} 
  \includegraphics[width=0.9\textwidth,clip]{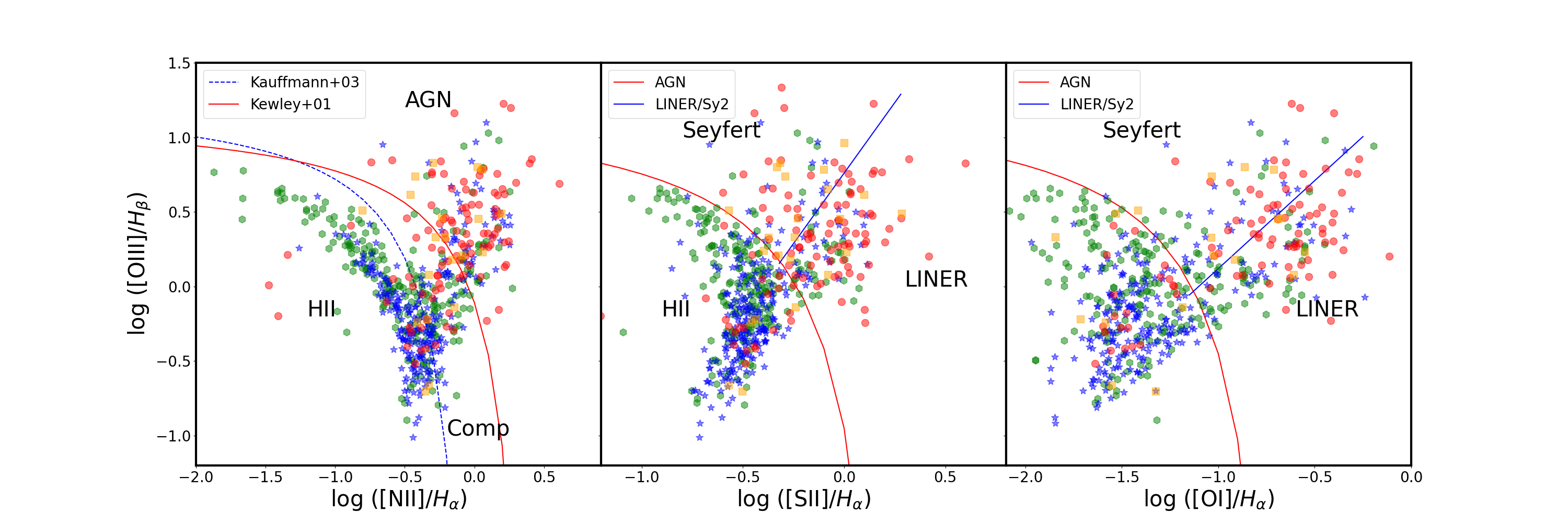}
  \includegraphics[width=0.3\textwidth,clip]{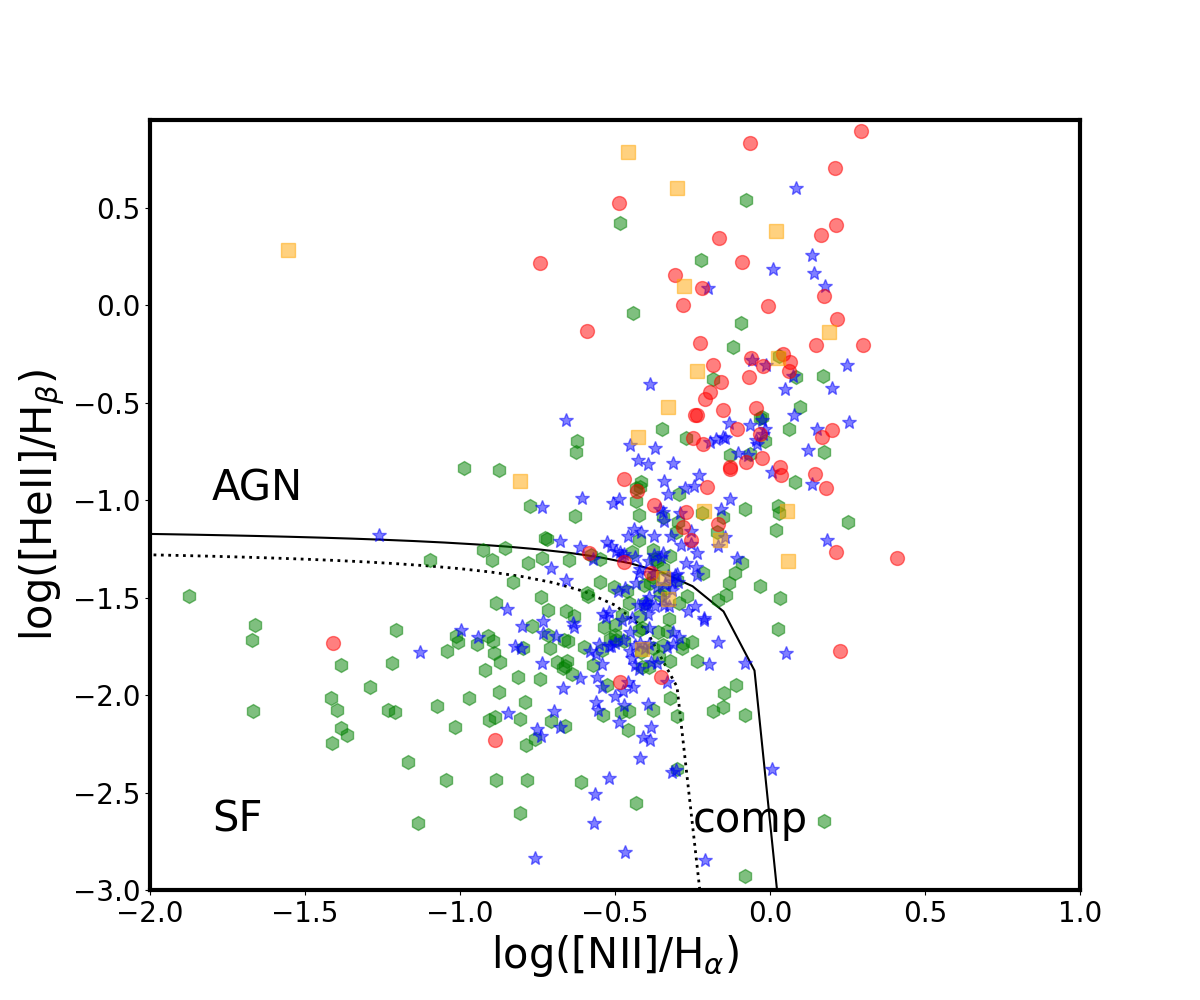}
  \includegraphics[width=0.15\textwidth,clip]{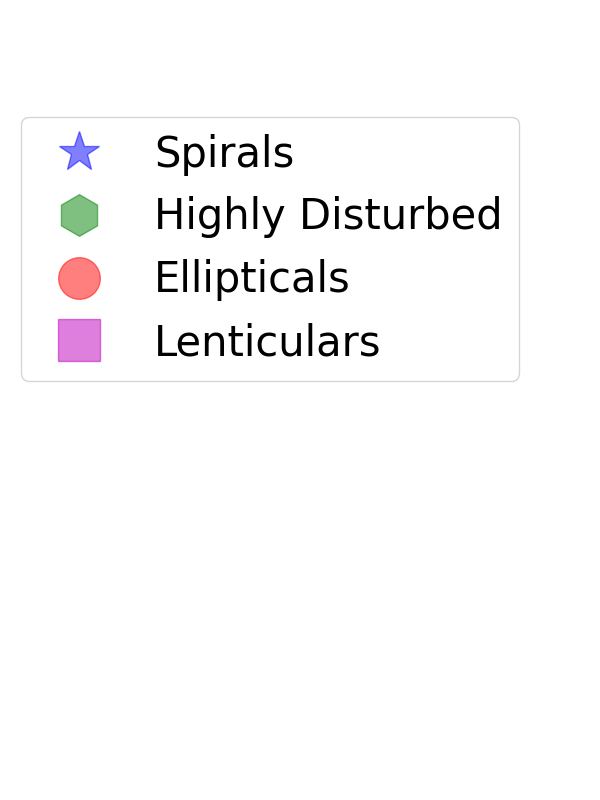}  
  \includegraphics[width=0.4\textwidth,clip]{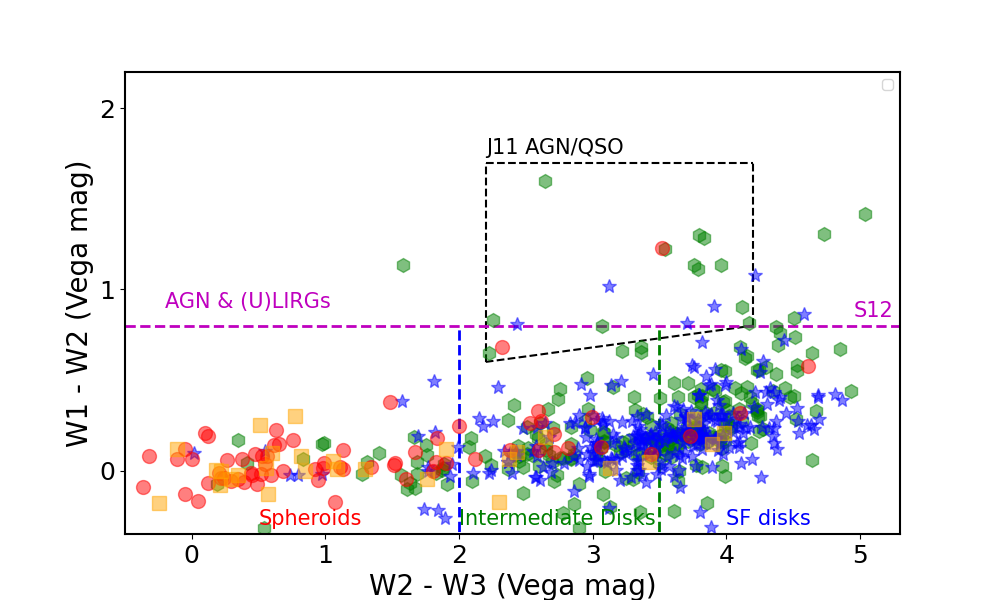}  
   \caption{ $Top$: BPT diagnostic diagrams of the sub-sample that shows emission lines. 
   Separations are shown in the legends. 
   $Bottom-left$: Diagnostic diagram based on the HeII emission line. Separations defined by SB12.
   $Bottom-right$: WISE colour-colour diagram. Separations defined by J11 and S12. 
   Mergers in the different panels are coloured by their morphology as shown in the legend. 
}  
\label{fig:AGNs_IDs_Morph}
\end{center}
\end{figure*}

In order to analyse any dependence on the stellar mass ratio between the components, we 
separated the panels in Fig. \ref{fig:SFmode_PrimSec_histos} into major and minor mergers. 
These are shown in the top (major mergers) and bottom (minor mergers) panels of Fig. \ref{fig:SFmode_PrimSec_histos_MajorMinor}.

Now considering primary spirals, we observe that the median SF mode of the 
primary is higher when the secondary is a HD (+0.5 dex) than if the secondary 
is any other type of galaxy (median SF mode of 0.3 dex).  
A similar scenario is observed for primary lenticulars, where higher SF modes are shown when this type 
of galaxy interacts with HD galaxies in minor mergers, compared to when interacting with spirals in 
major mergers. This could suggest that in minor mergers, the primary lenticular galaxy more strongly disturbs the 
secondary component, transforming it into a HD, and perhaps causing the transfer of some of the cold gas from 
the secondary to the primary lenticular, resulting in bursts of SF. However, this is not seen in major mergers, 
where the spiral galaxy is less strongly affected by the lenticular because of their more similar stellar mass. 
Without highly disturbing the spiral, there may be less opportunity for a significant transfer of cold gas to the 
lenticular galaxy, hence no rise in SFR. 
These dependencies on the morphology of the companion agree with \citet{Hwangetal11}'s work, where they 
show that LIRGs have a dependency on the morphology (late- and early-type) of their companions. 

Comparing major and minor mergers, spiral and HD galaxies (filled-blue and -green distributions) as primary 
or secondary galaxies show similar trends independent of stellar mass ratio. On the other hand, even considering the 
low number statistics, it appears that most of the early-type galaxies (filled-red and -orange 
distributions) show higher SF enhancement if they are in a minor merger compared to major mergers, with the single 
exception of primary ellipticals, which show the same trend for both major and minor mergers. 

We have also separated the sample in different stellar mass bins to study any dependence on 
stellar mass. This separation did not show any clear SF mode dependency on the stellar mass of the merger. 
The SF mode comparisons separated by stellar mass bin are shown in Fig. 
\ref{fig:SFmode_histos_Masses}. 

In order to study the plausibility of a transfer of gas between the merging components, we check whether the SF mode of the early-type component increases when the companion is a gas-rich galaxy. Thus, we separated the sample into merging systems with one early- plus one late-type galaxy (a `mixed' type pair) and we compare their measured SF modes to a control sample where both components are early-type galaxies. As our sub-samples are small, we use the bootstrapping approach (1 million realisations) to look for a difference in the mean values of the SF mode between the mixed pairs and the same type pairs. We make this measure twice - once for the case where the early-type galaxy is the primary component, and second when it is the secondary component.

We find that, when the early-type is the secondary component, this means we have a late-type (gas-rich) primary and early-type (gas-poor) galaxy as a secondary, we find that the mean SF mode of the mixed pairs is larger than that of the control sample in 99\% of the bootstrap realisations. On the other hand, if the early-type (gas-poor) galaxy is the primary and the secondary is the late-type (gas-rich) galaxy, there is no significant difference between the SF mode between the early-type galaxies (the mean SF mode was higher for mixed pairs in only 40\% of the bootstraps). Thus, there is a tentative evidence (albeit with low number statistics) that a gas-rich primary component appears able to transfer some of its gas to the secondary component which enables the latter to significantly enhance its star formation, but the same is not visible if the secondary component is the gas-rich one. This could be because the secondary does not transfer its gas to the primary galaxy efficiently, or because the amount of gas transferred is not sufficient to significantly enhance the early-type primary component's SF.

\subsection{Evolution during the merger process: SMBH activation}

To identify the AGNs in our merger sample, we have used three different methods. The first one is based 
on the emission line diagnostic diagrams (hereafter BPTs, \citealt{BPT81}), which uses the ratio between 
forbidden emission lines to identify galaxies with either star formation activity, active nuclei features, or 
galaxies with both type signatures (composite galaxies). 
The BPT that compares [OIII]/H$_{\beta}$ and [NII]/H$_{\alpha}$ will be referred to as BPT-NII. 
The second BPT diagram that we use is the one that compares [OIII]/H$_{\beta}$ and [SII]/H$_{\alpha}$, 
separating Seyfert galaxies from star-forming and LINER galaxies. This BPT will be referred to as BPT-SII.

\begin{figure}[h!]
\begin{center} 
  \includegraphics[width=0.5\textwidth,clip]{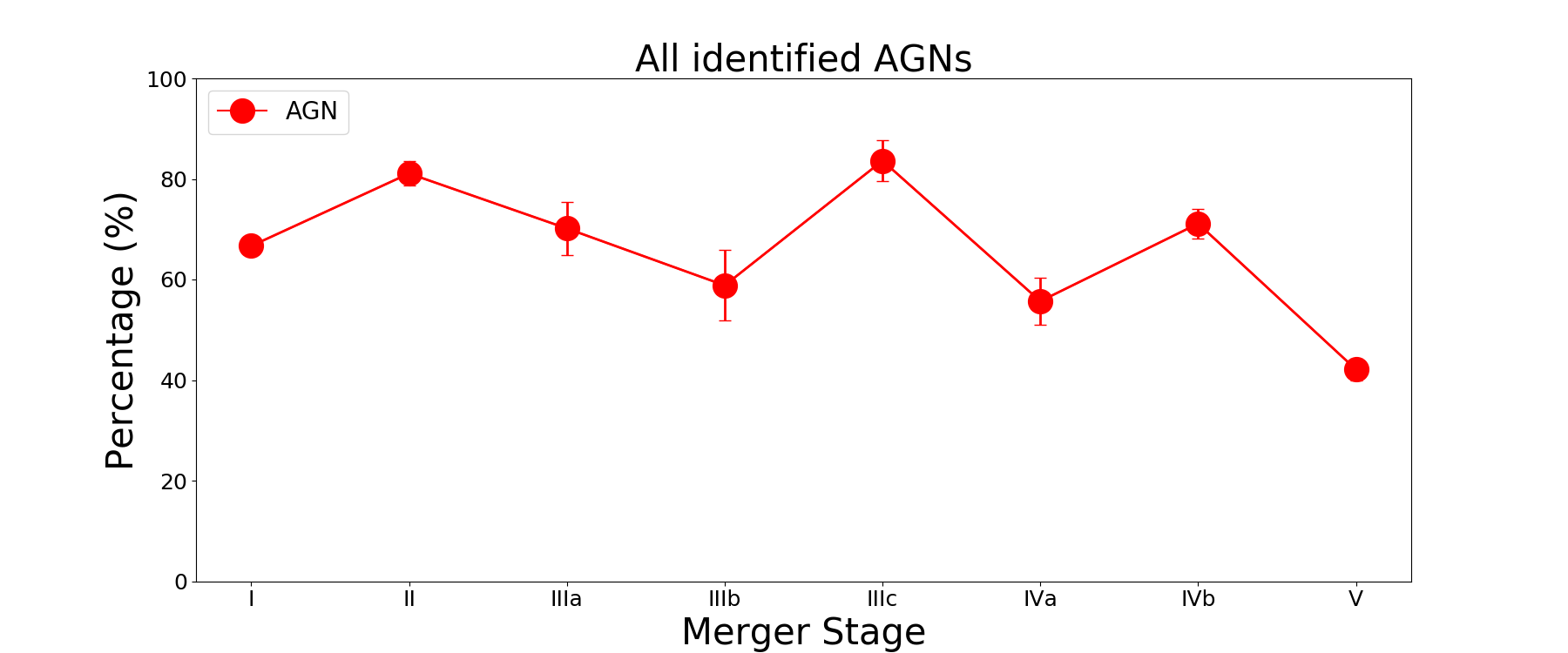}
   \caption{ Percentage of galaxies containing an AGN identified by either 
   BPT-NII, BPT-SII, HeII or WISE diagram in each merger stage. 
}  
\label{fig:AGN_Percentages_Total}
\end{center}
\end{figure}

The second method we use is based on a different emission line, HeII, which is observed only if there is a 
source with hard ionization radiation. This allows us to detect fainter AGNs and/or AGNs embedded in a 
galaxy with strong SF. This method compares HeII/H$_{\beta}$ and [NII]/H$_{\alpha}$ 
as shown by \citet[hereafter SB12]{ShirazinBrinchmann12}. We will refer to this method as HeII diagram.

For these first two methods, we have matched our sample to the OSSY \citep{OSSY11} 
catalogue. This catalogue has improved the emission line measurements from SDSS DR7, see details in 
\citet{OSSY11}. 
Sixty percent of the sample show emission lines. This sub-sample shows a merger stage and a M$_*$ 
distribution similar to the parent sample (see Figure \ref{fig:AGNsample}. 

The third method we use is based on the NIR WISE colours, which compares W1-W2 and W2-W3. 
This colour-colour diagram (hereafter WISE diagram) can identify an AGN in a galaxy as shown by 
\citet[hereafter J11]{Jarrettetal11} and \citet[hereafter S12]{Sternetal12}.

Figure \ref{fig:AGNs_IDs_Morph}  
shows the different methods coloured by morphology. 
The three top panels show the different BPTs, 
BPT-NII at the left and BPT-SII in the middle. Each panel shows the separations found in the literature, 
summarised by \citet{Kewleyetal06}. The bottom-left panel shows the HeII diagram. The separation lines 
are as shown by SB12. The bottom-right panel shows the 
WISE diagram, and the lines separating AGN, (U)LIRGs, spheroids and intermediate and star-forming disks, 
as shown by J11 and S12.

When analysing the diagrams coloured by morphology (Fig. \ref{fig:AGNs_IDs_Morph}), we observe that 
early-type mergers are likely to be in the AGN region and the SF region is mainly populated by spirals, 
as expected. The WISE diagram shows mostly HD galaxies in the AGN \& (U)LIRG region, this is expected 
as (U)LIRGs are generally very perturbed objects. 

Figure \ref{fig:AGN_Percentages_Total} shows the percentage of galaxies with an AGN in each merger stage, 
identified by any of the methods shown previously. Galaxies within the regions: AGN, composite, Seyfert, and 
LINER were all combined together inn the category of galaxies containing an AGN. Errors of the percentages were calculated 
following 
\citet{Sartorietal15}'s work, using binominal statistics for number of objects (N$_{tot}$) in each stage higher 
than 20, and for N$_{tot} \leq $ 20 following the confidence limits shown by \citet{Gehrels86}. 

AGNs seem to decrease overall during the merging process.  
The datapoints at merger stage IIIc could be considered an outlier because at this merger stage there are 
two nuclei visible within the same perturbed object. Thus, the measurements may be contaminated by the 
companion's nucleus.

\begin{figure}[h!]
\begin{center} 
  \includegraphics[width=0.5\textwidth,clip]{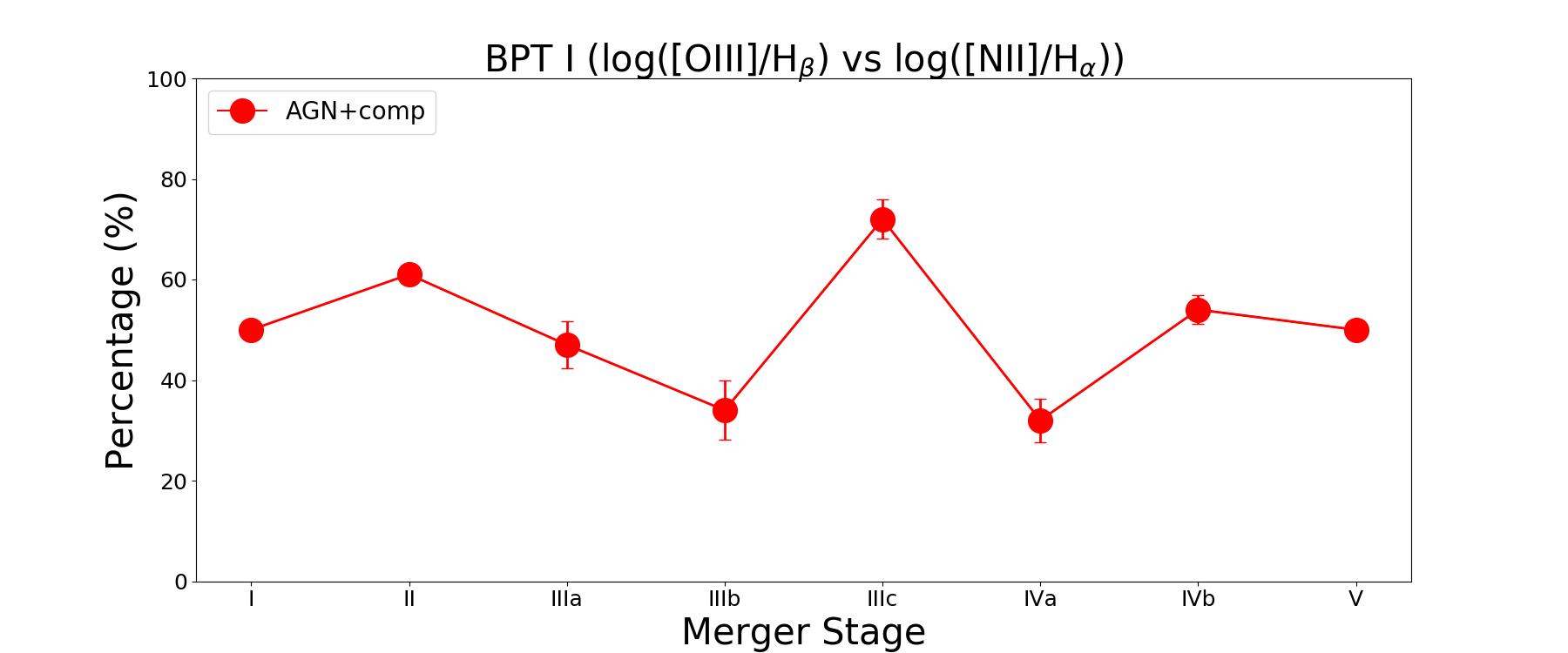}
  \includegraphics[width=0.5\textwidth,clip]{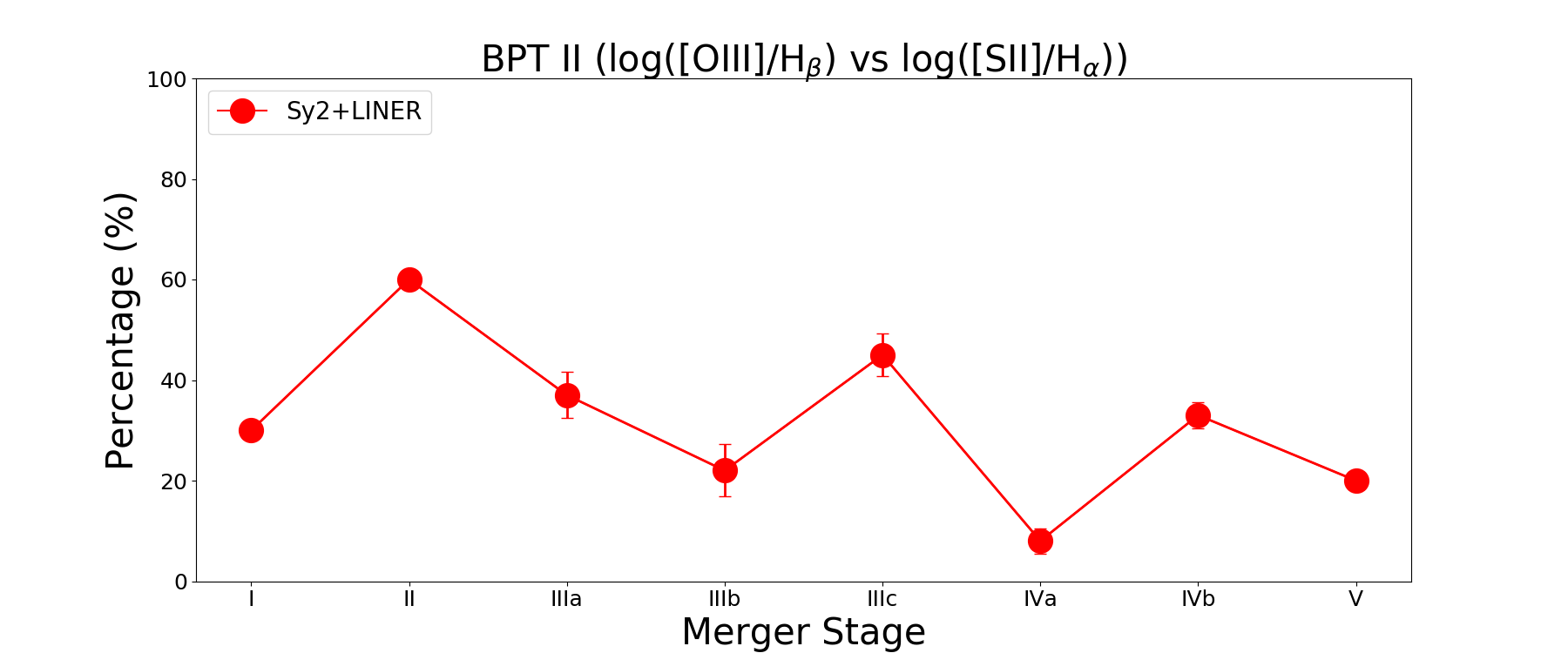}
  \includegraphics[width=0.5\textwidth,clip]{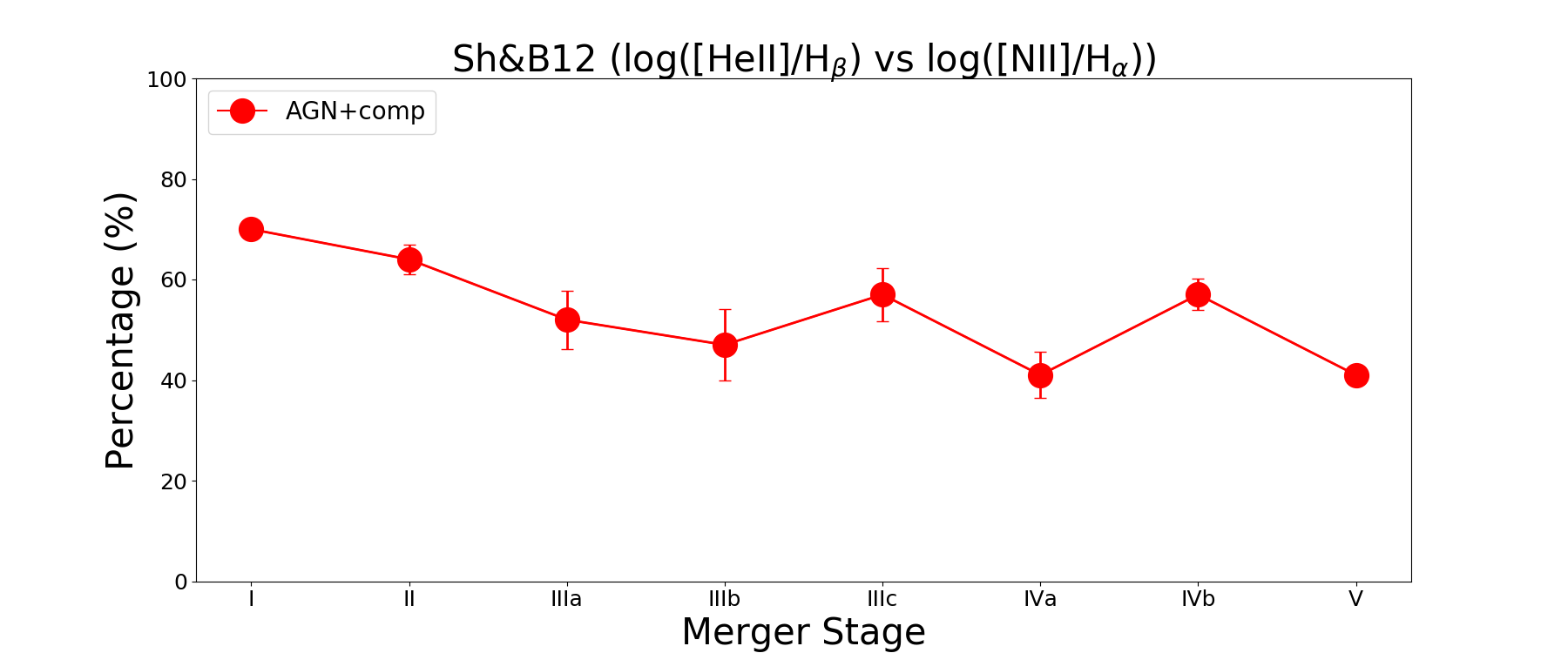}
  \includegraphics[width=0.5\textwidth,clip]{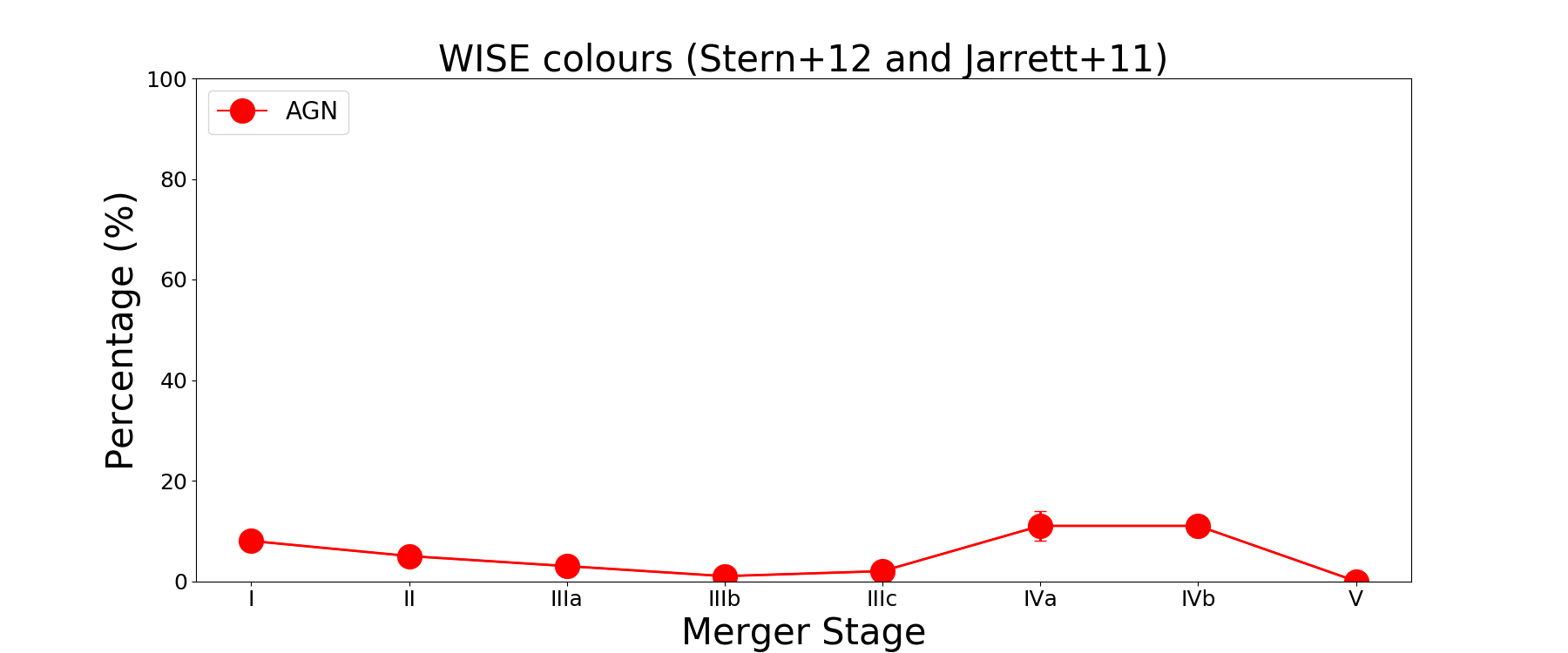}
   \caption{ Percentages of galaxies containing an AGN in each merger stage. 
   From top to bottom: AGNs identified using the BPTI, BPTII, HeII emission line, and WISE diagram. 
}  
\label{fig:AGN_Percentages}
\end{center}
\end{figure}

The decline of AGN fraction towards late merger stages seem to agree and disagree with different simulations. 
For example, \citep{Parketal17} shows that the accretion rate of gas into the central black hole declines for 
merging systems with stellar mass ratios 
higher than 1:3, but peaks before coalescence for pairs with stellar mass ratios of 1:1. This is similar to 
\citep{DiMatteoetal05}'s simulations, showing that the AGN accretion rates tend to peak at coalescence and rise 
from early stages depending on the virial velocity of the merging galaxies. On the other hand, observations 
show that the AGN fractions are higher when the distance between the galaxies is smaller, and peak for 
post-mergers \citep{Ellisonetal13, Satyapaletal14, Hwangetal10}. It's difficult to compare directly with their results as separation may alternately increase and decrease as we move through the merger stages. 

It is also important to consider some of the caveats of using emission line diagnostics to identify AGN. 
In particular, dust obscuration can effectively reduce the spectral lines used to indicate AGN. Hence, 
in principle, it might become more difficult to identify AGN at late merger stages, when the gas and dust 
may become more centrally concentrated. Fortunately, the WISE method should be more sensitive to dusty AGN, as long as they are sufficiently luminous. We note that the apparent bump in the WISE AGN fraction at stage IVa and IVb, and the corresponding high SFRs seen in these merger stages (see bottom panel of Figure 13), may be painting a consistent picture where radially infalling material fuels both nuclear star formation and accretion onto a young, obscured AGN. Also, a common source of data for AGN identification is the spectra 
of the SDSS, in which the spectral fiber is 3" in diameter. 
Therefore, the spectra will tend to include more light from the galaxy as a galaxy's distance increases. 
This latter issue was addressed in \citet{Hwangetal10} by restricting their IRGs sample to galaxies with redshifts z<0.04. 
We attempt to address these issues by checking for any dependency on redshift and dust fraction of our sample. 
We do not observe any clear dependency on these properties, and therefore do not consider these sources 
to strongly bias our data. Also, by following our approach, we are able to more fairly compare our results with other 
measurements of AGN fraction in the literature who follow the same approach 
\citep{Ellisonetal11, Ellisonetal13, Satyapaletal14}. 
Although to better address this issue, we believe that identifying AGN using X-ray emission (which is not affected by 
dust emission) could be of great benefit, and we will consider this topic in a future publication. 

We also analyse the number of AGNs identified by the different methods separately 
(Fig. \ref{fig:AGN_Percentages}) and show where the AGNs identified by one method are located in another 
method's diagram (App. \ref{AGNs_methods}). Figure \ref{fig:AGN_Percentages} shows the percentages, 
from top to bottom, of AGNs identified using the BPT-NII, BPT-SII, HeII diagram, and the WISE diagram. 
First, we notice that the decline of AGNs is more clearly seen with AGNs identified using the BPT-SII and the 
HeII diagram.  
Meanwhile, the AGNs identified using the BPT-NII 
show a more constant distribution with a large scatter. On the other hand, the AGNs identified using 
the WISE diagram show low numbers of AGNs, as these AGNs might not outshine the galaxy light 
at these wavelengths. Thus, disk and spheroidal galaxies dominate at all merger stages. 
The observed AGNs using the WISE diagram are mostly seen at stage IV ($a$ and $b$), where the galaxy is very 
perturbed near coalescence.

The decline in AGN numbers through the merger stages of the merging process suggests that SMBHs 
are more likely to be activated at early stages, when there might be more gas available in the system.

\section{Conclusions} \label{CONCLUSIONS}

Several simulations and observations have shown that mergers can raise their star formation rates (SFR) by 
10-100 times the SFR of unperturbed galaxies at the same stellar mass (M$_*$). We decided to study this 
enhancement using a more timeline-like approach, to try to better capture the chronological order of the merging process. 
For this, we used the M$_*$, SFR, and merger stages from the new catalogue of isolated merging systems (Calder\'on-Castillo et al. 2024). These were estimated applying the SED fitting code MAGPHYS. The photometry used was measured with a semi-automated approach which allows us to extract the entire light of merging galaxies, including faint tidal tails and 
bright star-forming regions, which were excluded by catalogues applying automated photometry. 
We then compared the SFR/M$_*$ values of our galaxies to galaxies on the star-forming main sequence (MS). The distance of galaxies from the MS is described as the `SF mode' and we use this parameter to study the 
SF enhancement during the merging process. 

Our main results are as follows:
\renewcommand{\labelitemi}{$\bullet$}
\begin{itemize}

\item Overall, merging galaxies show an SF mode according to their own morphology. 
For example, spiral and HD galaxies show high SF enhancement, with the highest values reached by the 
HD galaxies, which show a median value of 1 dex above the MS limit.  
On the other hand, elliptical and lenticular galaxies show the lowest SF mode values. However, even the  
elliptical and lenticular galaxies show enhanced SF compared to their unperturbed counterparts. This could suggest that the merging process induces the 
transfer of cold gas between the merging galaxies before coalescence, or enhances SF in any gas that exists prior to the collision. 

\item In general, secondary merging galaxies show higher SF mode compared to their primary galaxy, whether they 
are involved in major or minor mergers. This suggests that the lower mass galaxy seems to be more affected 
by the tidal forces during the interaction. 

\item In minor mergers, when a primary spiral has a HD galaxy as a secondary, the spiral's SF mode is generally higher 
than if the secondary is a spiral. This might suggest that, if the tidal interaction is sufficiently violent to convert a 
secondary spiral into a HD, the interaction is also strong enough to cause a more significant enhancement in the 
primary spiral itself.

\item There is a weak trend increasing SF along the merger sequence, peaking at merger stage IVa. However, overall, 
the SF is enhanced since early merger stages throughout the merging process. 

\item We don't see a clear signal that the AGN fraction changes with merger stage across various AGN diagnostics, and sometimes the trends can be very noisy. However, there is a small hint for a decrease in one or two cases (HeII diagram and BPT-SII).
\end{itemize}

Our results suggest that the transfer of cold gas may possibly occur since very early stages in the merger sequence. 
This is observed in simulations, where a bridge of gas can temporarily link both galaxies during close passages 
(e.g. \citealt{Wenigeretal09, Mosteretal11, HwangnPark15}). However, this bridge of gas can be compressed or even 
ionised depending on the presence of a hot halo in one or both of the galaxies involved in the merger 
\citep{Mosteretal11, HwangnPark15}. If this really is gas transfer, then it raises the possibility that interacting pairs 
could be used to constrain the hot gas halo content of galaxies. However, another possibility is that the tidal 
interactions may simply be inducing or enhancing star formation in any small quantities of gas that exist within the 
pre-merger early-types.  

In addition to the SF enhancement being observed since early merger stages, it is also interesting to note 
that this enhancement can be seen to last for the entire merging process.
This is often not what is seen in many previous simulations of interacting galaxies, where SF is mildly enhanced after first passage and then more strongly enhanced approaching coalescence but the SFR drops rapidly afterwards \citep{Parketal17, Morenoetal19}. This might be related to the sub-grid physics recipes used to model star formation and 
feedback which can act to rapidly suppress strong starbursts. It might also be related to the fact that 
many of these simulations were not conducted in a fully cosmological context, and thus could lack the 
cosmological inflow of gas necessary to sustain SF for longer periods of time. Perhaps, the new generation 
of cosmological simulations will have sufficient resolution to enable us to better understand the merger 
process, and to study the transfer of gas between interacting galaxies. 

Nevertheless, both simulations and observations tend to agree that the SF can be enhanced significantly close to the moment of coalescence \citep{Rubinuretal24}. Although the strength of the starburst during this time is likely sensitive to the presence of a bulge, mass ratio of the merger, and the response to AGN feedback \citep{Parketal17}.

If all merging galaxies go through an extended period of time of substantial SF enhancement, such galaxies may significantly increase their stellar mass through the merger process. For example, low-M$_*$ galaxies may nearly double their mass. Thus, these results show that mergers can play a crucial role in galaxy evolution. 

\begin{acknowledgements}
We would like to thank the referee for their encouragement and constructive comments. 
RS acknowledges financial support from FONDECYT Regular 2023 project No. 1230441. And Iniciativa PROY-0000001125 - MINGAL (Millennium Nucleus for GALaxies)
\end{acknowledgements}

\bibliographystyle{aa} 
\bibliography{MergingSystemsinIsolatedEnvironmentsII} 

\begin{thebibliography}{72}
\expandafter\ifx\csname natexlab\endcsname\relax\def\natexlab#1{#1}\fi

\bibitem[{{Arp}(1966)}]{Arp96}
{Arp}, H. 1966, \apjs, 14, 1

\bibitem[{{Baldwin} {et~al.}(1981){Baldwin}, {Phillips}, \& {Terlevich}}]{BPT81}
{Baldwin}, J.~A., {Phillips}, M.~M., \& {Terlevich}, R. 1981, \pasp, 93, 5

\bibitem[{{Byrne-Mamahit} {et~al.}(2023){Byrne-Mamahit}, {Hani}, {Ellison}, {Quai}, \& {Patton}}]{ByrneMamahitetal23}
{Byrne-Mamahit}, S., {Hani}, M.~H., {Ellison}, S.~L., {Quai}, S., \& {Patton}, D.~R. 2023, \mnras, 519, 4966

\bibitem[{{Calder{\'o}n-Castillo} {et~al.}(2024){Calder{\'o}n-Castillo}, {Nagar}, {Yi}, {Chang}, {Leiton}, \& {Hughes}}]{PauCC24}
{Calder{\'o}n-Castillo}, P., {Nagar}, N.~M., {Yi}, S.~K., {et~al.} 2024, \aap, 686, A151

\bibitem[{{Cano-D{\'\i}az} {et~al.}(2019){Cano-D{\'\i}az}, {{\'A}vila-Reese}, {S{\'a}nchez}, {Hern{\'a}ndez-Toledo}, {Rodr{\'\i}guez-Puebla}, {Boquien}, \& {Ibarra-Medel}}]{CanoDiazetal2019}
{Cano-D{\'\i}az}, M., {{\'A}vila-Reese}, V., {S{\'a}nchez}, S.~F., {et~al.} 2019, \mnras, 488, 3929

\bibitem[{{Catinella} {et~al.}(2018){Catinella}, {Saintonge}, {Janowiecki}, {Cortese}, {Dav{\'e}}, {Lemonias}, {Cooper}, {Schiminovich}, {Hummels}, {Fabello}, {Ger{\'e}b}, {Kilborn}, \& {Wang}}]{Catinellaetal18}
{Catinella}, B., {Saintonge}, A., {Janowiecki}, S., {et~al.} 2018, \mnras, 476, 875

\bibitem[{{Chang} {et~al.}(2015){Chang}, {van der Wel}, {da Cunha}, \& {Rix}}]{Changetal15}
{Chang}, Y.-Y., {van der Wel}, A., {da Cunha}, E., \& {Rix}, H.-W. 2015, \apjs, 219, 8

\bibitem[{{da Cunha} {et~al.}(2008){da Cunha}, {Charlot}, \& {Elbaz}}]{daCunhaetal2008}
{da Cunha}, E., {Charlot}, S., \& {Elbaz}, D. 2008, \mnras, 388, 1595

\bibitem[{{Daddi} {et~al.}(2007){Daddi}, {Dickinson}, {Morrison}, {Chary}, {Cimatti}, {Elbaz}, {Frayer}, {Renzini}, {Pope}, {Alexander}, {Bauer}, {Giavalisco}, {Huynh}, {Kurk}, \& {Mignoli}}]{Daddietal07}
{Daddi}, E., {Dickinson}, M., {Morrison}, G., {et~al.} 2007, \apj, 670, 156

\bibitem[{{Darg} {et~al.}(2010){Darg}, {Kaviraj}, {Lintott}, {Schawinski}, {Sarzi}, {Bamford}, {Silk}, {Andreescu}, {Murray}, {Nichol}, {Raddick}, {Slosar}, {Szalay}, {Thomas}, \& {Vandenberg}}]{Dargetal10}
{Darg}, D.~W., {Kaviraj}, S., {Lintott}, C.~J., {et~al.} 2010, \mnras, 401, 1552

\bibitem[{{Davies} {et~al.}(2015){Davies}, {Robotham}, {Driver}, {Alpaslan}, {Baldry}, {Bland-Hawthorn}, {Brough}, {Brown}, {Cluver}, {Drinkwater}, {Foster}, {Grootes}, {Konstantopoulos}, {Lara-L{\'o}pez}, {L{\'o}pez-S{\'a}nchez}, {Loveday}, {Meyer}, {Moffett}, {Norberg}, {Owers}, {Popescu}, {De Propris}, {Sharp}, {Tuffs}, {Wang}, {Wilkins}, {Dunne}, {Bourne}, \& {Smith}}]{Daviesetal15}
{Davies}, L.~J.~M., {Robotham}, A.~S.~G., {Driver}, S.~P., {et~al.} 2015, \mnras, 452, 616

\bibitem[{{Di Matteo} {et~al.}(2008){Di Matteo}, {Bournaud}, {Martig}, {Combes}, {Melchior}, \& {Semelin}}]{DiMatteoetal08}
{Di Matteo}, P., {Bournaud}, F., {Martig}, M., {et~al.} 2008, \aap, 492, 31

\bibitem[{{Di Matteo} {et~al.}(2005){Di Matteo}, {Springel}, \& {Hernquist}}]{DiMatteoetal05}
{Di Matteo}, T., {Springel}, V., \& {Hernquist}, L. 2005, \nat, 433, 604

\bibitem[{{Elbaz} {et~al.}(2007){Elbaz}, {Daddi}, {Le Borgne}, {Dickinson}, {Alexander}, {Chary}, {Starck}, {Brandt}, {Kitzbichler}, {MacDonald}, {Nonino}, {Popesso}, {Stern}, \& {Vanzella}}]{Elbazetal07}
{Elbaz}, D., {Daddi}, E., {Le Borgne}, D., {et~al.} 2007, \aap, 468, 33

\bibitem[{{Elbaz} {et~al.}(2011){Elbaz}, {Dickinson}, {Hwang}, {D{\'{\i}}az-Santos}, {Magdis}, {Magnelli}, {Le Borgne}, {Galliano}, {Pannella}, {Chanial}, {Armus}, {Charmandaris}, {Daddi}, {Aussel}, {Popesso}, {Kartaltepe}, {Altieri}, {Valtchanov}, {Coia}, {Dannerbauer}, {Dasyra}, {Leiton}, {Mazzarella}, {Alexander}, {Buat}, {Burgarella}, {Chary}, {Gilli}, {Ivison}, {Juneau}, {Le Floc'h}, {Lutz}, {Morrison}, {Mullaney}, {Murphy}, {Pope}, {Scott}, {Brodwin}, {Calzetti}, {Cesarsky}, {Charlot}, {Dole}, {Eisenhardt}, {Ferguson}, {F{\"o}rster Schreiber}, {Frayer}, {Giavalisco}, {Huynh}, {Koekemoer}, {Papovich}, {Reddy}, {Surace}, {Teplitz}, {Yun}, \& {Wilson}}]{Elbazetal11}
{Elbaz}, D., {Dickinson}, M., {Hwang}, H.~S., {et~al.} 2011, \aap, 533, A119

\bibitem[{{Ellison} {et~al.}(2013){Ellison}, {Mendel}, {Patton}, \& {Scudder}}]{Ellisonetal13}
{Ellison}, S.~L., {Mendel}, J.~T., {Patton}, D.~R., \& {Scudder}, J.~M. 2013, \mnras, 435, 3627

\bibitem[{{Ellison} {et~al.}(2011){Ellison}, {Patton}, {Mendel}, \& {Scudder}}]{Ellisonetal11}
{Ellison}, S.~L., {Patton}, D.~R., {Mendel}, J.~T., \& {Scudder}, J.~M. 2011, \mnras, 418, 2043

\bibitem[{{Ellison} {et~al.}(2008){Ellison}, {Patton}, {Simard}, \& {McConnachie}}]{Ellisonetal08}
{Ellison}, S.~L., {Patton}, D.~R., {Simard}, L., \& {McConnachie}, A.~W. 2008, \aj, 135, 1877

\bibitem[{{Ellison} {et~al.}(2010){Ellison}, {Patton}, {Simard}, {McConnachie}, {Baldry}, \& {Mendel}}]{Ellisonetal10}
{Ellison}, S.~L., {Patton}, D.~R., {Simard}, L., {et~al.} 2010, \mnras, 407, 1514

\bibitem[{{Engel} {et~al.}(2010){Engel}, {Tacconi}, {Davies}, {Neri}, {Smail}, {Chapman}, {Genzel}, {Cox}, {Greve}, {Ivison}, {Blain}, {Bertoldi}, \& {Omont}}]{Engeletal10}
{Engel}, H., {Tacconi}, L.~J., {Davies}, R.~I., {et~al.} 2010, \apj, 724, 233

\bibitem[{{Gardu{\~n}o} {et~al.}(2021){Gardu{\~n}o}, {Lara-L{\'o}pez}, {L{\'o}pez-Cruz}, {Hopkins}, {Owers}, {Pimbblet}, \& {Holwerda}}]{Gardunoetal21}
{Gardu{\~n}o}, L.~E., {Lara-L{\'o}pez}, M.~A., {L{\'o}pez-Cruz}, O., {et~al.} 2021, \mnras, 501, 2969

\bibitem[{{Gehrels}(1986)}]{Gehrels86}
{Gehrels}, N. 1986, \apj, 303, 336

\bibitem[{{Holincheck} {et~al.}(2016){Holincheck}, {Wallin}, {Borne}, {Fortson}, {Lintott}, {Smith}, {Bamford}, {Keel}, \& {Parrish}}]{Holinchecketal16}
{Holincheck}, A.~J., {Wallin}, J.~F., {Borne}, K., {et~al.} 2016, \mnras, 459, 720

\bibitem[{{Hopkins} {et~al.}(2008){Hopkins}, {Hernquist}, {Cox}, {Younger}, \& {Besla}}]{Hopkinsetal08}
{Hopkins}, P.~F., {Hernquist}, L., {Cox}, T.~J., {Younger}, J.~D., \& {Besla}, G. 2008, \apj, 688, 757

\bibitem[{{Hwang} {et~al.}(2011){Hwang}, {Elbaz}, {Dickinson}, {Charmandaris}, {Daddi}, {Le Borgne}, {Buat}, {Magdis}, {Altieri}, {Aussel}, {Coia}, {Dannerbauer}, {Dasyra}, {Kartaltepe}, {Leiton}, {Magnelli}, {Popesso}, \& {Valtchanov}}]{Hwangetal11}
{Hwang}, H.~S., {Elbaz}, D., {Dickinson}, M., {et~al.} 2011, \aap, 535, A60

\bibitem[{{Hwang} {et~al.}(2010){Hwang}, {Elbaz}, {Lee}, {Jeong}, {Park}, {Lee}, \& {Lee}}]{Hwangetal10}
{Hwang}, H.~S., {Elbaz}, D., {Lee}, J.~C., {et~al.} 2010, \aap, 522, A33

\bibitem[{{Hwang} \& {Park}(2015)}]{HwangnPark15}
{Hwang}, J.-S. \& {Park}, C. 2015, \apj, 805, 131

\bibitem[{{Jarrett} {et~al.}(2011){Jarrett}, {Cohen}, {Masci}, {Wright}, {Stern}, {Benford}, {Blain}, {Carey}, {Cutri}, {Eisenhardt}, {Lonsdale}, {Mainzer}, {Marsh}, {Padgett}, {Petty}, {Ressler}, {Skrutskie}, {Stanford}, {Surace}, {Tsai}, {Wheelock}, \& {Yan}}]{Jarrettetal11}
{Jarrett}, T.~H., {Cohen}, M., {Masci}, F., {et~al.} 2011, \apj, 735, 112

\bibitem[{{Karim} {et~al.}(2011){Karim}, {Schinnerer}, {Mart{\'{\i}}nez-Sansigre}, {Sargent}, {van der Wel}, {Rix}, {Ilbert}, {Smol{\v c}i{\'c}}, {Carilli}, {Pannella}, {Koekemoer}, {Bell}, \& {Salvato}}]{Karimetal11}
{Karim}, A., {Schinnerer}, E., {Mart{\'{\i}}nez-Sansigre}, A., {et~al.} 2011, \apj, 730, 61

\bibitem[{{Kewley} {et~al.}(2006){Kewley}, {Groves}, {Kauffmann}, \& {Heckman}}]{Kewleyetal06}
{Kewley}, L.~J., {Groves}, B., {Kauffmann}, G., \& {Heckman}, T. 2006, \mnras, 372, 961

\bibitem[{{Larson} \& {Tinsley}(1978)}]{LarsonnTinsley78}
{Larson}, R.~B. \& {Tinsley}, B.~M. 1978, \apj, 219, 46

\bibitem[{{Li} {et~al.}(2023){Li}, {Ho}, \& {Shangguan}}]{Lietal23}
{Li}, Y.~A., {Ho}, L.~C., \& {Shangguan}, J. 2023, \apj, 953, 91

\bibitem[{{Mihos} \& {Hernquist}(1994{\natexlab{a}})}]{MihosnHernquist94a}
{Mihos}, J.~C. \& {Hernquist}, L. 1994{\natexlab{a}}, \apjl, 425, L13

\bibitem[{{Mihos} \& {Hernquist}(1994{\natexlab{b}})}]{MihosnHernquist94b}
{Mihos}, J.~C. \& {Hernquist}, L. 1994{\natexlab{b}}, \apjl, 431, L9

\bibitem[{{Moreno} {et~al.}(2021){Moreno}, {Torrey}, {Ellison}, {Patton}, {Bottrell}, {Bluck}, {Hani}, {Hayward}, {Bullock}, {Hopkins}, \& {Hernquist}}]{Morenoetal21}
{Moreno}, J., {Torrey}, P., {Ellison}, S.~L., {et~al.} 2021, \mnras, 503, 3113

\bibitem[{{Moreno} {et~al.}(2019){Moreno}, {Torrey}, {Ellison}, {Patton}, {Hopkins}, {Bueno}, {Hayward}, {Narayanan}, {Kere{\v{s}}}, {Bluck}, \& {Hernquist}}]{Morenoetal19}
{Moreno}, J., {Torrey}, P., {Ellison}, S.~L., {et~al.} 2019, \mnras, 485, 1320

\bibitem[{{Moster} {et~al.}(2011){Moster}, {Macci{\`o}}, {Somerville}, {Naab}, \& {Cox}}]{Mosteretal11}
{Moster}, B.~P., {Macci{\`o}}, A.~V., {Somerville}, R.~S., {Naab}, T., \& {Cox}, T.~J. 2011, \mnras, 415, 3750

\bibitem[{{Noeske} {et~al.}(2007){Noeske}, {Weiner}, {Faber}, {Papovich}, {Koo}, {Somerville}, {Bundy}, {Conselice}, {Newman}, {Schiminovich}, {Le Floc'h}, {Coil}, {Rieke}, {Lotz}, {Primack}, {Barmby}, {Cooper}, {Davis}, {Ellis}, {Fazio}, {Guhathakurta}, {Huang}, {Kassin}, {Martin}, {Phillips}, {Rich}, {Small}, {Willmer}, \& {Wilson}}]{Noeskeetal07}
{Noeske}, K.~G., {Weiner}, B.~J., {Faber}, S.~M., {et~al.} 2007, \apjl, 660, L43

\bibitem[{{Oh} {et~al.}(2011){Oh}, {Sarzi}, {Schawinski}, \& {Yi}}]{OSSY11}
{Oh}, K., {Sarzi}, M., {Schawinski}, K., \& {Yi}, S.~K. 2011, \apjs, 195, 13

\bibitem[{{Pannella} {et~al.}(2009){Pannella}, {Carilli}, {Daddi}, {McCracken}, {Owen}, {Renzini}, {Strazzullo}, {Civano}, {Koekemoer}, {Schinnerer}, {Scoville}, {Smol{\v c}i{\'c}}, {Taniguchi}, {Aussel}, {Kneib}, {Ilbert}, {Mellier}, {Salvato}, {Thompson}, \& {Willott}}]{Pannellaetal09}
{Pannella}, M., {Carilli}, C.~L., {Daddi}, E., {et~al.} 2009, \apjl, 698, L116

\bibitem[{{Pannella} {et~al.}(2015){Pannella}, {Elbaz}, {Daddi}, {Dickinson}, {Hwang}, {Schreiber}, {Strazzullo}, {Aussel}, {Bethermin}, {Buat}, {Charmandaris}, {Cibinel}, {Juneau}, {Ivison}, {Le Borgne}, {Le Floc'h}, {Leiton}, {Lin}, {Magdis}, {Morrison}, {Mullaney}, {Onodera}, {Renzini}, {Salim}, {Sargent}, {Scott}, {Shu}, \& {Wang}}]{Pannellaetal15}
{Pannella}, M., {Elbaz}, D., {Daddi}, E., {et~al.} 2015, \apj, 807, 141

\bibitem[{{Park} {et~al.}(2017){Park}, {Smith}, \& {Yi}}]{Parketal17}
{Park}, J., {Smith}, R., \& {Yi}, S.~K. 2017, \apj, 845, 128

\bibitem[{{Patton} {et~al.}(2016){Patton}, {Qamar}, {Ellison}, {Bluck}, {Simard}, {Mendel}, {Moreno}, \& {Torrey}}]{Pattonetal16}
{Patton}, D.~R., {Qamar}, F.~D., {Ellison}, S.~L., {et~al.} 2016, \mnras, 461, 2589

\bibitem[{{Patton} {et~al.}(2013){Patton}, {Torrey}, {Ellison}, {Mendel}, \& {Scudder}}]{Pattonetal13}
{Patton}, D.~R., {Torrey}, P., {Ellison}, S.~L., {Mendel}, J.~T., \& {Scudder}, J.~M. 2013, \mnras, 433, L59

\bibitem[{{Pearson} {et~al.}(2019){Pearson}, {Wang}, {Alpaslan}, {Baldry}, {Bilicki}, {Brown}, {Grootes}, {Holwerda}, {Kitching}, {Kruk}, \& {van der Tak}}]{Pearsonetal19}
{Pearson}, W.~J., {Wang}, L., {Alpaslan}, M., {et~al.} 2019, \aap, 631, A51

\bibitem[{{Renaud} {et~al.}(2022){Renaud}, {Segovia Otero}, \& {Agertz}}]{Renaudetal22}
{Renaud}, F., {Segovia Otero}, {\'A}., \& {Agertz}, O. 2022, \mnras, 516, 4922

\bibitem[{{Renzini} \& {Peng}(2015)}]{RenzininPeng15}
{Renzini}, A. \& {Peng}, Y.-j. 2015, \apjl, 801, L29

\bibitem[{{Rodighiero} {et~al.}(2011){Rodighiero}, {Daddi}, {Baronchelli}, {Cimatti}, {Renzini}, {Aussel}, {Popesso}, {Lutz}, {Andreani}, {Berta}, {Cava}, {Elbaz}, {Feltre}, {Fontana}, {F{\"o}rster Schreiber}, {Franceschini}, {Genzel}, {Grazian}, {Gruppioni}, {Ilbert}, {Le Floch}, {Magdis}, {Magliocchetti}, {Magnelli}, {Maiolino}, {McCracken}, {Nordon}, {Poglitsch}, {Santini}, {Pozzi}, {Riguccini}, {Tacconi}, {Wuyts}, \& {Zamorani}}]{Rodighieroetal11}
{Rodighiero}, G., {Daddi}, E., {Baronchelli}, I., {et~al.} 2011, \apjl, 739, L40

\bibitem[{{Rodighiero} {et~al.}(2014){Rodighiero}, {Renzini}, {Daddi}, {Baronchelli}, {Berta}, {Cresci}, {Franceschini}, {Gruppioni}, {Lutz}, {Mancini}, {Santini}, {Zamorani}, {Silverman}, {Kashino}, {Andreani}, {Cimatti}, {S{\'a}nchez}, {Le Floch}, {Magnelli}, {Popesso}, \& {Pozzi}}]{Rodighieroetal14}
{Rodighiero}, G., {Renzini}, A., {Daddi}, E., {et~al.} 2014, \mnras, 443, 19

\bibitem[{{Rodriguez-Gomez} {et~al.}(2015){Rodriguez-Gomez}, {Genel}, {Vogelsberger}, {Sijacki}, {Pillepich}, {Sales}, {Torrey}, {Snyder}, {Nelson}, {Springel}, {Ma}, \& {Hernquist}}]{RodriguezGomezetal15}
{Rodriguez-Gomez}, V., {Genel}, S., {Vogelsberger}, M., {et~al.} 2015, \mnras, 449, 49

\bibitem[{{Rubinur} {et~al.}(2024){Rubinur}, {Das}, {Kharb}, {Yadav}, {Mondal}, \& {Rahna}}]{Rubinuretal24}
{Rubinur}, K., {Das}, M., {Kharb}, P., {et~al.} 2024, \mnras, 528, 4432

\bibitem[{{Sanders} {et~al.}(2003){Sanders}, {Mazzarella}, {Kim}, {Surace}, \& {Soifer}}]{Sandersetal03}
{Sanders}, D.~B., {Mazzarella}, J.~M., {Kim}, D.-C., {Surace}, J.~A., \& {Soifer}, B.~T. 2003, \aj, 126, 1607

\bibitem[{{Sanders} \& {Mirabel}(1996)}]{Sandersetal96}
{Sanders}, D.~B. \& {Mirabel}, I.~F. 1996, \araa, 34, 749

\bibitem[{{Sartori} {et~al.}(2015){Sartori}, {Schawinski}, {Treister}, {Trakhtenbrot}, {Koss}, {Shirazi}, \& {Oh}}]{Sartorietal15}
{Sartori}, L.~F., {Schawinski}, K., {Treister}, E., {et~al.} 2015, \mnras, 454, 3722

\bibitem[{{Satyapal} {et~al.}(2014){Satyapal}, {Ellison}, {McAlpine}, {Hickox}, {Patton}, \& {Mendel}}]{Satyapaletal14}
{Satyapal}, S., {Ellison}, S.~L., {McAlpine}, W., {et~al.} 2014, \mnras, 441, 1297

\bibitem[{{Schreiber} {et~al.}(2015){Schreiber}, {Pannella}, {Elbaz}, {B{\'e}thermin}, {Inami}, {Dickinson}, {Magnelli}, {Wang}, {Aussel}, {Daddi}, {Juneau}, {Shu}, {Sargent}, {Buat}, {Faber}, {Ferguson}, {Giavalisco}, {Koekemoer}, {Magdis}, {Morrison}, {Papovich}, {Santini}, \& {Scott}}]{Schreiberetal15}
{Schreiber}, C., {Pannella}, M., {Elbaz}, D., {et~al.} 2015, \aap, 575, A74

\bibitem[{{Schreiber} {et~al.}(2017){Schreiber}, {Pannella}, {Leiton}, {Elbaz}, {Wang}, {Okumura}, \& {Labb{\'e}}}]{Schreiberetal17}
{Schreiber}, C., {Pannella}, M., {Leiton}, R., {et~al.} 2017, \aap, 599, A134

\bibitem[{{Scudder} {et~al.}(2015){Scudder}, {Ellison}, {Momjian}, {Rosenberg}, {Torrey}, {Patton}, {Fertig}, \& {Mendel}}]{Scudderetal15}
{Scudder}, J.~M., {Ellison}, S.~L., {Momjian}, E., {et~al.} 2015, \mnras, 449, 3719

\bibitem[{{Scudder} {et~al.}(2012){Scudder}, {Ellison}, {Torrey}, {Patton}, \& {Mendel}}]{Scudderetal12}
{Scudder}, J.~M., {Ellison}, S.~L., {Torrey}, P., {Patton}, D.~R., \& {Mendel}, J.~T. 2012, \mnras, 426, 549

\bibitem[{{Shirazi} \& {Brinchmann}(2012)}]{ShirazinBrinchmann12}
{Shirazi}, M. \& {Brinchmann}, J. 2012, \mnras, 421, 1043

\bibitem[{{Springel} \& {Hernquist}(2005)}]{SpringelnHernquist05}
{Springel}, V. \& {Hernquist}, L. 2005, \apjl, 622, L9

\bibitem[{{Stern} {et~al.}(2012){Stern}, {Assef}, {Benford}, {Blain}, {Cutri}, {Dey}, {Eisenhardt}, {Griffith}, {Jarrett}, {Lake}, {Masci}, {Petty}, {Stanford}, {Tsai}, {Wright}, {Yan}, {Harrison}, \& {Madsen}}]{Sternetal12}
{Stern}, D., {Assef}, R.~J., {Benford}, D.~J., {et~al.} 2012, \apj, 753, 30

\bibitem[{{Toomre} \& {Toomre}(1972)}]{TnT72}
{Toomre}, A. \& {Toomre}, J. 1972, \apj, 178, 623

\bibitem[{{van de Voort} {et~al.}(2018){van de Voort}, {Davis}, {Matsushita}, {Rowlands}, {Shabala}, {Allison}, {Ting}, {Sansom}, \& {van der Werf}}]{vandeVoortetal18}
{van de Voort}, F., {Davis}, T.~A., {Matsushita}, S., {et~al.} 2018, \mnras, 476, 122

\bibitem[{{Veilleux} {et~al.}(2002){Veilleux}, {Kim}, \& {Sanders}}]{Veilleuxetal02}
{Veilleux}, S., {Kim}, D.-C., \& {Sanders}, D.~B. 2002, \apjs, 143, 315

\bibitem[{{Vorontsov-Velyaminov} {et~al.}(2001){Vorontsov-Velyaminov}, {Noskova}, \& {Arkhipova}}]{VVetal01}
{Vorontsov-Velyaminov}, B.~A., {Noskova}, R.~I., \& {Arkhipova}, V.~P. 2001, Astronomical and Astrophysical Transactions, 20, 717

\bibitem[{{Weinzirl}(2015)}]{Weinzirl15}
{Weinzirl}, T. 2015, {Probing Galaxy Evolution by Unveiling the Structure of Massive Galaxies Across Cosmic Time and Diverse Environments}

\bibitem[{{Weniger} {et~al.}(2009){Weniger}, {Theis}, \& {Harfst}}]{Wenigeretal09}
{Weniger}, J., {Theis}, C., \& {Harfst}, S. 2009, Astronomische Nachrichten, 330, 1019

\bibitem[{{Whitaker} {et~al.}(2014){Whitaker}, {Franx}, {Leja}, {van Dokkum}, {Henry}, {Skelton}, {Fumagalli}, {Momcheva}, {Brammer}, {Labb{\'e}}, {Nelson}, \& {Rigby}}]{Whitakeretal14}
{Whitaker}, K.~E., {Franx}, M., {Leja}, J., {et~al.} 2014, \apj, 795, 104

\bibitem[{{Whitaker} {et~al.}(2012){Whitaker}, {van Dokkum}, {Brammer}, \& {Franx}}]{Whitakeretal12}
{Whitaker}, K.~E., {van Dokkum}, P.~G., {Brammer}, G., \& {Franx}, M. 2012, \apjl, 754, L29

\bibitem[{{Wright} {et~al.}(2016){Wright}, {Robotham}, {Bourne}, {Driver}, {Dunne}, {Maddox}, {Alpaslan}, {Andrews}, {Bauer}, {Bland-Hawthorn}, {Brough}, {Brown}, {Clarke}, {Cluver}, {Davies}, {Grootes}, {Holwerda}, {Hopkins}, {Jarrett}, {Kafle}, {Lange}, {Liske}, {Loveday}, {Moffett}, {Norberg}, {Popescu}, {Smith}, {Taylor}, {Tuffs}, {Wang}, \& {Wilkins}}]{LAMBDAR}
{Wright}, A.~H., {Robotham}, A.~S.~G., {Bourne}, N., {et~al.} 2016, \mnras, 460, 765

\bibitem[{{Wuyts} {et~al.}(2011){Wuyts}, {F{\"o}rster Schreiber}, {van der Wel}, {Magnelli}, {Guo}, {Genzel}, {Lutz}, {Aussel}, {Barro}, {Berta}, {Cava}, {Graci{\'a}-Carpio}, {Hathi}, {Huang}, {Kocevski}, {Koekemoer}, {Lee}, {Le Floc'h}, {McGrath}, {Nordon}, {Popesso}, {Pozzi}, {Riguccini}, {Rodighiero}, {Saintonge}, \& {Tacconi}}]{Wuytsetal11}
{Wuyts}, S., {F{\"o}rster Schreiber}, N.~M., {van der Wel}, A., {et~al.} 2011, \apj, 742, 96

\end{thebibliography}

\begin{appendix}
    
\section{Stellar Mass bins} \label{App:Mstellarbins}

To identify any dependence of the SF mode on stellar mass, we have separated the mergers into three 
stellar mass bins. The stellar mass bins are calculated from the stellar mass of the system 
(primary's plus secondary's stellar masses). The low-mass bin shows stellar masses:  
$log(M_*/M_{\odot}) < 9.5$, the medium-mass bin shows stellar masses within the range: 
$9.5 < log(M_*/M_{\odot}) < 10.5$, and the high-mass bin shows stellar masses: 
$10.5 < log(M_*/M_{\odot})$. 

\begin{figure}[H]
\begin{center} 
\textbf{ Low-M$_*$}\par\smallskip
  \includegraphics[bb=50 0 800 415, width=0.4\textwidth,clip]{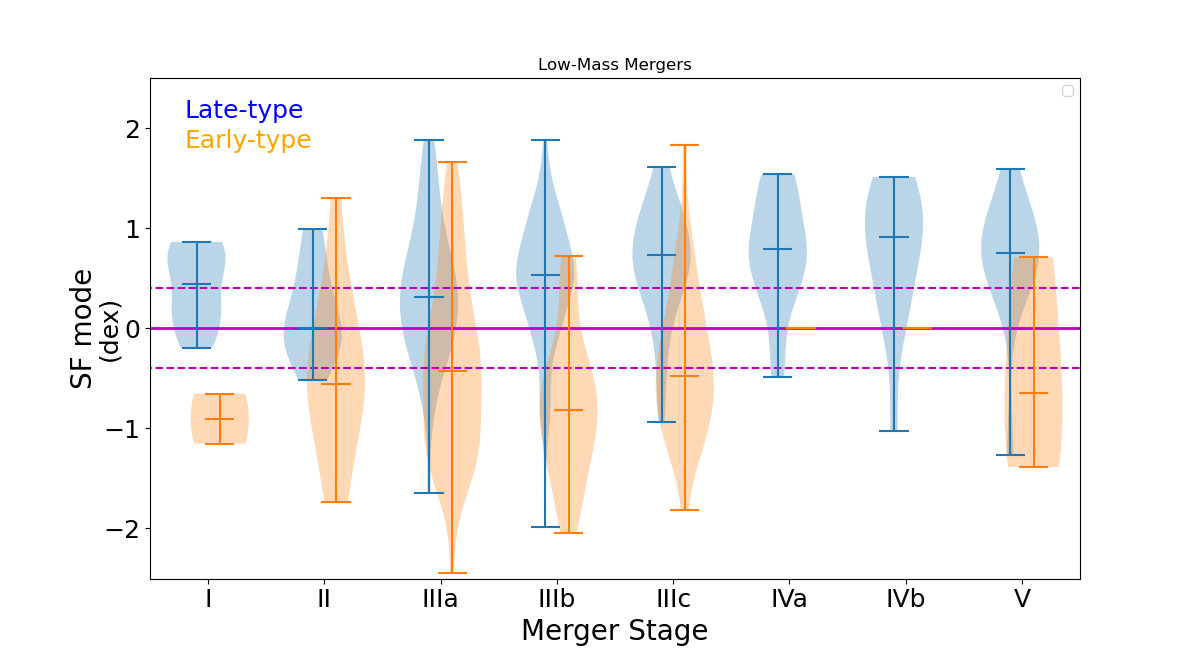}
  
\textbf{ Med-M$_*$}\par\smallskip
  \includegraphics[bb=50 0 800 415, width=0.4\textwidth,clip]{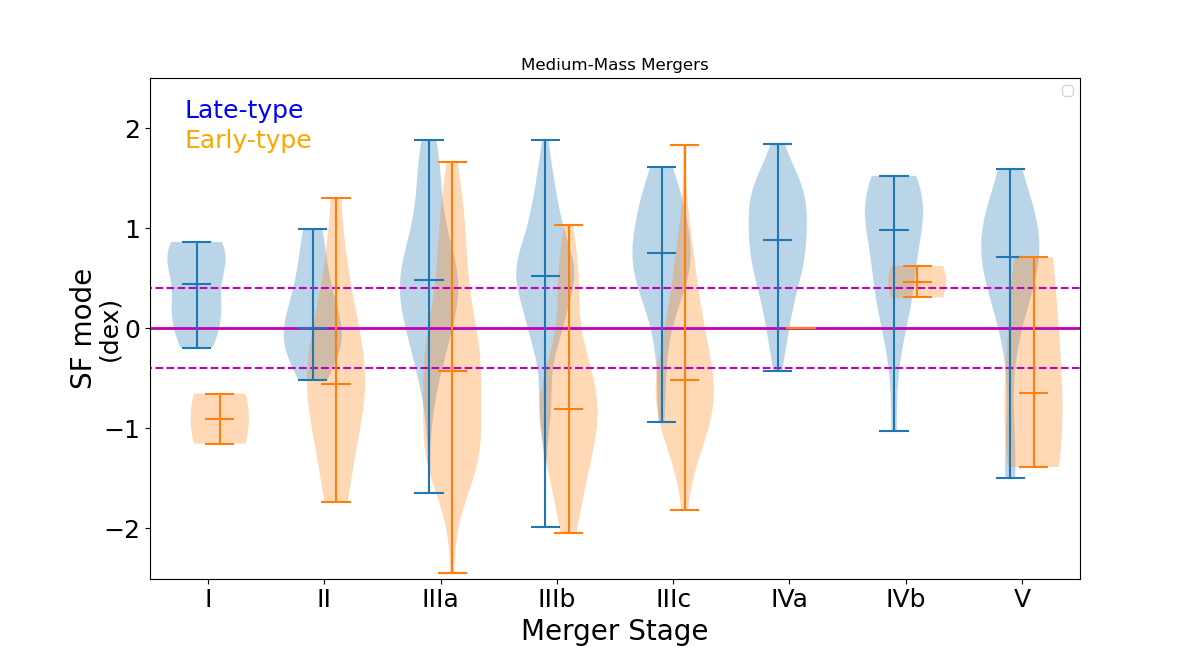}
  
\textbf{ Hi-M$_*$}\par\smallskip
  \includegraphics[bb=50 0 800 415, width=0.4\textwidth,clip]{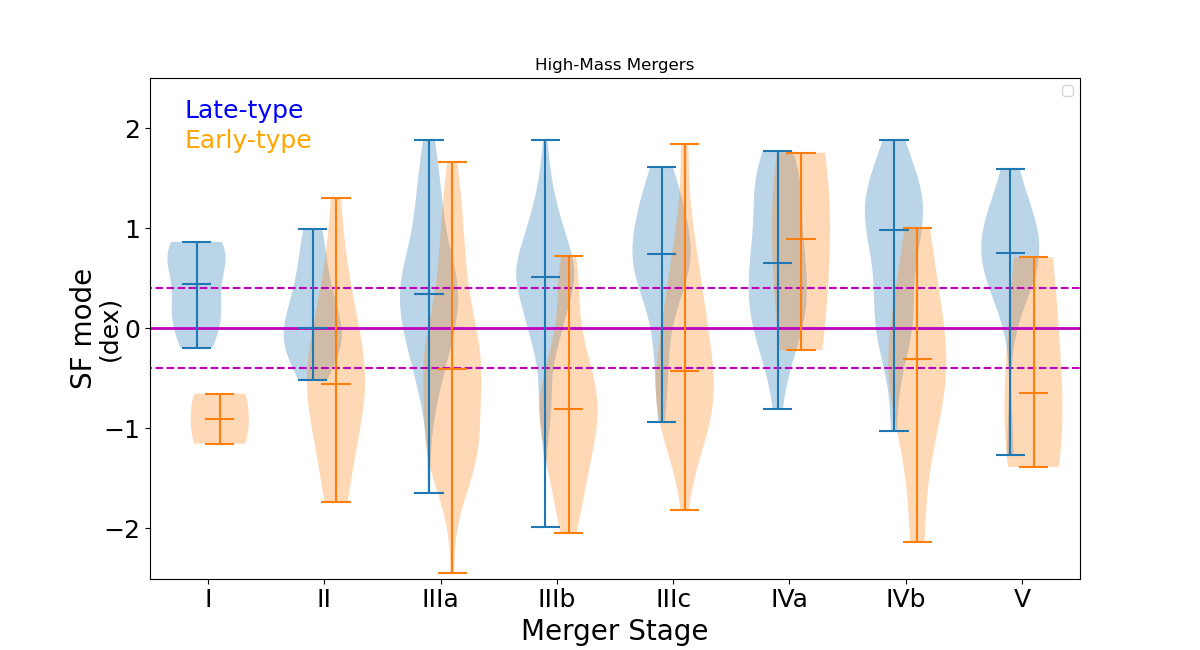}
   \caption{ SF mode distribution for each merger stage, coloured by late-(blue) and early-(orange) type. 
   The different sub-panels show the mergers separated by stellar mass bin: low-M$_*$ (top), medium-M$_*$ 
   (middle), and high-M$_*$ (bottom). 
}  
\label{fig:SFmode_dists_Mstellarbins}
\end{center}
\end{figure}

Similar to the bottom panel of Fig. \ref{fig:SFmode_Stg_violin_gral_lateearly}, 
Fig. \ref{fig:SFmode_dists_Mstellarbins} shows the SF mode distribution for each merger stage coloured by 
late-(blue) and early-(orange) type, and separated by stellar mass bin. From top to bottom: low-, medium-, and 
high-stellar mass bin. 

We notice the lack of early-type galaxies at low-mass mergers, as there are not mergers with early-type 
galaxies at these stellar masses in our parent sample. As we increase the stellar mass bin, early-type 
mergers appear but there is no strong difference on the trends seen in either of the three stellar mass bins.

 \begin{figure}[H]
\begin{center} 
\textbf{ Low-M$_*$}\par\smallskip
  \includegraphics[bb=200 150 2600 2600, width=0.24\textwidth,clip]{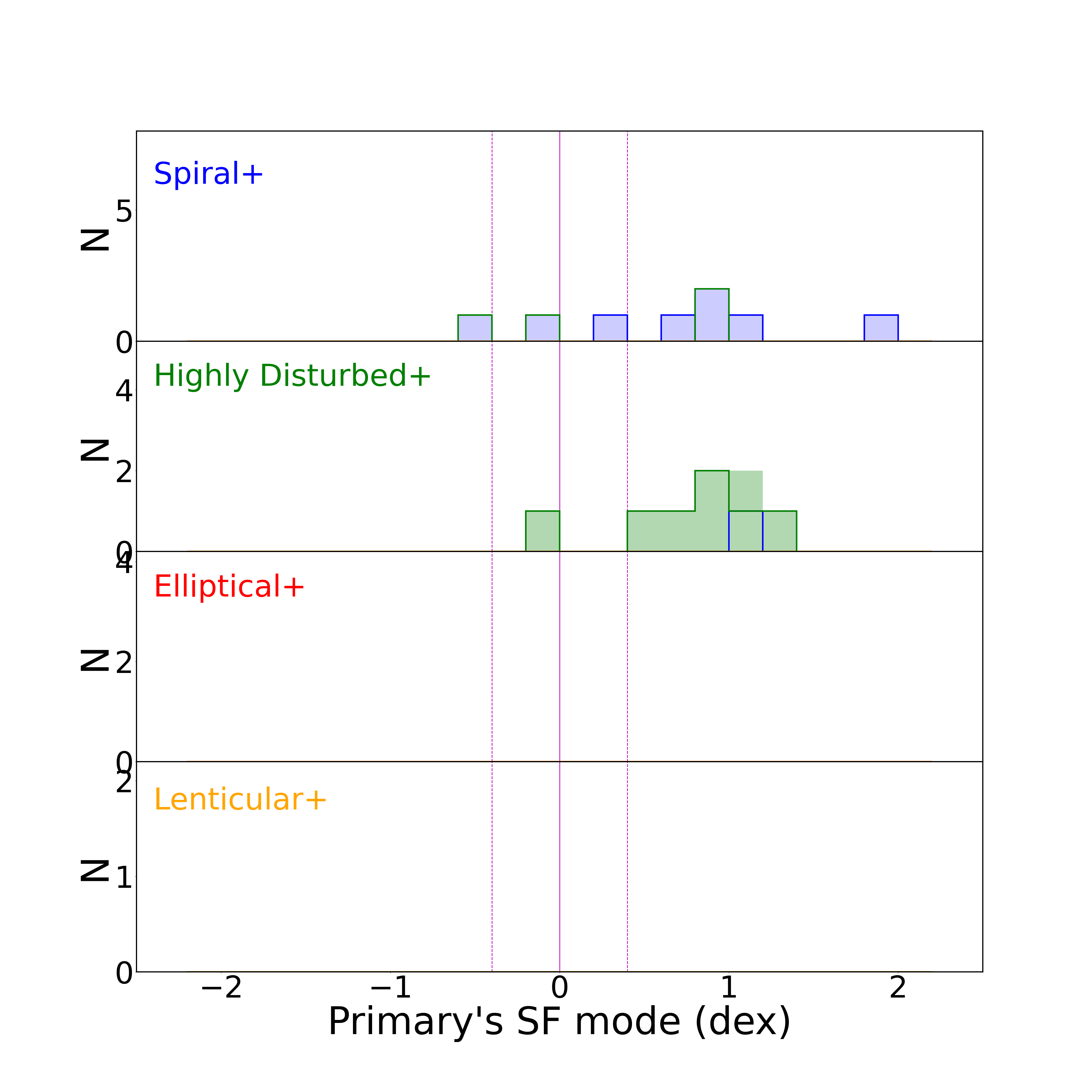}
  \includegraphics[bb=200 150 2600 2600, width=0.24\textwidth,clip]{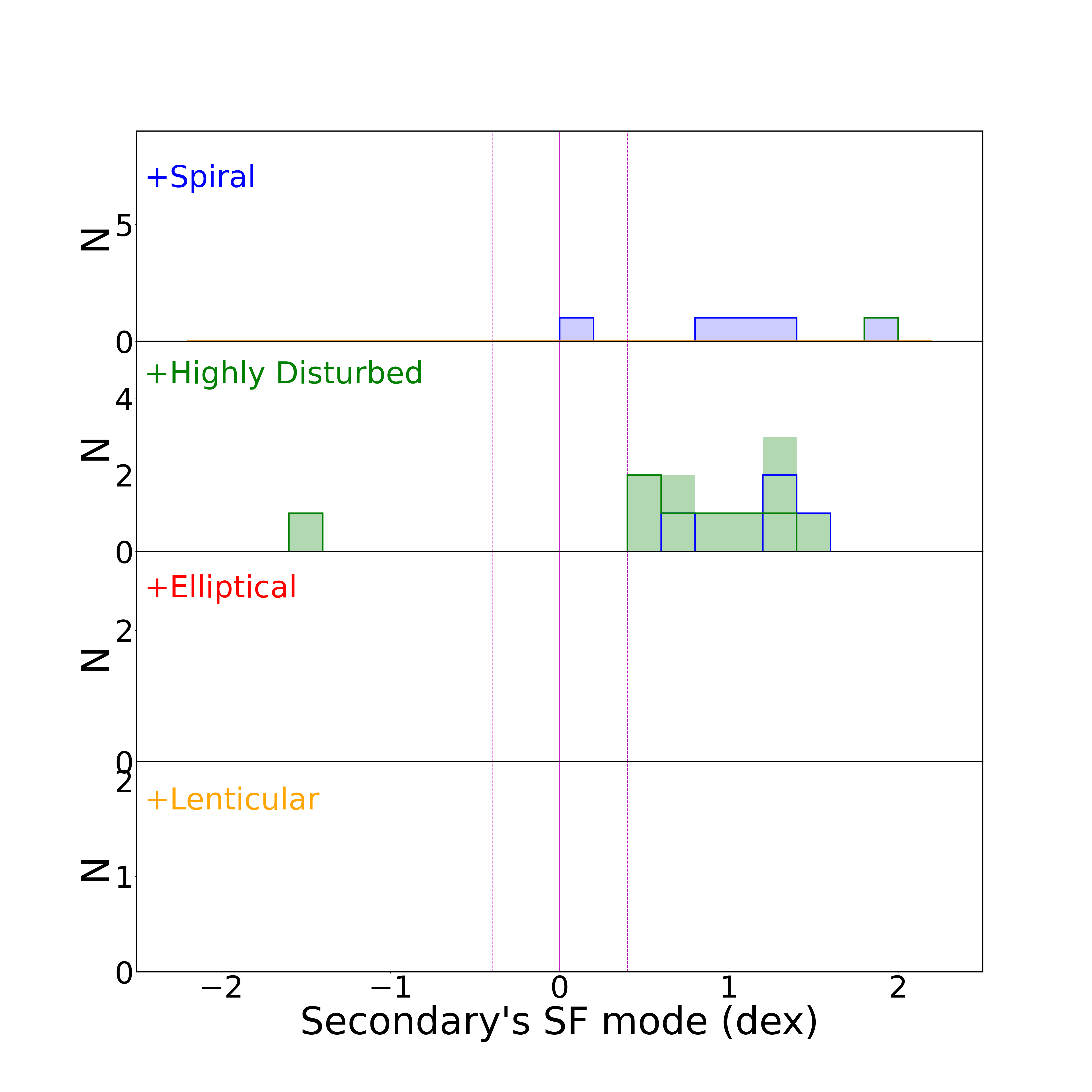}
\textbf{ Med-M$_*$}\par\smallskip
  \includegraphics[bb=200 150 2600 2600, width=0.24\textwidth,clip]{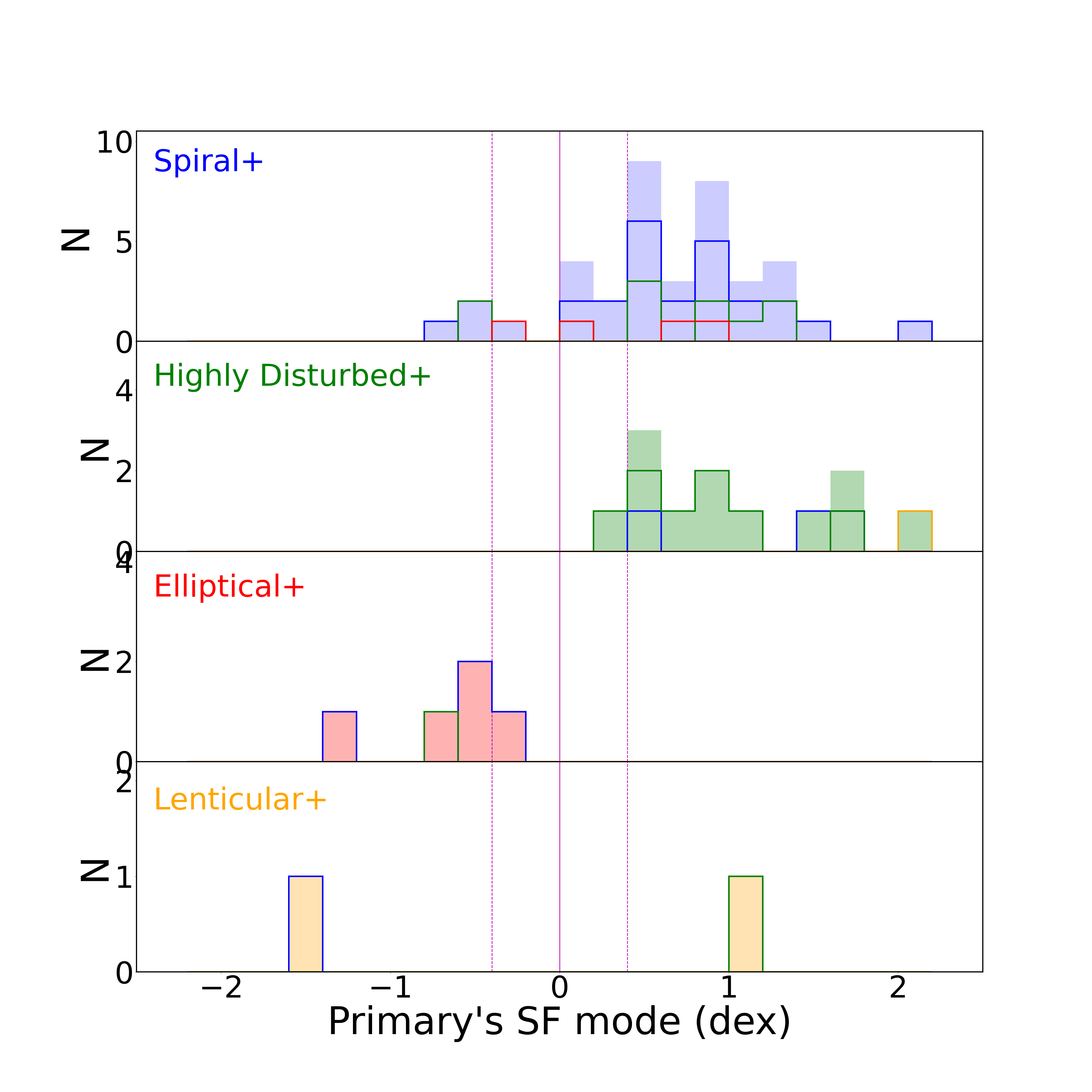}
  \includegraphics[bb=200 150 2600 2600, width=0.24\textwidth,clip]{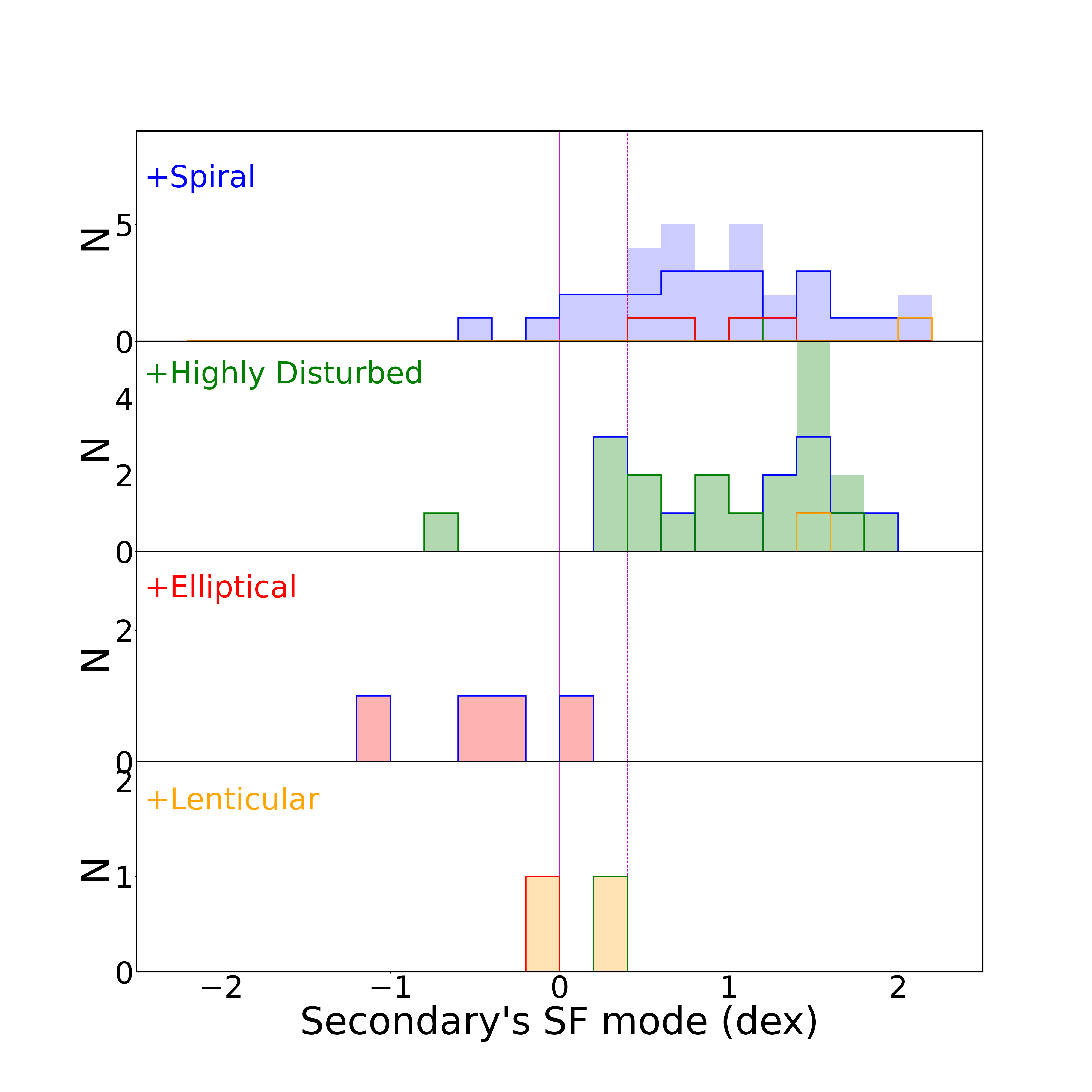}
\textbf{ Hi-M$_*$}\par\smallskip
  \includegraphics[bb=200 150 2600 2600, width=0.24\textwidth,clip]{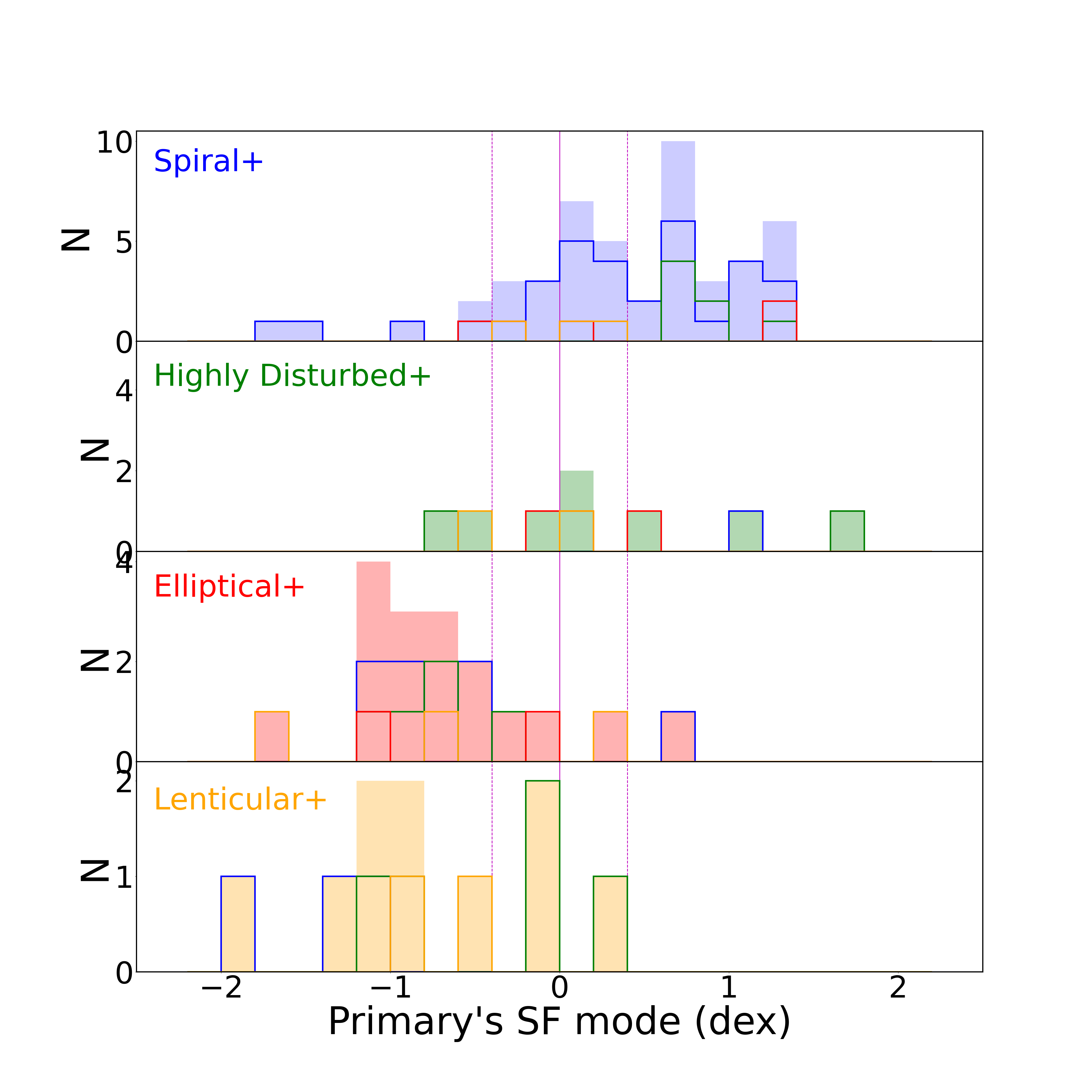}
  \includegraphics[bb=200 150 2600 2600, width=0.24\textwidth,clip]{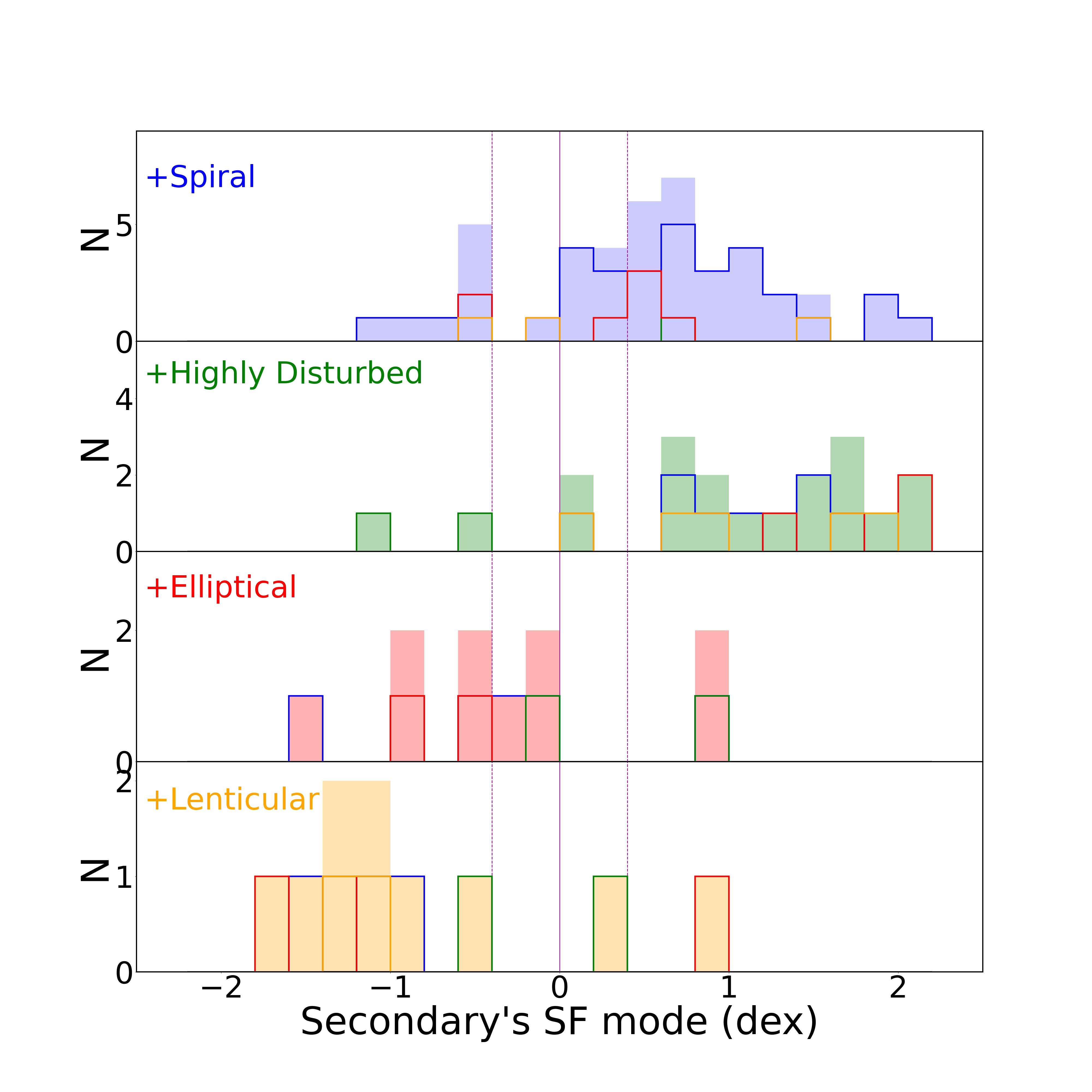}
   \caption{ SF mode distribution for the primary (left panels) and secondary (right panels) component, 
   separated by morphology. Top panels show the mergers with low-M$_*$, middle panels show the mergers 
   with medium-M$_*$, and the bottom panels show the mergers with high-M$_*$. 
}  
\label{fig:SFmode_histos_Masses}
\end{center}
\end{figure}

\clearpage

\section{Major and Minor Mergers} \label{App:MajorandMinor}

\begin{figure}[ht!]
\begin{center} 
  \includegraphics[width=0.45\textwidth,clip]{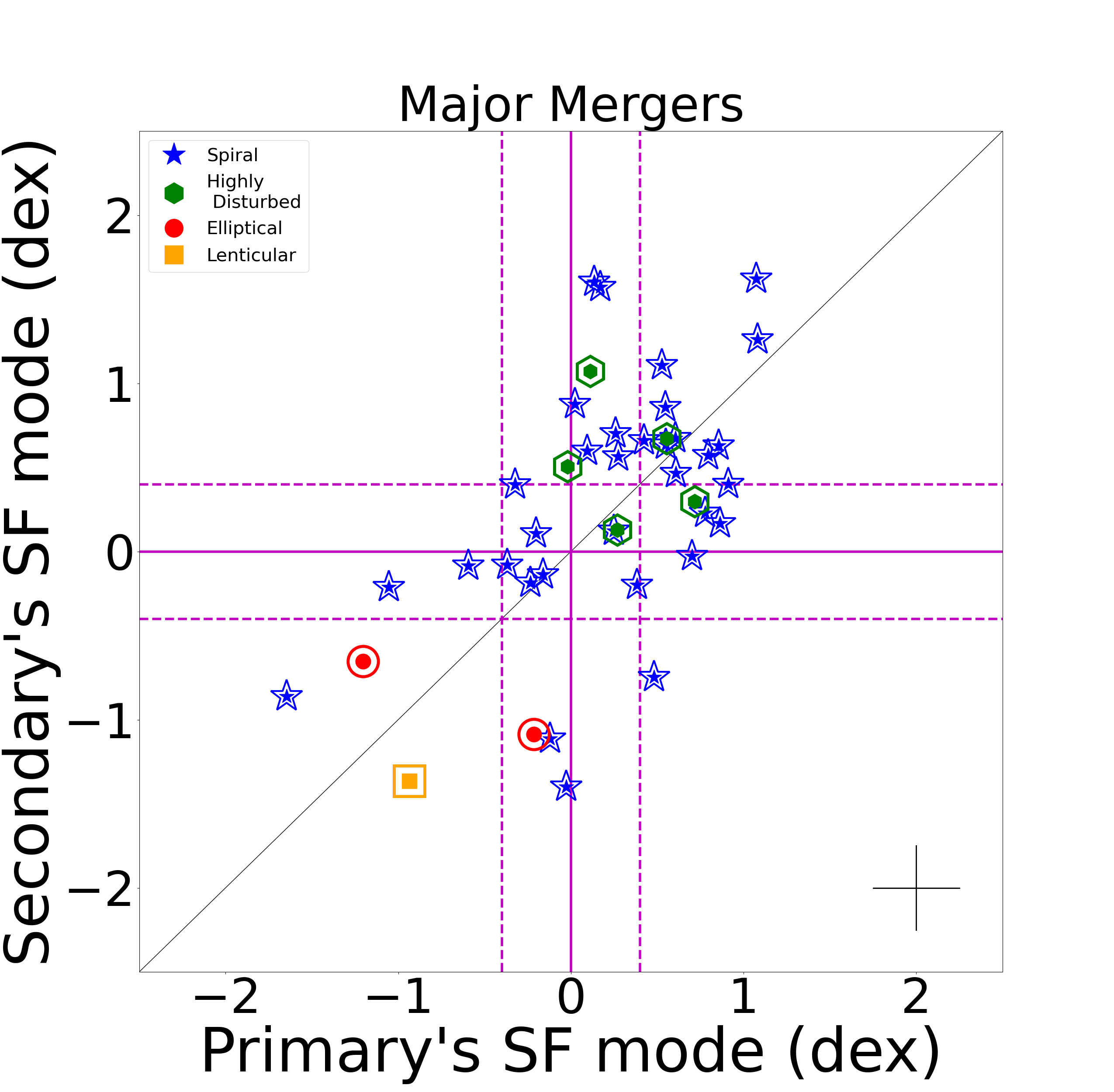}
  \includegraphics[width=0.45\textwidth,clip]{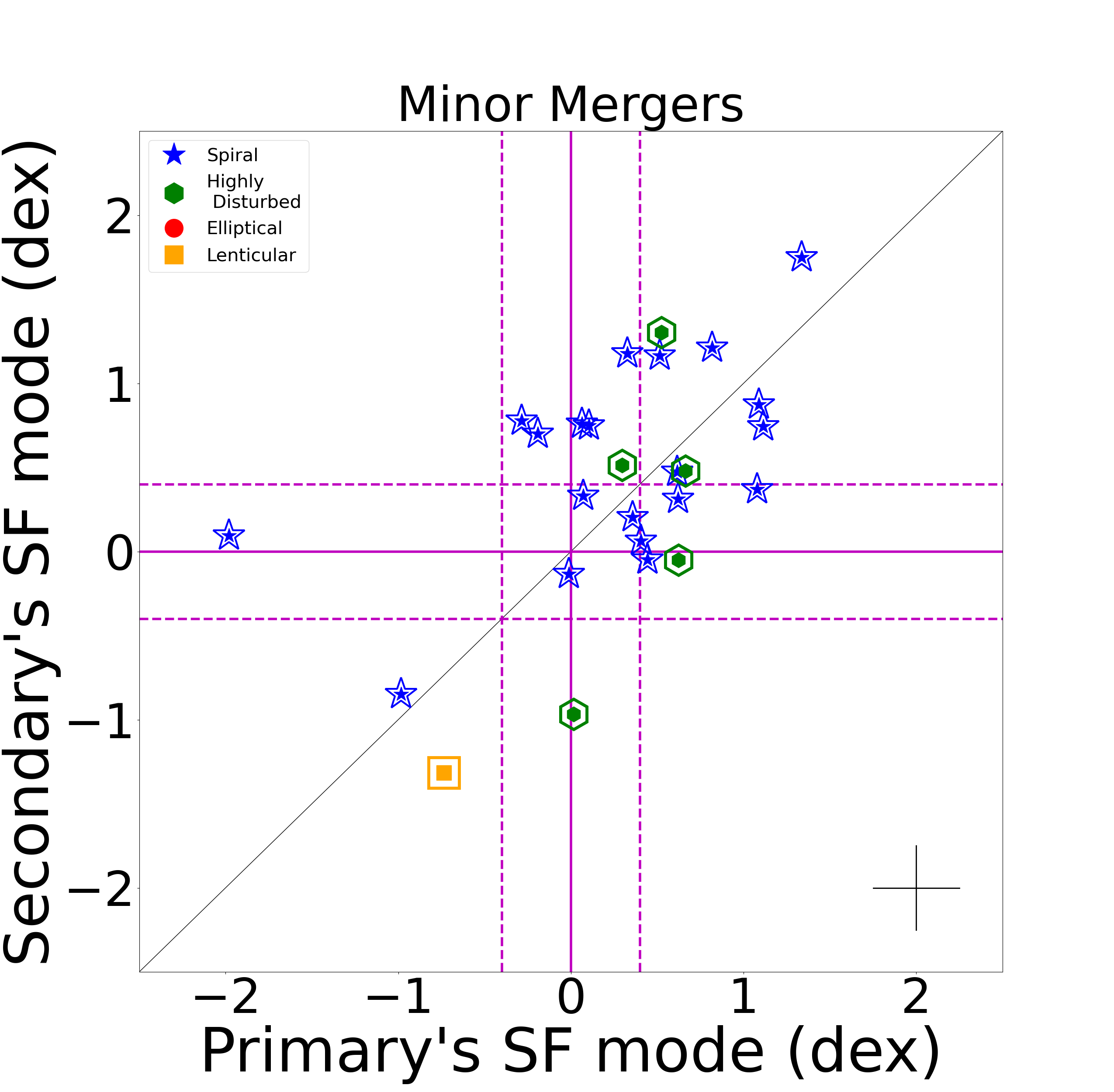}
   \caption{ Comparison of the SF mode of the primary and the SF mode of the secondary of mergers with 
   components of the same morphology. Top and bottom panels show major and minor mergers, respectively.  
   Coloured symbols show the morphology of the primary as unfilled symbols and the morphology of the 
   secondary as filled symbols. Morphologies are presented in the legend. The solid- and dashed-magenta lines 
   show the MS and the black line shows the one-to-one relation. 
}  
\label{fig:SFmode_PrimSec_MajorMinor}
\end{center}
\end{figure}

Similar to Fig: \ref{fig:SFmode_PrimSec_SameMorph} but separated by major and minor mergers. 
Figure. \ref{fig:SFmode_PrimSec_MajorMinor} shows the comparison between the SF mode of the primary and the 
SF mode of the secondary component. Big-unfilled symbols show the morphology of the primary and the small-filled 
symbols show the morphology of the secondary component as shown in the legend. Major and minor mergers are 
shown in the top and bottom panels, respectively. 
There is no clear dependency on the stellar mass ratio 
between the components affecting the SF mode for mergers with components of the same morphology. 
Thus, our results in Sec. \ref{sec:SFmode_Morphdep} are not affected by the stellar mass ratio between the 
merger components. 

\section{Comparison of AGN Identifiers } \label{AGNs_methods}

\begin{figure}[!htb]
\begin{center} 
  \includegraphics[width=0.4\textwidth,clip]{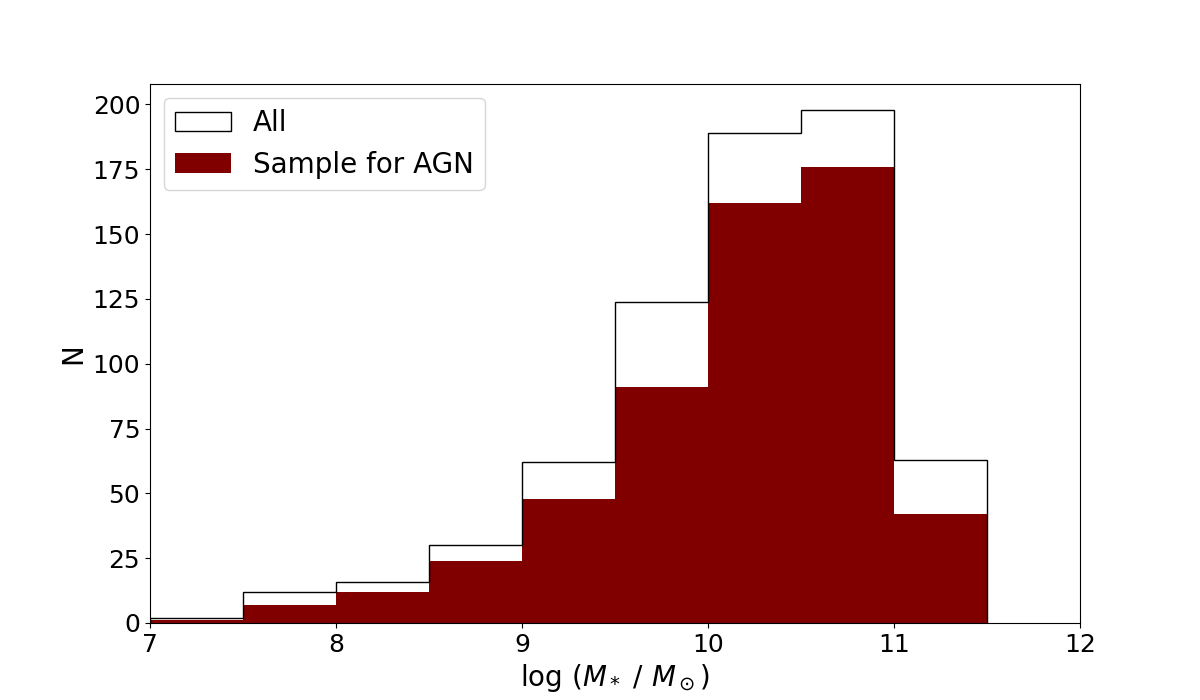}
  \includegraphics[width=0.4\textwidth,clip]{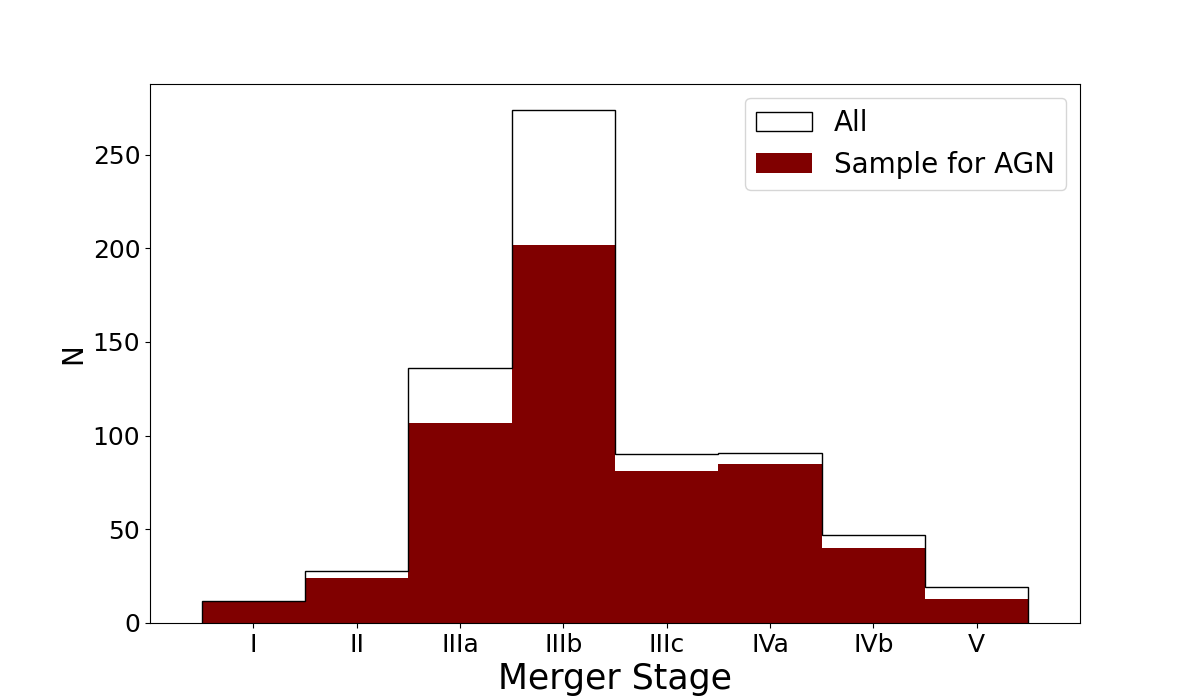}
   \caption{ Distribution of stellar mass (top panel) and merger stage (bottom panel) of the parent sample (in black) and the sub-sample used to classify AGNs (in maroon).
}  
\label{fig:AGNsample}
\end{center}
\end{figure}

Figure \ref{fig:AGNsample} shows the distribution of the sample with enough information to classify AGNs (in maroon) and the parent sample (in black). 

\begin{figure}[!htb]
\begin{center} 
  \includegraphics[width=0.3\textwidth,clip]{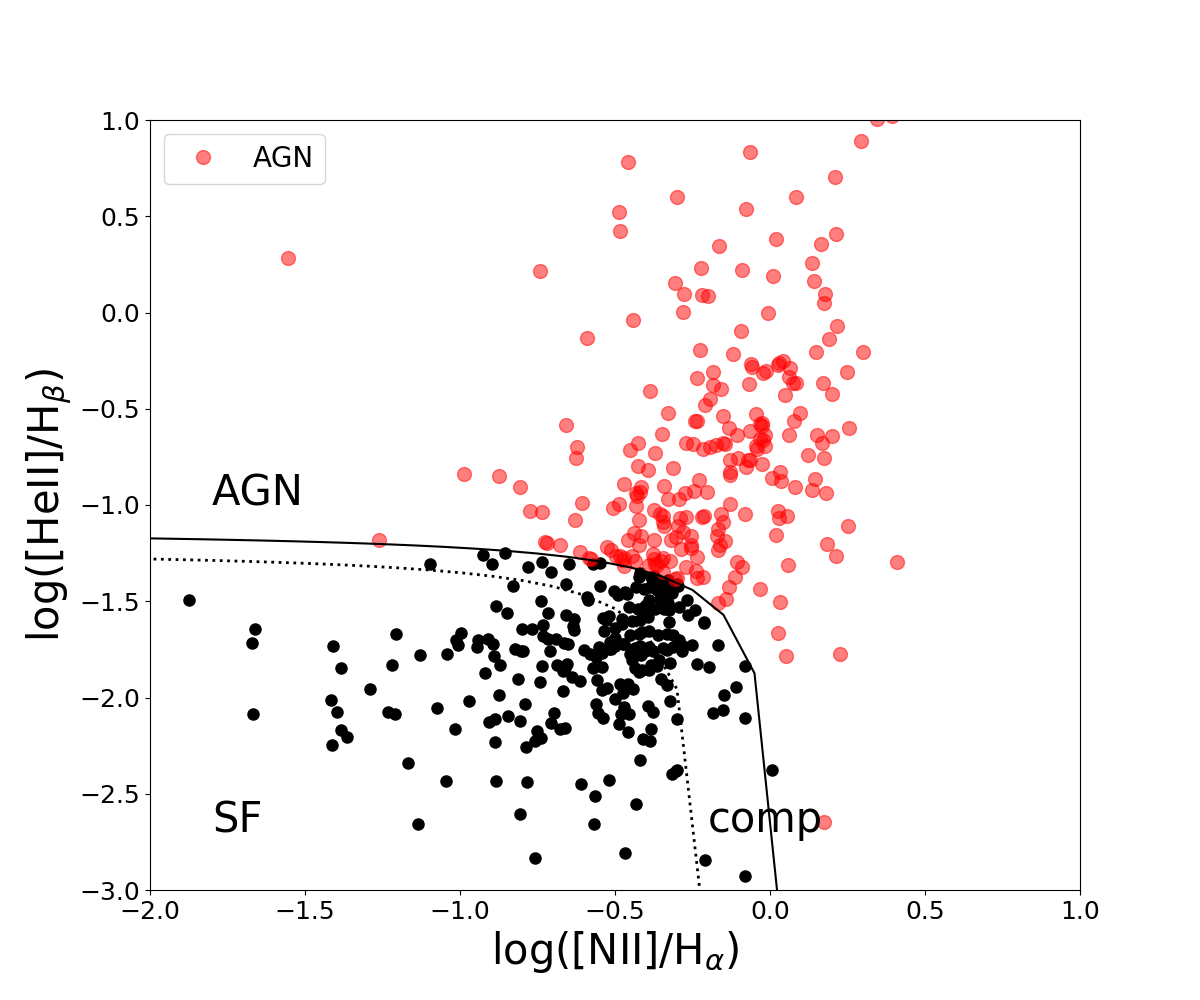}
  \includegraphics[width=0.3\textwidth,clip]{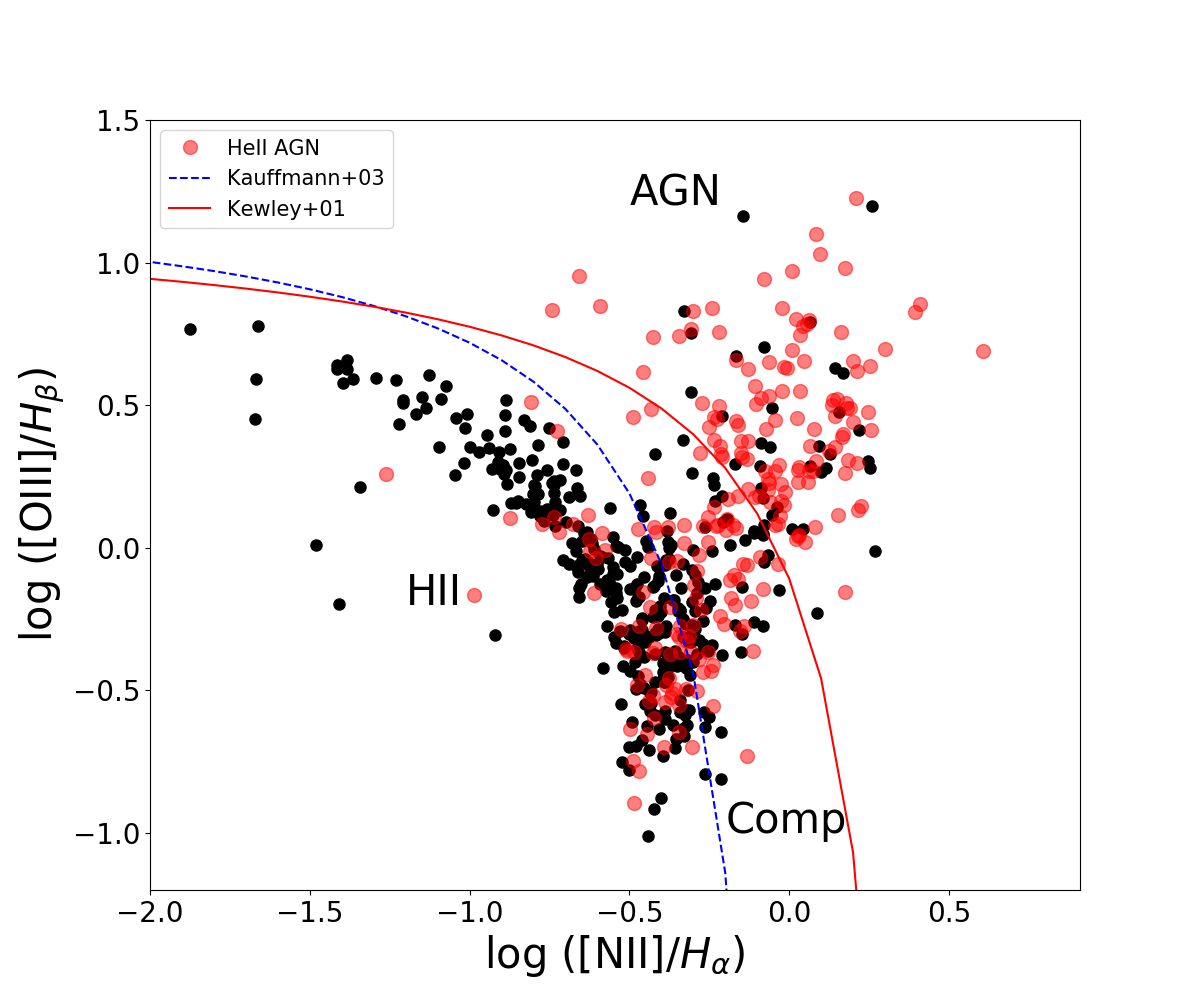}
  \includegraphics[width=0.3\textwidth,clip]{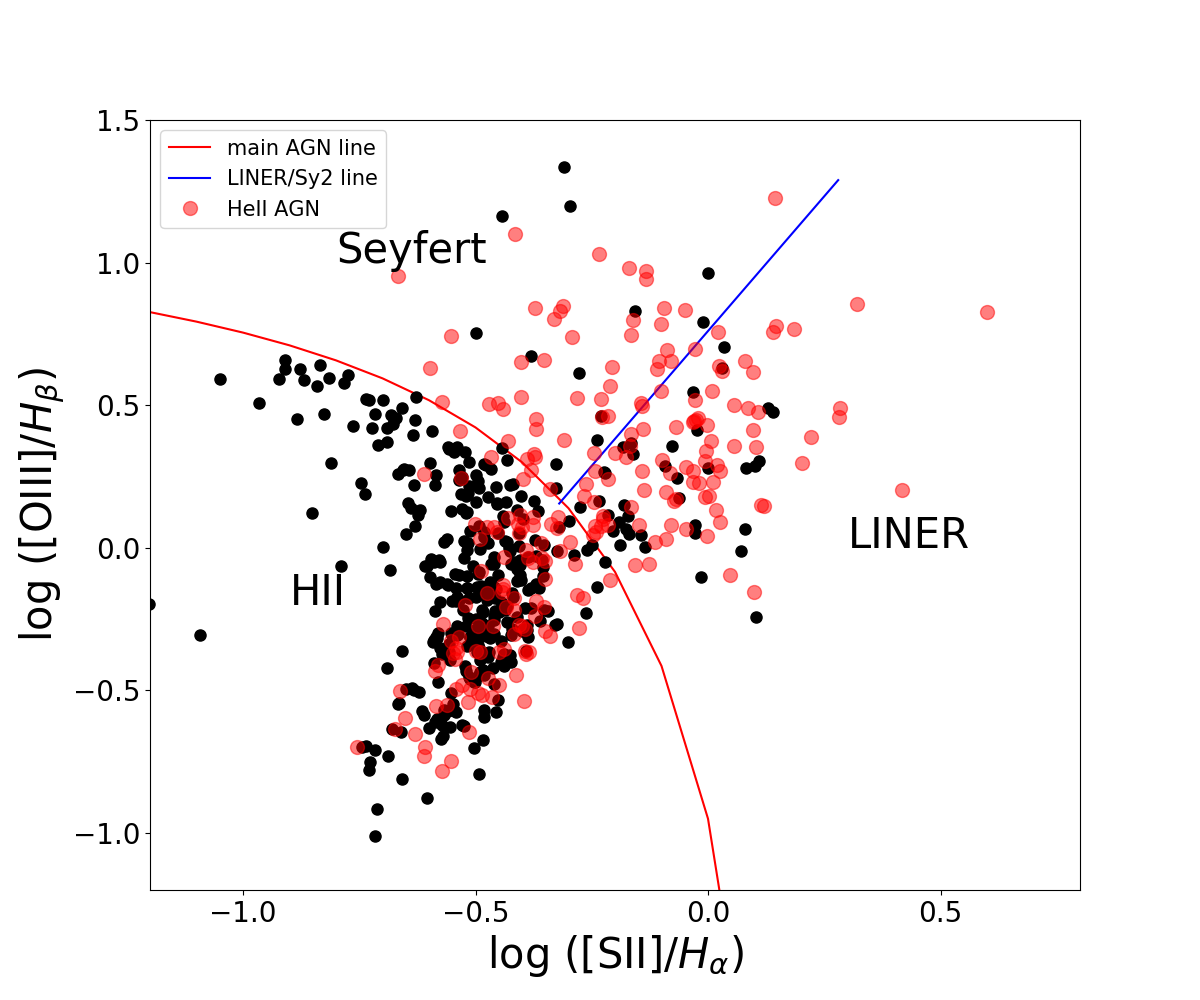}
  \includegraphics[width=0.295\textwidth,clip]{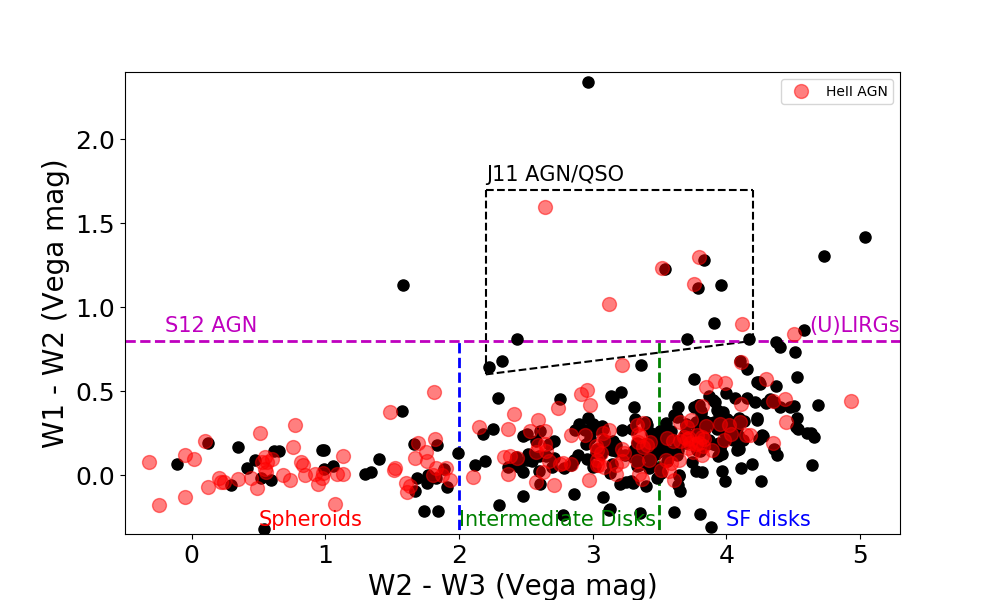}
   \caption{ HeII-classified AGNs (red circles) shown in the HeII (top), BPT-NII (second), BPT-SII 
   (third), and in the WISE (bottom) diagram.
}  
\label{fig:HeII_AGNs}
\end{center}
\end{figure}

As mentioned previously, we have used three different methods to identify AGNs in mergers. 
These three methods do not identify the same mergers as they measure different properties of the AGN. 
In order to show how different these methods perform for mergers, we show where the AGNs identified 
by one method locate in the other methods diagrams. In all four panels of Figure \ref{fig:HeII_AGNs}, red dots represent AGN classified using the HeII-diagram (from top to bottom: HeII, BPT-NII, BPT-SII, and WISE). We se3 that AGNs classified using the HeII diagram are very scattered across the other diagrams, and so would be frequently classified as non-AGN using those methods.

\end{appendix}

\end{document}